\pdfoutput=1
\documentclass[11pt,twoside,a4paper,cmspaper,final,collab]{cms-tdr}
\def\svnVersion{bfd8cbe}\def\svnDate{2022/03/10}\def\cmsCernNoTag{CERN-EP-2021-219}\def\cmsCernDate{\today}\def\cmsMessage{Published in Physical Review D as \href{http://dx.doi.org/10.1103/PhysRevD.105.052003}{\doi{10.1103/PhysRevD.105.052003}.}}
\begin{document}\cmsNoteHeader{SMP-20-005}

\newlength\cmsFigWidthZ
\newlength\cmsFigWidthY
\newlength\cmsFigWidthX
\newlength\cmsFigWidthW
\newlength\cmsFigWidthV
\newlength\cmsFigWidthU
\newlength\cmsFigWidthT
\newlength\cmsFigWidthS
\newlength\cmsFigWidthR
\newlength\cmsFigWidthQ
\newlength\cmsFigWidthP
\ifthenelse{\boolean{cms@external}}{\setlength\cmsFigWidthZ{0.20\textwidth}}{\setlength\cmsFigWidthZ{0.28\textwidth}}
\ifthenelse{\boolean{cms@external}}{\setlength\cmsFigWidthY{0.23\textwidth}}{\setlength\cmsFigWidthY{0.33\textwidth}}
\ifthenelse{\boolean{cms@external}}{\setlength\cmsFigWidthX{0.40\textwidth}}{\setlength\cmsFigWidthX{0.60\textwidth}}
\ifthenelse{\boolean{cms@external}}{\setlength\cmsFigWidthW{0.41\textwidth}}{\setlength\cmsFigWidthW{0.49\textwidth}}
\ifthenelse{\boolean{cms@external}}{\setlength\cmsFigWidthV{0.45\textwidth}}{\setlength\cmsFigWidthV{0.60\textwidth}}
\ifthenelse{\boolean{cms@external}}{\setlength\cmsFigWidthU{0.41\textwidth}}{\setlength\cmsFigWidthU{0.48\textwidth}}
\ifthenelse{\boolean{cms@external}}{\setlength\cmsFigWidthT{0.90\textwidth}}{\setlength\cmsFigWidthT{0.98\textwidth}}
\ifthenelse{\boolean{cms@external}}{\setlength\cmsFigWidthS{0.49\textwidth}}{\setlength\cmsFigWidthS{0.70\textwidth}}
\ifthenelse{\boolean{cms@external}}{\setlength\cmsFigWidthR{0.49\textwidth}}{\setlength\cmsFigWidthR{0.80\textwidth}}
\ifthenelse{\boolean{cms@external}}{\setlength\cmsFigWidthQ{0.80\textwidth}}{\setlength\cmsFigWidthQ{0.98\textwidth}}
\ifthenelse{\boolean{cms@external}}{\setlength\cmsFigWidthP{0.60\textwidth}}{\setlength\cmsFigWidthP{0.80\textwidth}}

\ifthenelse{\boolean{cms@external}}{\providecommand{\cmsLeft}{upper\xspace}}{\providecommand{\cmsLeft}{left\xspace}}
\ifthenelse{\boolean{cms@external}}{\providecommand{\cmsRight}{lower\xspace}}{\providecommand{\cmsRight}{right\xspace}}
\providecommand{\cmsTable}[1]{\resizebox{\textwidth}{!}{#1}}
\ifthenelse{\boolean{cms@external}}{\providecommand{\NA}{\ensuremath{\cdots}\xspace}}{\providecommand{\NA}{\ensuremath{\text{---}}\xspace}}
\ifthenelse{\boolean{cms@external}}{\providecommand{\CL}{C.L.\xspace}}{\providecommand{\CL}{CL\xspace}}
\ifthenelse{\boolean{cms@external}}{}{%
\renewenvironment{scotch}[1]{\protect\centering\begin{tabular}{#1}\\[-3ex]\hline}{\\[-2.5ex]\hline\end{tabular}}
}
\hyphenation{ATLAS}
\newcommand{\fbns}{\ensuremath{\text{fb}}\xspace}
\newcommand{\WGamma}{\ensuremath{\PWpm\PGg}\xspace}
\newcommand{\WWGamma}{\ensuremath{\PW\PW\PGg}\xspace}
\newcommand{\OWWW}{\ensuremath{\mathcal{O}_{3W}}\xspace}
\newcommand{\WWV}{\ensuremath{\PW\PW\PV}\xspace}
\newcommand{\WPlusGamma}{\ensuremath{\PWp\PGg}\xspace}
\newcommand{\PhiGen}{\ensuremath{\phi^{\text{gen}}}\xspace}
\newcommand{\PhiTrue}{\ensuremath{\phi^{\text{true}}}\xspace}
\newcommand{\mTcluster}{\ensuremath{m_{\mathrm{T}}^{\text{cluster}}}\xspace}
\newcommand{\Ich}{\ensuremath{I_{\text{ch}}}\xspace}
\newcommand{\sigeta}{\ensuremath{\sigma_{\eta\eta}}\xspace}
\newcommand{\MATRIX}{\textsc{matrix}\xspace}
\providecommand{\GENEVA}{\textsc{geneva}\xspace}
\newcommand{\MGSHORT}{\textsc{mg5}\_a\textsc{mc}\xspace}
\newcommand{\MGPYSHORT}{\MGSHORT{}+\textsc{py8}\xspace}
\newcommand{\pp}{\ensuremath{\Pp\Pp}\xspace}
\newcommand{\ptG}{\ensuremath{\pt^{\PGg}}\xspace}
\newcommand{\ptL}{\ensuremath{\pt^{\Pell}}\xspace}
\newcommand{\ptE}{\ensuremath{\pt^{\Pe}}\xspace}
\newcommand{\ptN}{\ensuremath{\pt^{\PGn}}\xspace}
\newcommand{\ptVecN}{\ensuremath{\vec{p}_{\mathrm{T}}^{\PGn}}\xspace}
\newcommand{\aeta}{\ensuremath{\abs{\eta}}\xspace}
\newcommand{\etaG}{\ensuremath{\eta^{\PGg}}\xspace}
\newcommand{\etaL}{\ensuremath{\eta^{\Pell}}\xspace}
\newcommand{\etaE}{\ensuremath{\eta^{\Pe}}\xspace}
\newcommand{\etaN}{\ensuremath{\eta^{\PGn}}\xspace}
\newcommand{\etaM}{\ensuremath{\eta^{\PGm}}\xspace}
\newcommand{\Deta}{\ensuremath{\Delta\eta}\xspace}
\newcommand{\DetaLG}{\ensuremath{\Deta(\Pell,\PGg)}\xspace}
\newcommand{\OPi}{\ensuremath{\mathcal{O}_{i}}\xspace}
\newcommand{\CWWW}{\ensuremath{C_{3W}}\xspace}
\newcommand{\sigSM}{\ensuremath{\sigma^{\text{SM}}}\xspace}
\newcommand{\sigInt}{\ensuremath{\sigma^{\text{int}}}\xspace}
\newcommand{\sigBSM}{\ensuremath{\sigma^{\text{BSM}}}\xspace}
\newcommand{\muInt}{\ensuremath{\mu^{\text{int}}}\xspace}
\newcommand{\muBSM}{\ensuremath{\mu^{\text{BSM}}}\xspace}
\newcommand{\mW}{\ensuremath{m_{\PW}}\xspace}
\newcommand{\fPlu}{\ensuremath{f_{+}}\xspace}
\newcommand{\fMin}{\ensuremath{f_{-}}\xspace}
\newcommand{\TwoToTwo}{\ensuremath{2\to2}\xspace}
\newcommand{\HoverE}{\ensuremath{H/E}\xspace}
\newcommand{\IchMax}{\ensuremath{I_{\text{ch}}^{\text{max}}}\xspace}
\newcommand{\ZtoLL}{\ensuremath{\PZ\to\Pell\Pell}\xspace}
\newcommand{\ZtoEE}{\ensuremath{\PZ\to\Pe\Pe}\xspace}
\newcommand{\ZtoMM}{\ensuremath{\PZ\to\PGm\PGm}\xspace}
\newcommand{\mLG}{\ensuremath{m_{\Pell\PGg}}\xspace}
\newcommand{\mTLmiss}{\ensuremath{\mT(\Pell,\ptmiss)}\xspace}
\newcommand{\LNG}{\ensuremath{\Pell\PGn\PGg}\xspace}
\newcommand{\pTLG}{\ensuremath{p_{\mathrm{T},\Pell\PGg}}\xspace}
\newcommand{\pTLGmiss}{\ensuremath{p_{\mathrm{T},\Pell\PGg\text{miss}}}\xspace}
\newcommand{\phif}{\ensuremath{\phi_{f}}\xspace}
\newcommand{\aphif}{\ensuremath{\abs{\phif}}\xspace}
\newcommand{\Nmisid}{\ensuremath{N_{\text{misid}}}\xspace}
\newcommand{\fGam}{\ensuremath{f_{\PGg}}\xspace}
\newcommand{\fL}{\ensuremath{f_{\Pell}}\xspace}
\newcommand{\Nprompt}{\ensuremath{N_{\text{sim}}^{\text{prompt-}\PGg}}\xspace}
\newcommand{\muR}{\ensuremath{\mu_{\mathrm{R}}}\xspace}
\newcommand{\muF}{\ensuremath{\mu_{\mathrm{F}}}\xspace}
\newcommand{\valpha}{\ensuremath{\vec{\alpha}}\xspace}
\newcommand{\vtheta}{\ensuremath{\vec{\theta}}\xspace}
\newcommand{\qStat}{\ensuremath{q(\valpha)}\xspace}
\newcommand{\hvalpha}{\ensuremath{\hat{\valpha}}\xspace}
\newcommand{\hvtheta}{\ensuremath{\hat{\vtheta}}\xspace}
\newcommand{\hvthetac}{\ensuremath{\hvtheta_{\valpha}}\xspace}
\newcommand{\tvtheta}{\ensuremath{\widetilde{\vtheta}}\xspace}
\newcommand{\sigj}{\ensuremath{\sigma_{j}}\xspace}
\newcommand{\Rij}{\ensuremath{R_{ij}}\xspace}
\newcommand{\sigjmiss}{\ensuremath{\sigj^{\text{miss}}}\xspace}
\newcommand{\sigooa}{\ensuremath{\sigma^{\text{OOA}}}\xspace}
\newcommand{\sigjmissmc}{\ensuremath{\sigma_{j}^{\text{miss}, \text{MC}}}\xspace}
\newcommand{\sigjmc}{\ensuremath{\sigj^{\text{MC}}}\xspace}
\newcommand{\sigtot}{\ensuremath{\sigma_{\text{tot}}}\xspace}
\newcommand{\fracj}{\ensuremath{f_{j}}\xspace}
\newcommand{\DRLG}{\ensuremath{\DR(\Pell,\PGg)}\xspace}
\newcommand{\qT}{\ensuremath{q_{\mathrm{T}}}\xspace}

\cmsNoteHeader{SMP-20-005}
\title{Measurement of \texorpdfstring{\WGamma}{Wgamma} differential cross sections in proton-proton collisions at \texorpdfstring{$\sqrt{s}=13\TeV$}{sqrt(s) = 13 TeV} and effective field theory constraints}

\date{\today}

\abstract{
Differential cross section measurements of \WGamma production in proton-proton collisions at $\sqrt{s} = 13\TeV$ are presented. The data set used in this study was collected with the CMS detector at the CERN LHC in 2016--2018 with an integrated luminosity of 138\fbinv. Candidate events containing an electron or muon, a photon, and missing transverse momentum are selected. The measurements are compared with standard model predictions computed at next-to-leading and next-to-next-to-leading orders in perturbative quantum chromodynamics. Constraints on the presence of \TeVns-scale new physics affecting the \WWGamma vertex are determined within an effective field theory framework, focusing on the \OWWW operator.  A simultaneous measurement of the photon transverse momentum and the azimuthal angle of the charged lepton in a special reference frame is performed. This two-dimensional approach provides up to a factor of ten more sensitivity to the interference between the standard model and the \OWWW contribution than using the transverse momentum alone.
}

\hypersetup{
pdfauthor={CMS Collaboration},
pdftitle={Measurement of Wgamma differential cross sections in proton-proton collisions at sqrt(s) = 13 TeV and effective field theory constraints},
pdfsubject={CMS},
pdfkeywords={CMS,  effective field theory, Wgamma production}}

\maketitle

\section{Introduction}\label{sec:intro}
The measurement of the properties of vector boson pair production is an important test of the electroweak sector of the standard model (SM).
The production cross sections of these processes are sensitive to the \WWV triple gauge couplings (TGCs), where $\PV = \PGg$ or \PZ.
Because the strengths of these TGCs are predicted precisely in the SM, the measurement of an anomalous value would indicate the presence of physics beyond the SM (BSM).

This paper reports an analysis of \WGamma production in proton-proton (\pp) collisions at $\sqrt{s}=13\TeV$ using data recorded with the CMS detector at the CERN LHC in 2016--2018 with an integrated luminosity of 138\fbinv.
This process has previously been studied by the CDF and D0 Collaborations~\cite{Acosta:2004it,Abazov:2005ni,Abazov:2008ad} at the Fermilab Tevatron in $\Pp\Pap$ collisions at $\sqrt{s} = 1.96\TeV$, and by the CMS~\cite{Chatrchyan:2013fya} and ATLAS~\cite{Aad:2013izg} Collaborations using \pp collision data collected at $\sqrt{s} = 7\TeV$.
The CMS Collaboration has also performed the first \WGamma measurement at $\sqrt{s} = 13\TeV$~\cite{CMS:2021foa}.
All of these studies found the  measured inclusive cross sections to be compatible with the SM predictions and set limits on the presence of anomalous TGCs.

\begin{figure*}[hbtp]
\centering
\includegraphics[width=\cmsFigWidthZ]{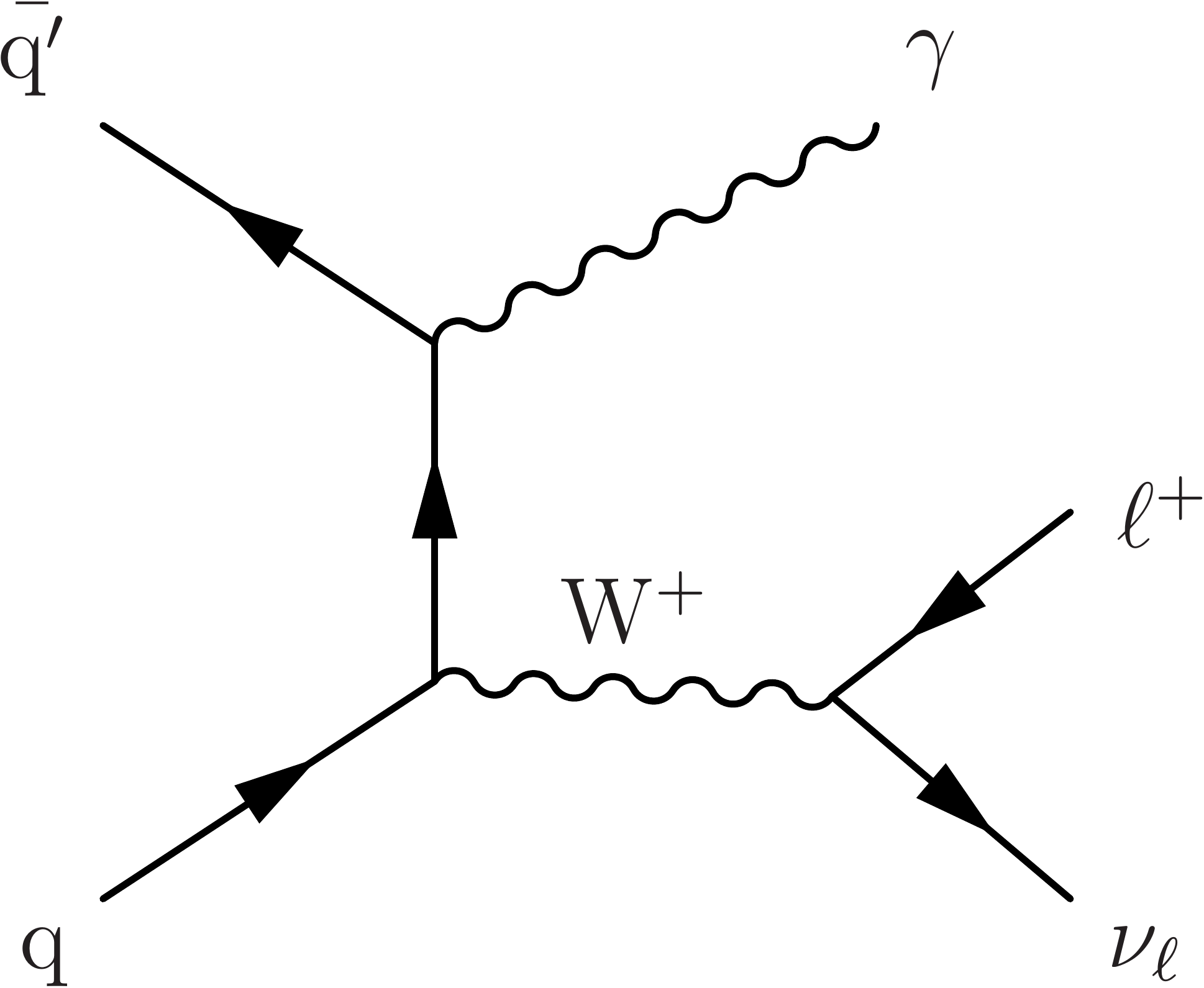}\quad
\includegraphics[width=\cmsFigWidthY]{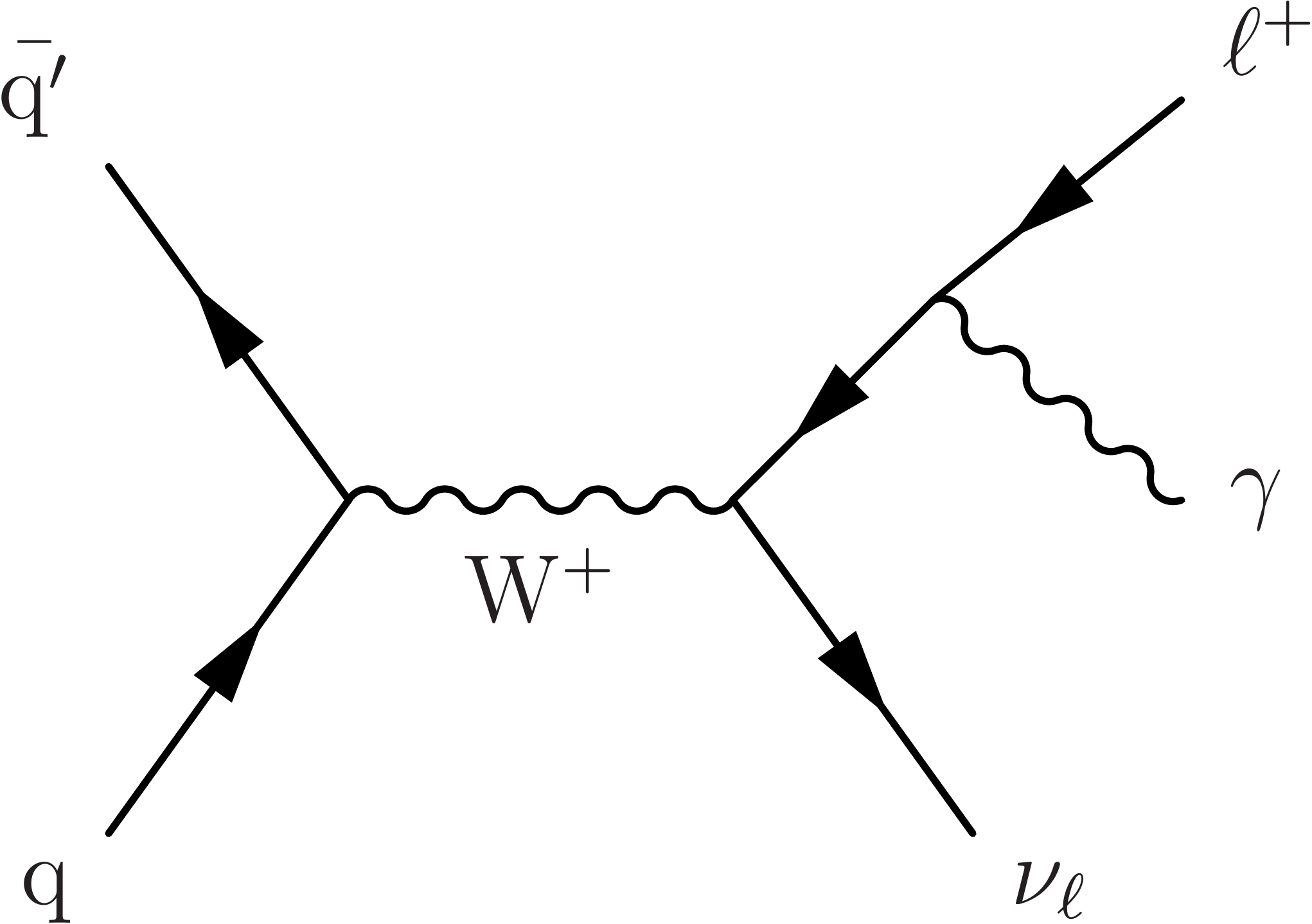}\quad
\includegraphics[width=\cmsFigWidthY]{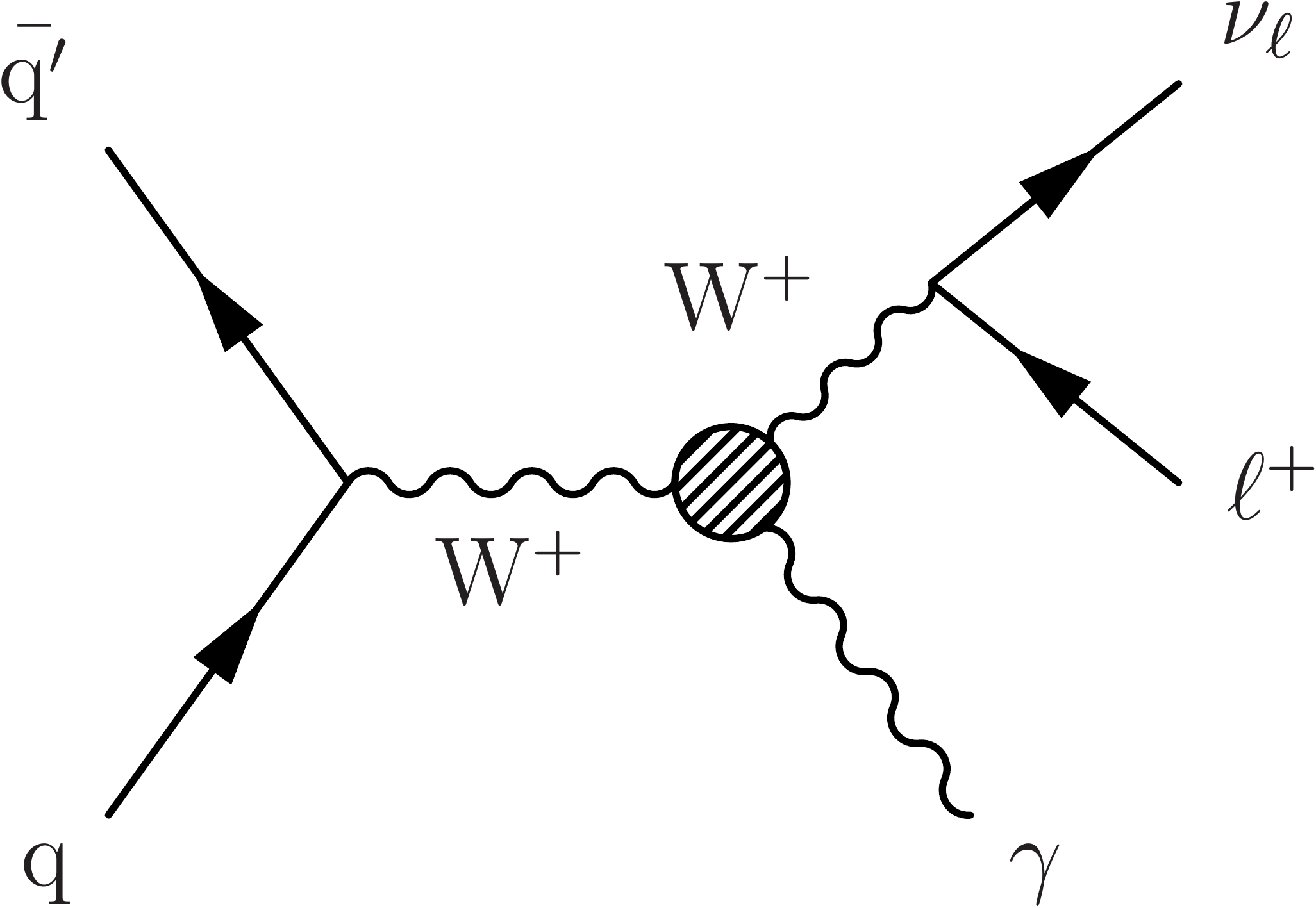}
\caption{LO Feynman diagrams for \WPlusGamma production showing initial-state (left) and final-state (center) radiation of the photon, and the \WWGamma TGC process (right).}\label{fig:feynman}
\end{figure*}

In the present analysis, \WGamma events are selected with one charged lepton (\Pell), which is either an electron or muon, one neutrino (\PGn), and one photon (\PGg) in the final state.
A few leading order (LO) Feynman diagrams are shown in Fig.~\ref{fig:feynman}.
The photon can be produced as a result of either initial- or final-state radiation, in addition to the contribution involving a TGC vertex.

Differential cross sections are measured for several observables, including the transverse momentum \ptG and pseudorapidity \etaG of the photon, and the transverse mass of the \LNG system, denoted \mTcluster. The cross section is also measured as a function of the number of additional jets.
The measured values are compared with the SM predictions at next-to-leading order (NLO) and next-to-next-to-leading order (NNLO) in perturbative quantum chromodynamics (QCD).
Interference between the LO \WGamma production diagrams results in a cross section that vanishes in specific phase space regions.
This effect is known as a radiation amplitude zero (RAZ)~\cite{Mikaelian:1979nr,Goebel:1980es,Brodsky:1982sh,Brown:1982xx,Baur:1994sa}.
A differential cross section measurement of the pseudorapidity difference between the lepton and the photon, \DetaLG, is used to explore this effect.

Constraints on BSM contributions to the \WWGamma vertex are determined in an effective field theory (EFT) framework.
For the first time in the study of this process, both \ptG and the angular properties of the final-state particles are exploited to increase the sensitivity to the interference between the SM and BSM amplitudes.
This novel approach is referred to as ``interference resurrection''~\cite{Panico:2017frx,Azatov:2017kzw}.

The paper is organized as follows.
The interference resurrection technique is described in Section~\ref{sec:intres}.
The CMS detector, data samples, and event simulation are summarized in Sections~\ref{sec:detector} and~\ref{sec:datasets}.
The object reconstruction and event selection are described in Sections~\ref{sec:event_reco} and~\ref{sec:event_sel}.
The estimation of the main backgrounds is given in Section~\ref{sec:backgrounds}.
The systematic uncertainties are discussed in Section~\ref{sec:systs}, the results presented in Section~\ref{sec:results}, and the paper summarized in Section~\ref{sec:summary}.

\section{Interference resurrection}\label{sec:intres}
An EFT approach can be used to study how new physics entering at an energy scale $\Lambda$, assumed to be much larger than the electroweak scale, leads to deviations from the SM at an energy regime accessible at the LHC.\@
The SM EFT is constructed by the addition of higher-dimensional operators, \OPi, to the SM Lagrangian,
\begin{linenomath*}\begin{equation}
\mathcal{L}_{\text{EFT}}=\mathcal{L}_{\text{SM}}+\sum_{i}C_{i}^{(6)}\OPi^{(6)}+\sum_{i}C_{i}^{(8)}\OPi^{(8)} + \cdots,
\end{equation}\end{linenomath*}
where $i$ enumerates the set of operators under consideration, $C_{i}^{(D)}$ are Wilson coefficients that scale as $\Lambda^{4-D}$ and $D$ denotes the operator dimension.
The leading deviations from the SM are generally expected to occur at $D=6$, since $D=5$ operators violate lepton number conservation~\cite{Weinberg:1979sa}.
Examples of the relationship between the operators affecting diboson production and specific BSM scenarios are described in Refs.~\cite{Liu:2016idz,Marzocca:2020jze}.

The dimension-six operator of interest in this analysis, \OWWW, is a $CP$-even modification of the \WWV TGC defined in the EFT basis of Ref.~\cite{Grzadkowski:2010es} as
\begin{linenomath*}\begin{equation}
\OWWW = \epsilon^{ijk}W_{\mu}^{i\nu}W_{\nu}^{j\rho}W_{\rho}^{k\mu},
\end{equation}\end{linenomath*}
where $W_{\mu}^{i\nu}$ is the weak isospin field strength tensor and $\epsilon^{ijk}$ is the totally antisymmetric tensor with $\epsilon^{123}=1$.\@
The cross section in the presence of this operator can be expressed as
\begin{linenomath*}\begin{equation}\label{eqn:xsec}
\sigma(\CWWW) = \sigSM + \CWWW\sigInt + \CWWW^{2}\sigBSM,
\end{equation}\end{linenomath*}
where \CWWW is the Wilson coefficient, \sigInt is the contribution from the interference between the SM and \OWWW, and \sigBSM is the pure BSM component.

However, it has been demonstrated~\cite{Panico:2017frx,Azatov:2017kzw} that in the high-energy limit, $E > \mW$, the \TwoToTwo amplitudes for transverse vector boson production, $ff\to\PW_{\mathrm{T}}\PV_{\mathrm{T}}$, have different final-state helicity configurations for the SM ($\pm\mp$) and BSM ($\pm\pm$) components.
This means the effect of the interference is typically not detectable when considering observables inclusive over the decay angles, for example, the \pt of the photon or \PWpm boson.
This narrows our sensitivity in such observables to just the pure BSM contribution at order $\CWWW^{2}$, which scales as $\Lambda^{-4}$.\@
In this scenario, the validity of any derived constraints can be limited by the unknown effect of the leading dimension-eight contributions, which also enter at order $\Lambda^{-4}$.
Therefore, increasing the sensitivity to the interference, which scales as $\Lambda^{-2}$, is important for improving the validity of the constraints in any global EFT interpretation~\cite{Azatov:2016sqh}.

\begin{figure}[hbtp]
\centering
\includegraphics[width=\cmsFigWidthX]{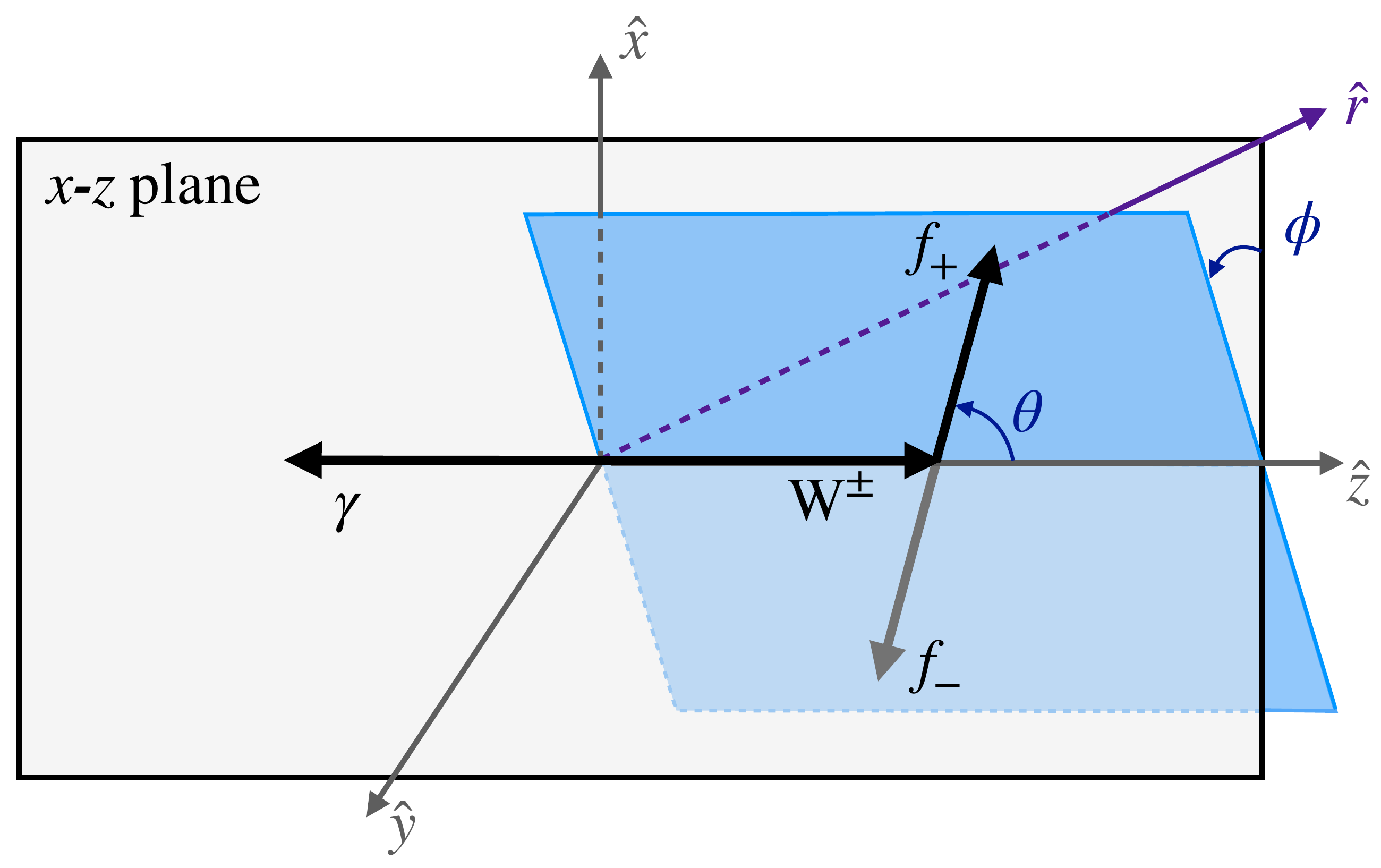}
\caption{Scheme of the special coordinate system for \WGamma production, defined by a Lorentz boost to the center-of-mass frame along the direction $\hat{r}$. The $z$ axis is chosen as the \PWpm boson direction in this frame, and $y$ is given by $\hat{z} \times \hat{r}$. The \PWpm boson decay plane is indicated in blue, where the labels \fPlu and \fMin refer to positive and negative helicity final-state fermions. The angles $\phi$ and $\theta$ are the azimuthal and polar angles of \fPlu.}\label{fig:angle_diagram}
\end{figure}

A method has been proposed~\cite{Panico:2017frx,Azatov:2019xxn} that gives sensitivity to the SM-BSM interference by measuring the decay angles of the final-state fermions.
A special coordinate system, illustrated in Fig.~\ref{fig:angle_diagram}, is defined event-by-event by a Lorentz boost to the center-of-mass frame of the \WGamma system, where the boost direction is denoted $\hat{r}$.
Since the longitudinal component of the neutrino momentum is not measurable, additional constraints are required in the calculation of the \WGamma four-momentum, described in Section~\ref{sec:phi_reco}.
In the boosted frame the boson momenta are back-to-back, and the $z$ axis is taken as the direction of the \PWpm boson, the $y$ axis direction is given by $\hat{z} \times \hat{r}$, and the $x$ axis given by $\hat{y} \times \hat{z}$.
The final-state fermions from the \PWpm boson decay are labeled as \fPlu and \fMin, referring to the positive and negative fermion helicity states, respectively.

In this frame the angle $\theta$ is the polar angle of the \PWpm boson decay in its rest frame, taken as the angle between the three-momenta of the \PWpm boson and fermion \fPlu.
The angle $\phi \in [-\pi,\pi]$ is the azimuthal angle of \fPlu: $\phi=\phi(\fPlu) = \phi(\fMin) + \pi$, modulo $2\pi$.

\begin{figure}[hbtp]
\centering
\includegraphics[width=\cmsFigWidthW]{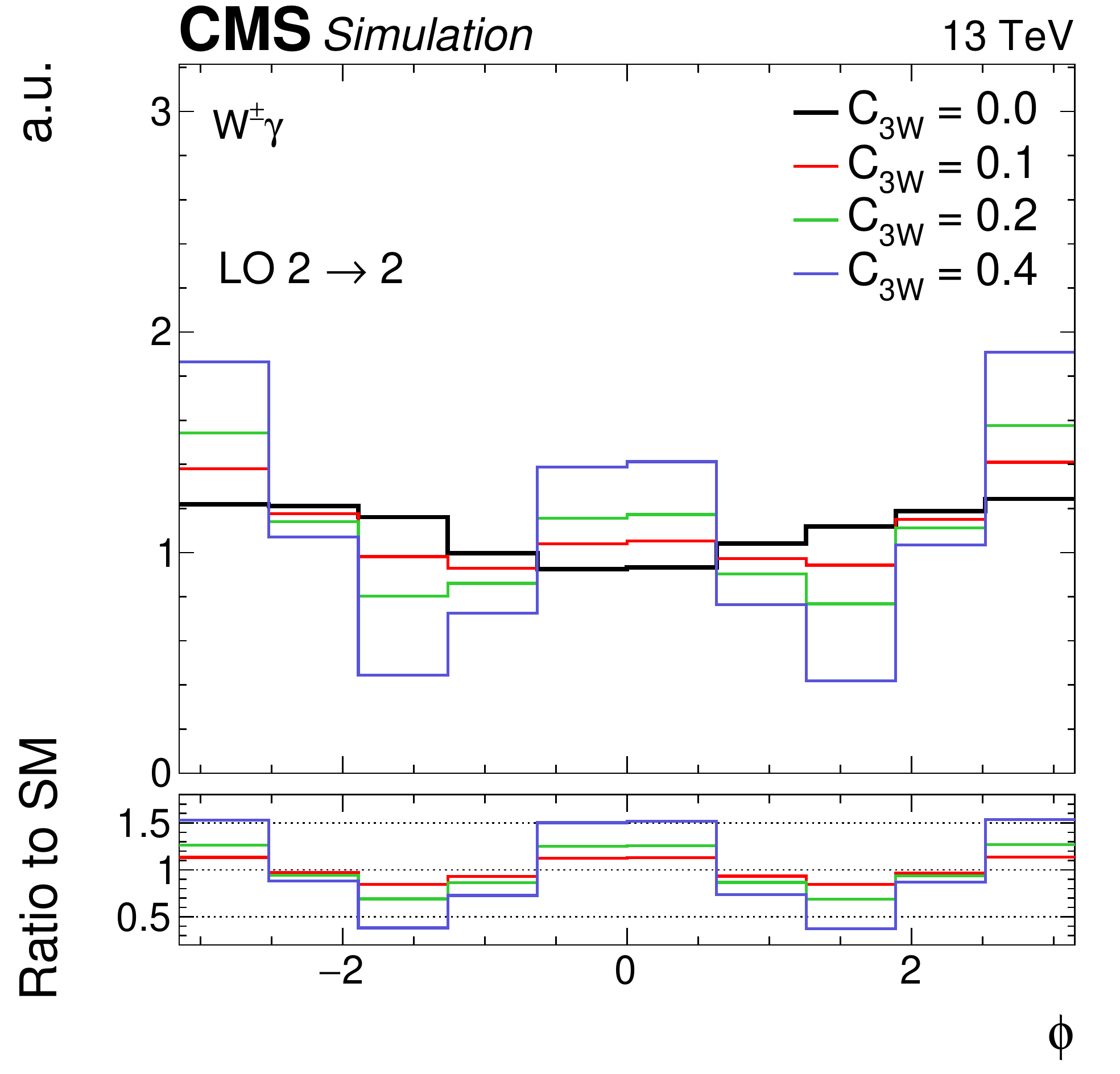}
\includegraphics[width=\cmsFigWidthW]{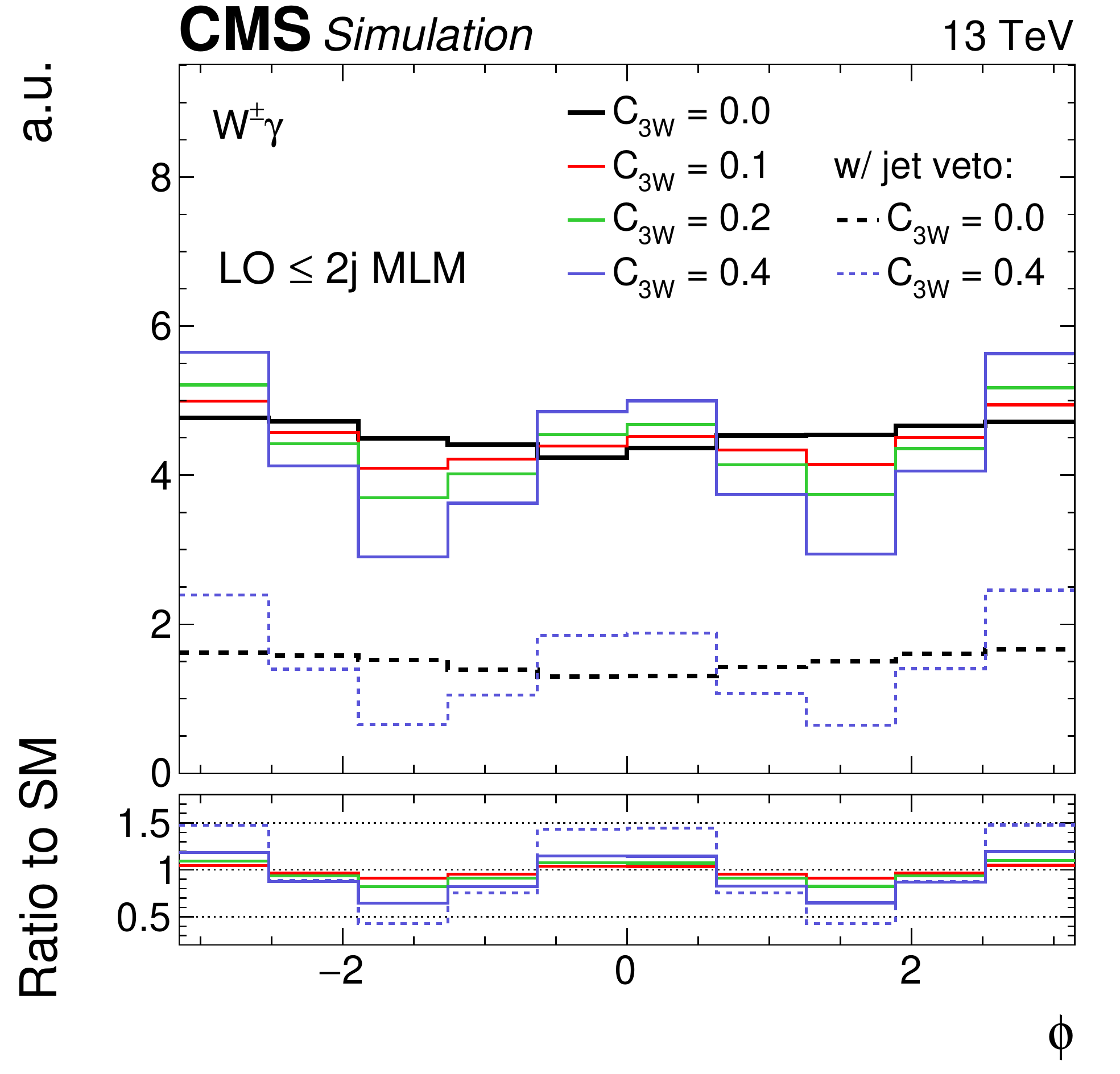}
\caption{Particle-level distributions (in arbitrary units) of the decay angle $\phi$, comparing the LO \TwoToTwo process (\cmsLeft) to the LO MLM-merged prediction with up to two additional jets in the matrix element calculations (\cmsRight). The black line gives the SM prediction ($\CWWW = 0$) and the red, green, and blue lines correspond to different nonzero values of \CWWW, for which only the interference contribution is shown. The black and blue dashed lines in the \cmsRight figure give the distributions in the presence of a jet veto, as described in the text.}\label{fig:lo_eft_effect_2}
\end{figure}

A measurement of $\phi$ is sensitive to the interference of the \OWWW term with the SM contribution.
Figure~\ref{fig:lo_eft_effect_2} (\cmsLeft) shows the particle-level distribution in $\phi$ in the LO \TwoToTwo scattering process for varying values of \CWWW, with only the inclusion of the interference term.
A clear modulation is present, which grows with increasing \CWWW.
Figure~\ref{fig:lo_eft_effect_2} (\cmsRight) shows that the magnitude of this modulation is reduced with the inclusion of additional jets in the matrix element calculations, with MLM merging~\cite{Alwall:2007fs} used in the matching to the parton shower.
However, a veto on the presence of additional jets in the event with $\pt>30\GeV$ and $\aeta<2.5$ is shown to substantially restore the effect.

\section{The CMS detector}\label{sec:detector}
The central feature of the CMS apparatus is a superconducting solenoid of 6\unit{m} internal diameter, providing a magnetic field of 3.8\unit{T}.
Within the solenoid volume are a silicon pixel and strip tracker, a lead tungstate crystal electromagnetic calorimeter (ECAL), and a brass and scintillator hadron calorimeter (HCAL), each composed of a barrel and two endcap sections.
Forward calorimeters extend the pseudorapidity coverage provided by the barrel and endcap detectors.
Muons are detected in gas-ionization chambers embedded in the steel flux-return yoke outside the solenoid.

In the barrel section of the ECAL, an energy resolution of about 1\% is achieved for unconverted or late-converting photons in the tens of $\GeV$ energy range.
The remaining barrel photons have a resolution of about 1.3\% up to $\aeta = 1$, rising to about 2.5\% at $\aeta = 1.4$.
In the endcaps, the resolution of unconverted or late-converting photons is about 2.5\%, while the remaining endcap photons have a resolution of 3--4\%~\cite{CMS:EGM-14-001}.

The electron momentum is estimated by combining the energy measurement in the ECAL with the momentum measurement in the tracker.
The momentum resolution for electrons with $\pt \approx 45\GeV$ from $\PZ\to\Pe\Pe$ decays ranges from 1.7 to 4.5\%.
It is generally better in the barrel region than in the endcaps, and also depends on the bremsstrahlung energy emitted by the electron as it traverses the material in front of the ECAL~\cite{Khachatryan:2015hwa}.

Muons are measured in the  range $\aeta < 2.4$, with detection planes made using three technologies: drift tubes, cathode strip chambers, and resistive plate chambers.
Matching muons to tracks measured in the silicon tracker results in a relative \pt resolution of 1\% in the barrel and 3\% in the endcaps for muons with \pt up to 100\GeV.
The \pt resolution in the barrel is better than 7\% for muons with \pt up to 1\TeV~\cite{Sirunyan:2018fpa}.

Events of interest are selected using a two-tiered trigger system.
The first level (L1), composed of custom hardware processors, uses information from the calorimeters and muon detectors to select events at a rate of around 100\unit{kHz} within a fixed latency of about 4\mus~\cite{Sirunyan:2020zal}.
The second level, known as the high-level trigger, consists of a farm of processors running a version of the full event reconstruction software optimized for fast processing, and reduces the event rate to around 1\unit{kHz} before data storage~\cite{Khachatryan:2016bia}.

A more detailed description of the CMS detector, together with a definition of the coordinate system used and the relevant kinematical variables, can be found in Ref.~\cite{Chatrchyan:2008zzk}.

\section{Data samples and event simulation}\label{sec:datasets}
This analysis uses \pp collision data collected by the CMS experiment between 2016 and 2018 at $\sqrt{s}=13\TeV$ with an integrated luminosity of 138\fbinv.
Simulated event samples are used to model both the \WGamma signal and several backgrounds.
The \WGamma signal, with up to one additional jet, and the $\PZ$+jets, $\PZ\PGg$, $\PGg\PGg$, single-$\PQt$+$\PGg$, $\PW\PW$, $\PW\PZ$, and $\PZ\PZ$ background samples are generated with \MGvATNLO{}~\cite{MadGraph} v2.3.3--2.6.5 at NLO in QCD.\@
Backgrounds from $\ttbar$+jets and single-$\PQt$ production are simulated with \POWHEG{}~v2.0~\cite{Nason:2004rx,Frixione:2007vw,Alioli:2010xd,Frixione:2007nw,Frederix:2012dh,Re:2010bp}.
The $\ttbar\PGg$ background is simulated with \MGvATNLO{} v2.6.0 at LO in QCD.\@
The effect of EFT operators on \WGamma distributions is modeled using a dedicated sample generated with \MGvATNLO{} v2.6.7 at LO with up to two additional jets.
Event weights for varying values of \CWWW are embedded, utilizing LO matrix element reweighting~\cite{Mattelaer:2016gcx} and the \textsc{SMEFTsim} UFO model~\cite{Brivio:2017btx,Brivio:2020onw}.

The NNPDF3.1~\cite{Ball:2017nwa} NNLO parton distribution functions (PDFs) are used in the \WGamma signal simulation.
The background samples use either the NNPDF3.0~\cite{Ball:2014uwa} NLO or NNPDF3.1 NNLO PDFs.
All samples are interfaced with \PYTHIA{}~v8.230~\cite{Sjostrand:2014zea} for parton showering and underlying event simulation.
Samples produced for the 2016 analysis use the CUETP8M1~\cite{Khachatryan:2015pea} underlying event tune, with the exception of the $\ttbar$ samples, which use CUETP8M2T4~\cite{CMS-PAS-TOP-16-021}.
The 2017 and 2018 samples use the CP5~\cite{Sirunyan:2019dfx} tune.
The CMS detector response is simulated with the \GEANTfour{} package~\cite{Agostinelli:2002hh}.
Additional \pp interactions occurring in both the same and adjacent bunch crossings (pileup) are simulated as a set of minimum bias interactions that are mixed with the hard scattering event.
In the analysis of each data-taking year the simulated events are weighted based on the number of pileup events to match the distribution measured in data.

\section{Event and object reconstruction}\label{sec:event_reco}
The \pp collision vertices in an event are reconstructed by grouping tracks consistent with originating at a common point in the luminous region.
The candidate vertex with the largest value of summed physics-object $\pt^2$ is taken to be the primary \pp interaction vertex.
The physics objects are the jets, clustered using the anti-\kt jet finding algorithm~\cite{Cacciari:2008gp,Cacciari:2011ma} with all the tracks assigned to candidate vertices as inputs, and the associated missing transverse momentum, taken as the negative vector \pt sum of those jets.

The full event reconstruction uses the particle-flow (PF) algorithm~\cite{CMS-PRF-14-001}, which aims to reconstruct and identify each individual particle in an event, with an optimized combination of information from the various elements of the CMS detector.
The energy of photons is obtained from the ECAL measurement.
The energy of electrons is determined from a combination of the electron momentum at the primary interaction vertex as determined by the tracker, the energy of the corresponding ECAL cluster, and the energy sum of all bremsstrahlung photons spatially compatible with originating from the electron track.
The energy of muons is obtained from the curvature of the corresponding track.
The energy of charged hadrons is determined from a combination of their momentum measured in the tracker and the matching ECAL and HCAL energy deposits, corrected for the response function of the calorimeters to hadronic showers.
Finally, the energy of neutral hadrons is obtained from the corresponding corrected ECAL and HCAL energies.

For each event, hadronic jets are clustered from these reconstructed particles using the infrared- and collinear-safe anti-\kt algorithm~\cite{Cacciari:2008gp, Cacciari:2011ma} with a distance parameter of 0.4.
The jet momentum is determined as the vector sum of all particle momenta in the jet, and is found from simulation to be, on average, within 5--10\% of the true momentum over the entire \pt spectrum and detector acceptance.
Pileup interactions can contribute additional tracks and calorimetric energy depositions to the jet momentum.
To mitigate this effect, charged particles identified to be originating from pileup vertices are discarded, and an offset correction is applied to subtract remaining contributions.
Jet energy corrections are derived from simulation to bring the measured response of jets to that of particle level jets on average.
In situ measurements of the momentum balance in dijet, photon+jet, $\PZ$+jet, and multijet events are used to correct any residual differences in the jet energy scale between data and simulation~\cite{Khachatryan:2016kdb}.

Events of interest in this analysis are characterized by the presence of a single isolated charged lepton, either an electron or a muon; an isolated photon; and missing transverse momentum.
Identification and isolation criteria are used to improve the purity of events containing prompt leptons or photons, while rejecting candidates that originate from jets, pileup interactions, or misreconstruction.

The identification criteria for photons~\cite{Sirunyan:2020ycc} include selections on the ratio of hadronic to electromagnetic energy deposited in calorimeters (\HoverE), a variable quantifying the lateral extension of the shower (\sigeta), and a requirement that the photon ECAL cluster is not matched to an electron track that has hits in the innermost tracker layers.
Electron candidates are required to pass identification criteria related to the impact parameters of the track with respect to the primary vertex, the matching between the track and ECAL cluster, the shower shape in the ECAL, and the same \HoverE variable as for photons.
A veto on electrons originating from photon conversions is applied~\cite{CMS:EGM-14-001}.
Muon candidates are reconstructed by a simultaneous track fit to hits in the tracker and in the muon chambers.
Identification criteria are applied to the impact parameters with respect to the primary vertex, to the number of spatial measurements in the silicon tracker and the muon system, and to the fit quality of the combined muon track.

Both lepton and photon candidates are required to be isolated from any other activity in the event.
For electrons and muons, a relative isolation variable is defined as
\begin{linenomath*}
\ifthenelse{\boolean{cms@external}}
{
\begin{multline}\label{eqn:iso}
I^{\Pell}_{\text{rel}} = \frac{1}{\ptL} \Bigg[ \sum_{\text{charged}} \pt + \max\bigg(0, \sum_{\text{neutral}} \pt  \\
 + \sum_{\text{photons}} \pt - \pt^{\text{pileup}}\bigg) \Bigg],
\end{multline}
}
{
\begin{equation}\label{eqn:iso}
I^{\Pell}_{\text{rel}} = \frac{1}{\ptL} \Bigg[ \sum_{\text{charged}} \pt + \max\bigg(0, \sum_{\text{neutral}} \pt + \sum_{\text{photons}} \pt - \pt^{\text{pileup}}\bigg) \Bigg],
\end{equation}
}
\end{linenomath*}
where the sums are over the PF candidates of the indicated type that are within $\DR < 0.4$ of the lepton, with $\DR = \sqrt{\smash[b]{(\Delta\eta)^{2} + (\Delta\phi)^{2}}}$.
To suppress the effect of pileup, only charged hadrons compatible with originating at the primary vertex are included.
For the neutral hadron and photon sums, an estimate of the expected pileup contribution is subtracted~\cite{Sirunyan:2020ycc,Sirunyan:2018fpa}.
Requirements of $I^{\Pgm}_{\text{rel}} < 0.15$ and $I^{\Pe}_{\text{rel}} < 0.05$--$0.09$ are applied, where the latter selection varies with \ptE and $\abs{\etaE}$.
For photons, isolation requirements are applied separately to each of the three components in Eq.~(\ref{eqn:iso}).
In addition, a selection is applied on the maximal charged particle isolation sum \IchMax computed with respect to any vertex.
A requirement of $\IchMax < \min(0.05\ptG, 6\GeV)$ is applied.
This more than halves the rate of photons from pileup jets, which constitute around 60\% of misidentified photons at low \ptG.

For leptons and photons, several corrections are applied to the simulation to improve the modeling of the data.
Correction factors are applied as event weights for the efficiencies of electrons and muons to be reconstructed, pass their respective identification and isolation criteria, and pass the single-lepton trigger requirements.
These are determined using the ``tag-and-probe'' (T\&P) technique~\cite{Khachatryan:2010xn} with \ZtoLL data.
Similarly, scale factors for the efficiencies of photons to pass identification and isolation criteria are applied.
These are measured with the T\&P technique in $\ZtoEE$ events, where the probe electron is reconstructed as a photon.
The muon momentum scale in data and simulation is corrected using a calibration derived in \ZtoMM events~\cite{Bodek:2012id,Sirunyan:2018fpa}.
In simulation this includes an additional resolution smearing to match the performance in data.
Similarly, scale and resolution corrections are applied to simulated electrons and photons to better match the data resolution.

The missing transverse momentum vector \ptvecmiss is computed as the negative vector \pt sum of all the PF candidates in an event, and its magnitude is denoted as \ptmiss~\cite{Sirunyan:2019kia}.
The \ptvecmiss vector is modified to account for corrections to the energy scale of the reconstructed jets.
The pileup-per-particle identification algorithm~\cite{Bertolini:2014bba} is applied to reduce the pileup dependence of the \ptvecmiss observable.
In the sum over PF candidates, each \pt is weighted by the probability for that particle to originate from the primary interaction vertex.

Anomalous high-\ptmiss events can be due to a variety of reconstruction failures, detector malfunctions, or noncollision backgrounds.
Such events are rejected by event filters that are designed to identify more than 85\% of the spurious high-\ptmiss events with a mistagging rate of less than 0.1\%~\cite{Sirunyan:2019kia}.

\section{Event selection}\label{sec:event_sel}
Single-lepton triggers are used to select data events.
These require that candidates pass loose identification and isolation criteria and have a \pt above thresholds of 27--32\GeV for electrons and 24--27\GeV for muons, with the exact values depending on the data-taking year.

The offline event selection requires the presence either of an electron with $\pt > 35\GeV$ and $\aeta < 2.5$ (electron channel) or a muon with $\pt > 30\GeV$ and $\aeta < 2.4$ (muon channel).
The lepton must pass the identification criteria outlined in Section~\ref{sec:event_reco} and be spatially matched to the object that fired the trigger.
Electron candidates falling in the transition region between the ECAL barrel and endcap calorimeters, $1.44 < \aeta < 1.57$, are rejected.
Events are vetoed if they contain additional leptons with $\pt > 10\GeV$ that pass looser versions of the identification criteria.
The same photon selection is used for both channels.
The photon is required to have $\pt > 30\GeV$ and $\aeta < 2.5$, not fall in the ECAL transition region, pass the identification and isolation requirements and be separated from the selected lepton by $\DR > 0.7$.
Events with more than one photon passing these requirements are rejected.
Hadronic jets in the event are counted if having $\pt > 30\GeV$, $\aeta < 2.5$, and a separation from both the lepton and photon of $\DR > 0.4$.

In both channels, events are required to have $\ptmiss > 40\GeV$.
This reduces a large source of background from processes that do not contain any intrinsic \ptmiss, for example \ZtoLL.
Events with \mLG, the invariant mass of the lepton-photon pair, close to $m_{\PZ}$ are rejected.
In the electron channel this veto is on events with $70 < m_{\Pe\PGg} < 110\GeV$, which reduces the remaining \ZtoEE contribution.
In the muon channel the range is $70 < m_{\PGm\PGg} < 100\GeV$, which rejects events where the photon originates as final-state radiation, $\PZ\to\Pgm\Pgm\to\Pgm\Pgm\PGg$, and takes a large fraction of the original muon momentum.

\begin{figure*}[hbtp]
\centering
\includegraphics[width=\cmsFigWidthW]{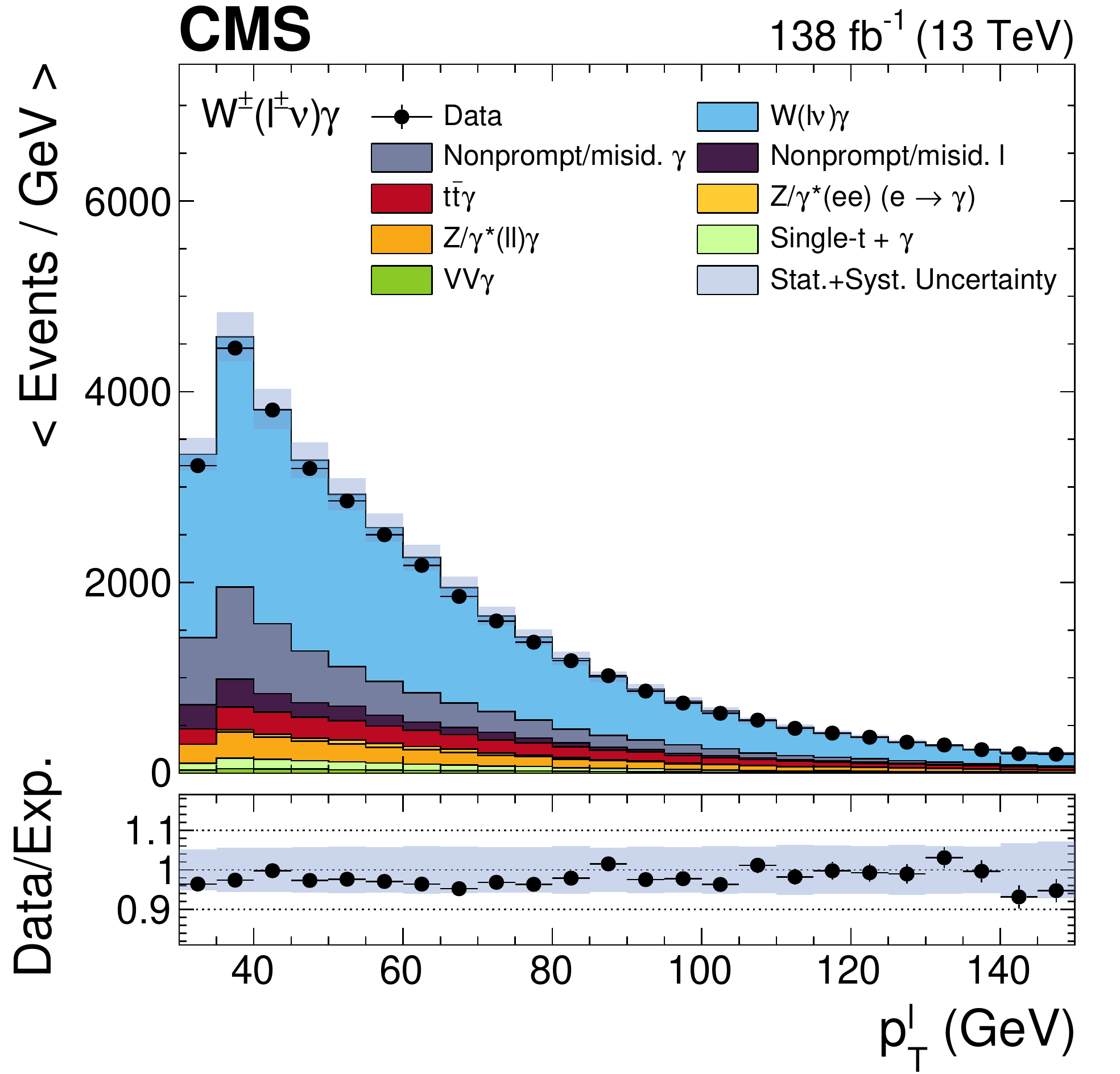}
\includegraphics[width=\cmsFigWidthW]{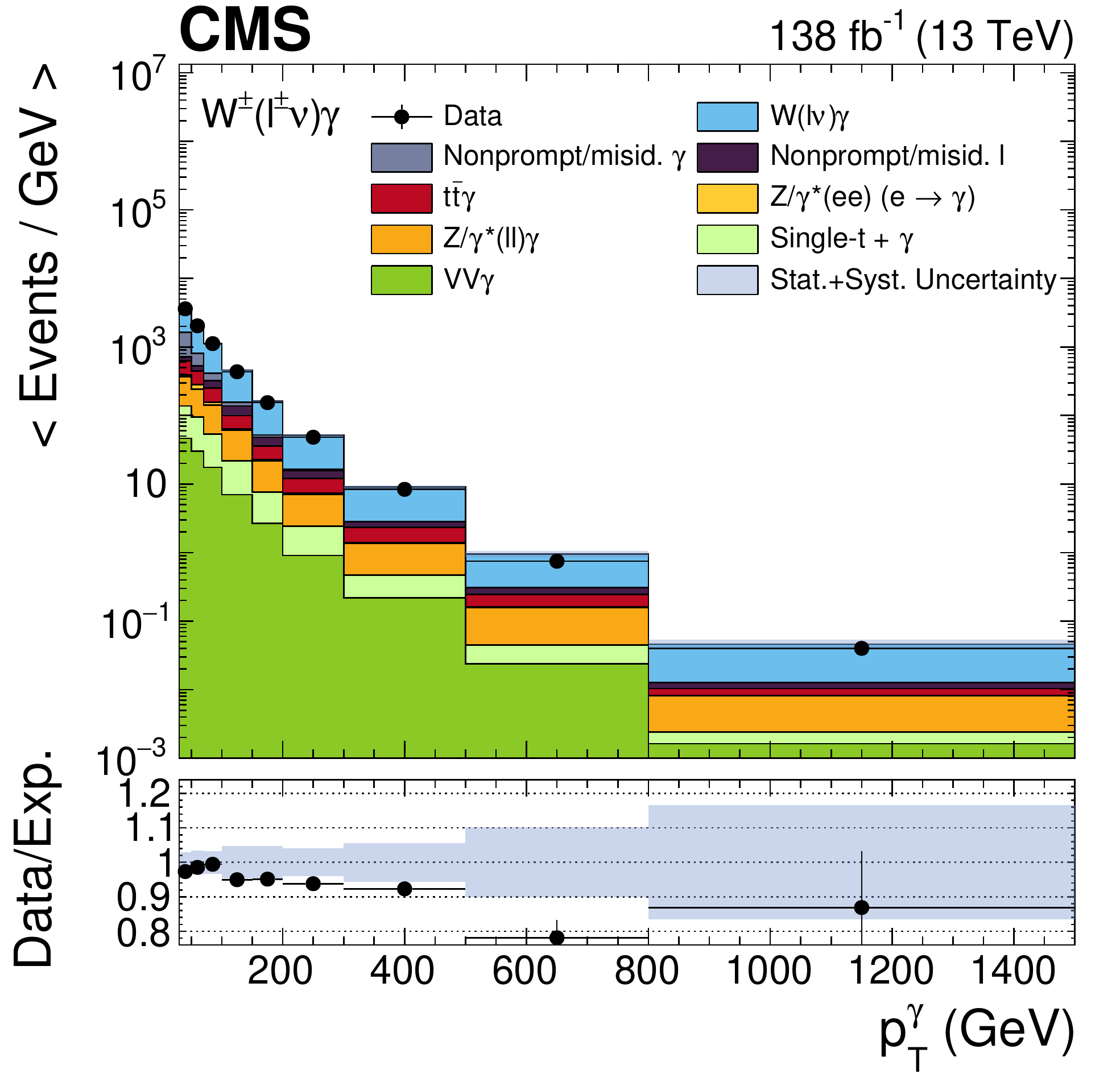} \\
\includegraphics[width=\cmsFigWidthW]{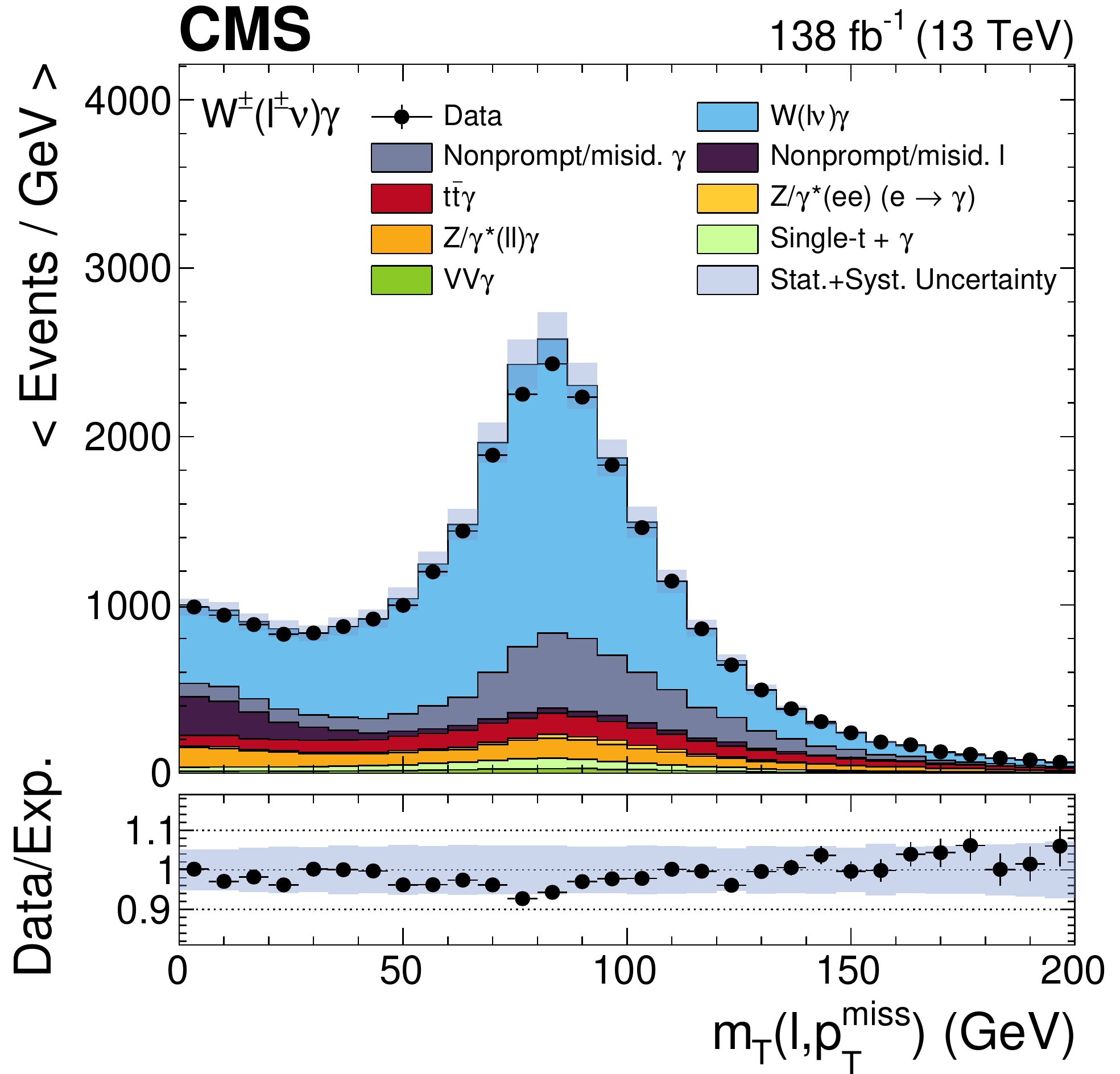}
\includegraphics[width=\cmsFigWidthW]{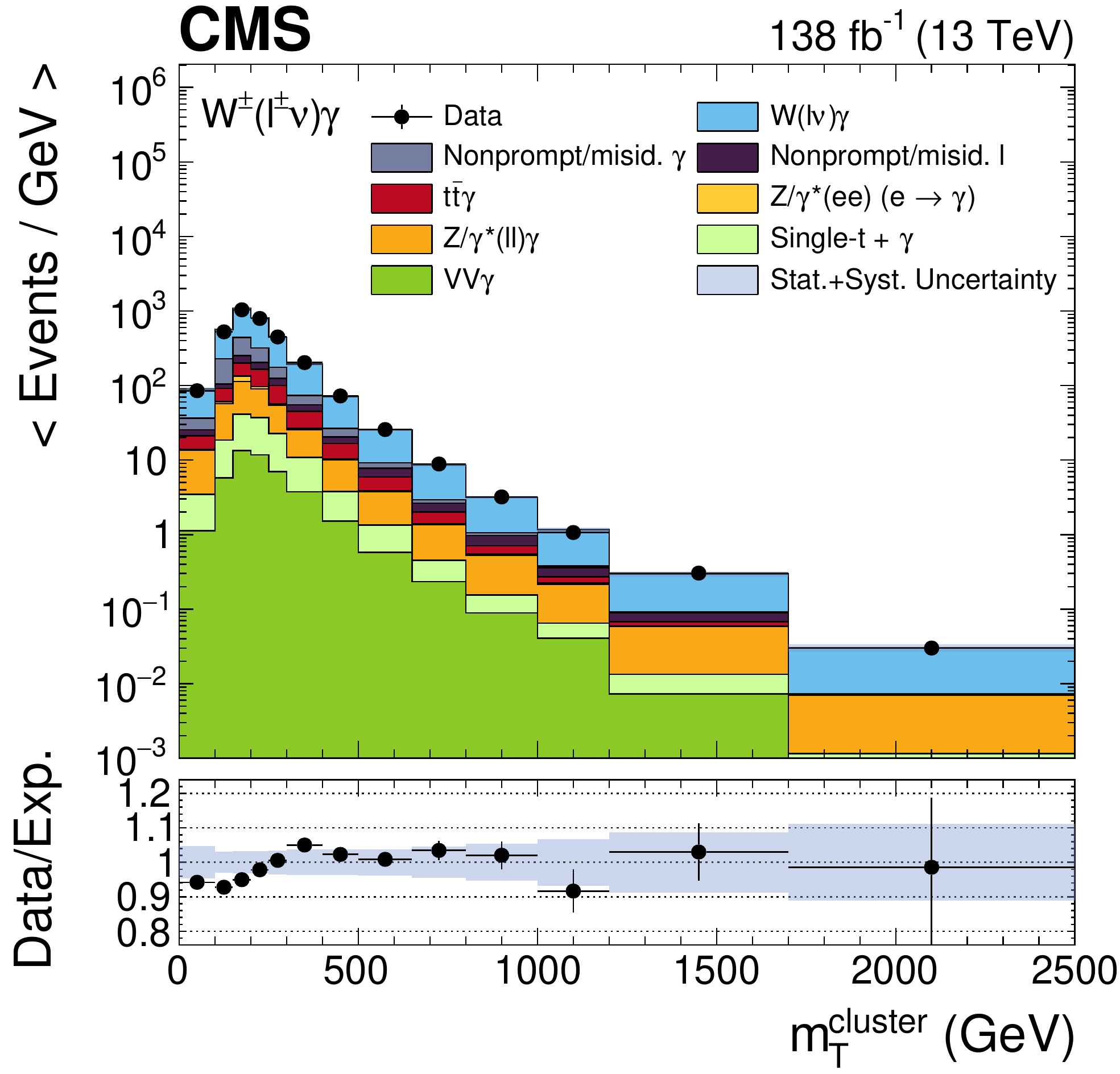}
\caption{Distributions of the lepton \pt (upper left), the photon \pt (upper right), \mTLmiss (lower left), and \mTcluster (lower right), combining the electron and muon channels. The horizontal and vertical bars associated to the data points correspond to the bin widths and statistical uncertainties, respectively. The shaded band represents the total statistical and systematic uncertainty in the signal plus background expectation.}\label{fig:ctl_baseline}
\end{figure*}

This baseline event selection results in the selection of $72\,798$ electron channel and $109\,669$ muon channel events, of which \WGamma production is expected to account for 53 and 58\%, respectively.
Figure~\ref{fig:ctl_baseline} shows the \ptL, \ptG, \mTLmiss, and \mTcluster distributions in data and simulation, summed over the electron and muon channels.
The \mTcluster observable is the transverse mass of the \LNG system, defined as
\begin{linenomath*}\begin{equation}
\mTcluster = \sqrt{\left[\sqrt{\mLG^{2} + \pTLG^{2}} + \ptmiss\right]^{2} - \pTLGmiss^{2}},
\end{equation}\end{linenomath*}
where \pTLG is the transverse momentum of the lepton and photon system, and \pTLGmiss is the transverse momentum of the lepton, photon, and \ptmiss system.
The baseline selection is used for the majority of the differential cross section measurements reported in Section~\ref{sec:results}.
For the EFT analysis a tighter selection, the EFT selection, with $\ptL > 80\GeV$ and $\ptG > 150\GeV$ is applied, and events containing hadronic jets, as defined above, are vetoed.

\subsection{Reconstruction of \texorpdfstring{$\phi$}{phi}}\label{sec:phi_reco}
In the EFT analysis the neutrino four-momentum is required to construct the \WGamma center-of-mass frame and determine the azimuthal angle $\phi$.
The reconstruction procedure described in Ref.~\cite{Panico:2017frx} is followed.
The \ptvecmiss vector in the event is assumed to be the transverse momentum of the neutrino, \ptVecN.
Then by requiring that the invariant mass of the charged lepton and neutrino system equals the \PWpm boson pole mass \mW, it is possible to form an equation for \etaN,
\begin{linenomath*}\begin{equation}\label{eqn:neutrino_eta}
\etaN = \etaL \pm \ln \left[ 1 + \Delta \sqrt{2 + \Delta^{2}} + \Delta^{2} \right],
\end{equation}\end{linenomath*}
where:
\begin{linenomath*}\begin{equation}\label{eqn:neutrino_eta_delta}
\Delta = \sqrt{\frac{\mW^{2} - \mT^{2}}{2\ptL\ptN}},
\end{equation}\end{linenomath*}
\etaL is the pseudorapidity of the lepton, and \mT is the transverse mass of the $\Pell\PGn$ system.
Of the two possible solutions for \etaN, only one will correspond to the unknown true value.
One of the two solutions is picked at random event-by-event.
In the limit of high \PWpm boson momentum, it can be demonstrated that the two solutions for $\phi$, $\phi^{+}$ and $\phi^{-}$, are related by $\phi^{+} = \pi - \phi^{-}$, modulo $2\pi$.
This intrinsic ambiguity does not, however, prevent the observation of the interference effect.
This is illustrated in Fig.~\ref{fig:lo_eft_effect_2}, where the deviation in the $\phi$ distribution is unaffected by a transformation $\phi \to \pi - \phi$.

\begin{figure}[hbtp]
\centering
\includegraphics[width=\cmsFigWidthV]{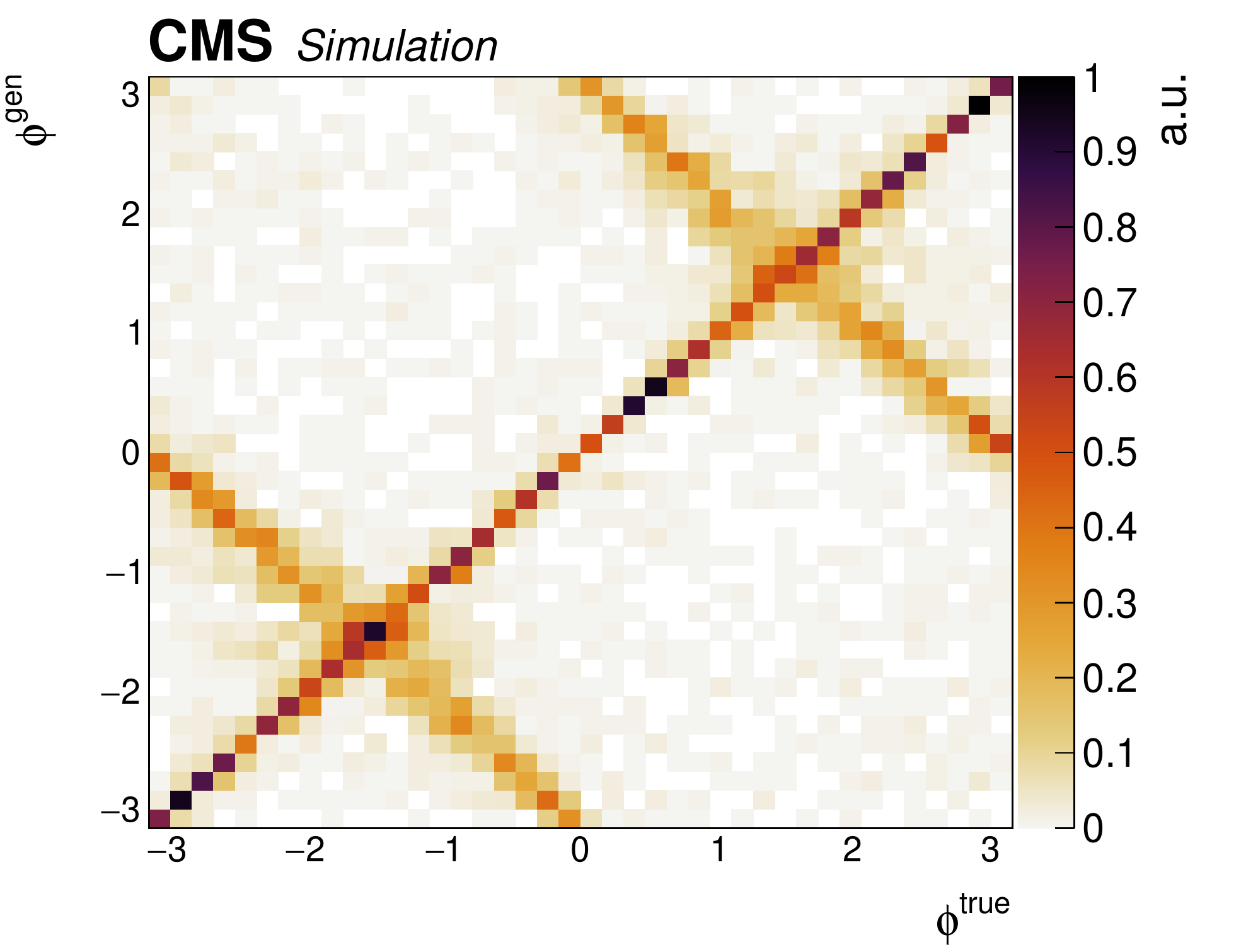}
\caption{The two-dimensional distribution of \PhiGen versus \PhiTrue, where the former is reconstructed using the particle-level lepton and photon momenta and \ptvecmiss. The off-diagonal components correspond to events where the incorrect solution for \etaN is chosen, as described in the text.}\label{fig:phi_gen_vs_true}
\end{figure}

One additional complication to this reconstruction is in cases where $\mT > \mW$, either because the \PWpm boson was produced off shell, or because of mismeasured \ptvecmiss in the reconstruction.
In this case Eq.~(\ref{eqn:neutrino_eta}) has no real solutions and instead \etaN is assumed to equal \etaL, which gives a lepton-neutrino mass as close as possible to \mW.
In these events $\phi$ is biased towards $\pm \pi/2$.
Figure~\ref{fig:phi_gen_vs_true} shows the two-dimensional (2D) distribution of \PhiGen versus \PhiTrue, where \PhiGen is the reconstructed $\phi$ angle using particle-level quantities.

Given this ambiguity in $\phi$, the variable \phif is used in the subsequent analysis, where
\begin{linenomath*}\begin{equation}
    \phif =
\begin{cases}
    -(\pi + \phi), & \text{for } \phi < -\frac{\pi}{2}; \\
    \phi, & \text{for } \abs{\phi} < \frac{\pi}{2}; \\
    \pi - \phi, & \text{for } \phi > \frac{\pi}{2}. \\
\end{cases}
\end{equation}\end{linenomath*}
Furthermore, since the distributions of \phif shown in Fig.~\ref{fig:lo_eft_effect_2} are symmetric about zero, the magnitude \aphif will be used.

\section{Background estimation}\label{sec:backgrounds}
The background from processes containing prompt leptons and photons is estimated from simulation.
This includes $\PZ/\PGg^{\ast}(\Pell\Pell)\PGg$, $\ttbar\PGg$, $\PW\PV\PGg$, and single-$\PQt$+$\PGg$ production, where $\PV = \PW, \PZ$.

An important background in the electron channel is from events with two prompt electrons, predominantly \ZtoEE, in which one is misidentified as the photon candidate.
The rate at which electrons, having passed the photon identification criteria described in Section~\ref{sec:event_reco}, are subsequently not rejected by the specific electron veto is measured in both data and simulation using the T\&P method.
The probability in data is typically 1--5\%, depending on the \pt and $\eta$ of the electron.
The resulting correction factors are applied in all simulated events where the photon candidate is matched to a prompt electron at the generator level.

The remaining backgrounds originate from other processes where one or both of the lepton and photon are misidentified, and these are discussed in the following sections.

\subsection{Jets misidentified as photons}\label{sec:jet_fakes}
After the baseline selection, the main background in both channels comes from $\PW$+jets events in which a jet is misidentified as a photon.
These jets tend to have high fractions of energy deposited in the ECAL, and may contain genuine photons, for example from $\Pgpz\to\PGg\PGg$ decays.
Since the probability for jets to pass the photon identification criteria is not guaranteed to be well modeled in simulation, this background is estimated using data.

Independent control regions, with a high purity in misidentified jets, are defined by inverting some of the photon identification requirements. Two variables are used: the charged isolation sum defined in Eq.~(\ref{eqn:iso}), denoted \Ich, and \sigeta.
This exploits the fact that misidentified jets tend to be less well isolated and have broader showers than prompt photons.
For a given selection and observable bin, the events in the inverted-\sigeta region are used to predict the number of misidentified jet events in the nominal region, \Nmisid.
Each inverted-\sigeta event is assigned an extrapolation factor, \fGam, corresponding to the expected inverted-to-nominal ratio given the \pt and \aeta of the photon candidate. This gives
\begin{linenomath*}\begin{equation}
\Nmisid = \sum_{\text{data}} \fGam(\ptG, \etaG) - \Nprompt,
\end{equation}\end{linenomath*}
where the sum runs over all data events in the inverted-\sigeta region, and \Nprompt is the estimated prompt photon contamination, determined by applying the same extrapolation factors to simulated events in this region.
The \fGam value for each \pt and \aeta bin is determined from a region where \Ich is inverted.
The \sigeta extrapolation factor, calculated after the subtraction of the expected prompt photon contribution, varies linearly with \Ich.
Therefore, a linear fit to the ratio is performed, with the result extrapolated to the nominal $\Ich$ region to give $\fGam$.

The statistical uncertainty in \fGam is 2--35\%, depending on the \pt and \aeta bin.
Systematic uncertainties  come from the normalization of the prompt photon contribution that is subtracted, and the bias in the extrapolated value of \fGam, compared with the true value, when the method is validated in simulation.
The former is typically 10\%, and the latter is up to 15\% in a given bin.
The total statistical and systematic uncertainty, which ranges from 10 to 45\%, is propagated through all subsequent results.

In the EFT selection, because of the limited number of control region events for $\ptG > 150\GeV$, this method is modified.
In the inverted-\sigeta region, the nominal requirement on \Ich is retained, whereas all other photon identification criteria are removed and replaced with only a loose requirement of $\HoverE < 0.15$.
The total uncertainty in \fGam is 20--60\% under this selection, and the statistical component is typically dominant.

\subsection{Misidentified leptons}\label{sec:lep_fakes}
Another contribution comes from events with a prompt photon but a nonprompt or misidentified lepton, \eg, within a jet containing a semileptonic meson decay.
The main identifying feature of such events is that the leptons will generally be less well isolated.
The contribution is estimated by applying an inverted lepton isolation requirement, and applying a per-event extrapolation factor \fL, binned in \pt and \aeta, to predict the number of events in the nominal isolated selection.

The \fL factors are measured in a control region in data designed to increase the purity of events with misidentified leptons.
This control region is obtained by applying the baseline selection in both channels, but vetoing the presence of an identified photon and instead requiring the presence of a jet with $\pt > 30\GeV$ and $\aeta < 2.5$.
A requirement of $\mTLmiss < 30\GeV$ is applied to reject the background from $\PW$+jets events.
The remaining contamination from $\PW$+jets and other prompt-lepton sources is estimated using simulation and subtracted.
The normalization uncertainty in this prompt subtraction dominates the uncertainties in the \fL factors, which are typically 10--40\%.

Both the misidentified lepton and misidentified photon background estimates predict the contribution from events where both objects are misidentified.
To avoid double-counting these events, their expected contribution is subtracted from the misidentified lepton background estimate.
This is estimated using events in a region where both the lepton isolation and photon \sigeta requirements are inverted, where the extrapolation factors for both methods are applied.

\section{Systematic uncertainties}\label{sec:systs}
The effects of systematic uncertainties in the signal and background expectations are incorporated for all measurements in this analysis.
The majority of the uncertainties affect both the shape and normalization of the predictions, whereas several affect only the normalization.
The shape variations are modeled by a smooth interpolation between the nominal and $\pm1$ standard deviation predictions in each bin and a linear extrapolation beyond this.
The systematic uncertainties and their typical effect on the expected yields are summarized in Table~\ref{tab:systs_summary}.
The experimental sources are as follows:

\begin{table*}[hbtp]
\centering
\topcaption{Summary of the systematic uncertainties affecting the signal and background predictions. The table notes whether each uncertainty affects the shape of the measured observable or just the normalization (Affects shape), and whether the effect is correlated between the data-taking years (Corr.\@ years). The normalization effect on the expected yield of the applicable processes is also given. For some shape uncertainties the values vary significantly across the observable distribution. In these cases the typical range and maximum values are given, where the former is the central 68\% interval considering all bins.}
\begin{scotch}{l p{1.0cm} p{1.0cm} l}
Uncertainty        &  Affects shape &  Corr.\@ years & Relative effect on expected yield \\
\hline
\multicolumn{4}{c}{\textit{Experimental}} \\
Integrated luminosity & \NA{} & Partial & 1.6\%  \\
Pileup modeling & \checkmark{} & \checkmark{} & 0.2--3.1\%  \\
L1 trigger & \checkmark{} & \checkmark{} & 0.3--1.1\%  \\
Electron ID & \checkmark{} & \checkmark{} & 0.7--2.8\%  \\
Electron ID ($\ptE > 200\GeV$) & \checkmark{} & \NA{} & 0.1--1.2\% \\
Electron trigger & \NA{} & \NA{} & 0.5\% \\
Muon ID (stat) & \checkmark{} & \NA{} & 0.1--0.6\% \\
Muon ID (syst) & \checkmark{} & \checkmark{} & 0.2--0.7\% \\
Muon trigger & \checkmark{} & \NA{} & 0.1--0.7\% \\
Photon ID & \checkmark{} & \checkmark{} & 0.6--6.0\% \\
Photon ID ($\ptG > 200\GeV$) & \checkmark{} & \NA{} & 2.1--4.7\% \\
Photon ID (high \pt extrapolation) & \checkmark{} & \NA{} & Typically 3.0--9.0\%, max. 14\% \\
Photon ($\Pe$ veto) & \NA{} & \NA{} & 1\% \\
Photon energy scale & \checkmark{} & \checkmark{} & Typically 0.1--4.8\%, max. 13\% \\
Jet energy scale & \checkmark{} & \NA{} & 1--4\% \\
\ptmiss scale & \checkmark{} & Partial & 0.1--10.1\% \\
$\Pe\to\PGg$ misidentification & \checkmark{} & \NA{} & Typically 6.7--18\%, max. 25\% \\
$\text{Jet}\to\PGg$ misidentification & \checkmark{} & \NA{} & 10--45\% \\
Misidentified $\Pe$ & \checkmark{} & \NA{} & Typically 13--36\%, max. 75\% \\
Misidentified $\Pgm$ & \checkmark{} & \NA{} & Typically 16--42\%, max. 70\% \\
\multicolumn{4}{c}{\textit{Theoretical}} \\
\WGamma acceptance (scale) & \checkmark{} & \checkmark{} & 0.3--1.7\% \\
\WGamma acceptance (PDF) & \checkmark{} & \checkmark{} & Typically 0.5--2.2\%, max. 7.6\% \\
\WGamma out-of-acceptance (scale) & \checkmark{} & \checkmark{} & 5.2--12\% \\
\WGamma parton shower modeling & \checkmark{} & \checkmark{} & 0.2--1.3\% \\
Background normalization (scale) & \NA{} & \checkmark{} & 2.0--16\% \\
Background normalization (PDF) & \NA{} & \checkmark{} & 4.2--4.8\% \\
\end{scotch}\label{tab:systs_summary}
\end{table*}

\begin{itemize}
 \item \textit{Integrated luminosity}: The 2016, 2017, and 2018 data set uncertainties are 1.2\%, 2.3\%, and 2.5\%~\cite{CMS:2021xjt,CMS-PAS-LUM-17-004,CMS-PAS-LUM-18-002}, respectively, applied to all processes estimated purely with simulation. The total integrated luminosity has an uncertainty of 1.6\%, the improvement in precision reflecting the (uncorrelated) time evolution of some systematic effects.
 \item \textit{Pileup modeling}: The uncertainty in the weighting of the pileup distribution in simulation is derived by varying the total inelastic cross section by $\pm5\%$ in the estimate of the distribution in data~\cite{Sirunyan:2018nqx}.
 \item \textit{L1 trigger}: During the 2016 and 2017 data-taking periods, a gradual shift in the timing of the inputs of the ECAL L1 trigger in the $\aeta > 2.4$ region led to a specific inefficiency. A correction amounting to 2--3\% is applied to the simulation along with the corresponding uncertainty in the inefficiency measurement.
 \item \textit{Lepton ID and trigger efficiencies}: The uncertainties in the measured correction factors are applied to the simulation, which include both statistical and systematic sources. The latter dominate at low \pt and the former in the higher \pt range probed in this analysis.
 \item \textit{Photon ID efficiency}: The uncertainties in the identification correction factors, derived in \ZtoEE events, are applied to the simulation. The highest \pt measurement bin ($\ptG > 200\GeV$) is dominated by the statistical uncertainty and so is uncorrelated from the uncertainty at lower \pt. An uncertainty in the extrapolation of the efficiencies to the maximum \ptG considered, 1.5\TeV, is also included.
 \item \textit{Photon energy scale}: The uncertainty from the calibration of the scale in data is determined using \ZtoEE events. Typical values are 0.1--3\% of the photon energy depending on \pt.
 \item \textit{Jet energy scale corrections}: The energies of jets in simulation are varied within the uncertainties of the scale correction measurements and propagated to all relevant observables. The magnitude varies with \pt and $\eta$, and is typically a few percent.
 \item \ptmiss \textit{scale}: The uncertainty in the jet energy scale also has a correlated effect on the \ptmiss calculation. An additional uncertainty is included for the energy scale of PF candidates not clustered in jets.
 \item \textit{Electrons misidentified as photons}: An uncertainty in the correction of the electron to photon misidentification rate, due to the subtraction of genuine photon events in the T\&P measurement, is derived.
 \item \textit{Jets misidentified as photons}: For jets misidentified as photons, the uncertainties from the method based on control samples in data described in Section~\ref{sec:jet_fakes} are applied, independently for each \ptG and $\abs{\etaG}$ bin that \fGam is measured in. These uncertainties are correlated between the electron and muon channels.
 \item \textit{Misidentified leptons}: The dominant source of uncertainty in the \fL measurement, described in Section~\ref{sec:lep_fakes}, comes primarily from the normalization of the subtracted prompt lepton contamination in the measurement region.
 \item \textit{Simulated sample size}: The statistical uncertainty due to the finite number of simulated events per bin is applied following the Barlow--Beeston method~\cite{Barlow:1993dm}.
\end{itemize}

Theoretical uncertainties in the \WGamma signal prediction because of missing higher orders in the perturbative cross section calculation are evaluated by varying the renormalization and factorization scales, denoted \muR and \muF, respectively, in simulated events by factors of 0.5 and 2.
The maximum variation among all possible pairings of \muR and \muF is considered, with the exception of the $(0.5, 2)$ and $(2, 0.5)$ combinations.
For constraints on \CWWW, the full uncertainty from this procedure is applied, which amounts to an 8--11\% normalization effect depending on the bin.
For the cross section measurements, the effect on the predicted cross sections is neglected, whereas the effect on the reconstruction-level acceptance is retained.
These values typically vary from 0.5 to 2\% and are treated as independent for each measured cross section bin.
Conversely, for the contamination of signal events from outside the fiducial region, the effect on both cross section and acceptance is included.
Uncertainties from the NNPDF3.1 PDF set eigenvectors are also evaluated following the PDF4LHC prescription~\cite{Butterworth:2015oua}, and affect the acceptance by up to 2\%.
For measurements that include a jet veto, an additional migration uncertainty between the 0-jet and $\geq$1-jet regions is applied, following the procedure described in Ref.~\cite{Stewart:2011cf}.
This results in an 8.6\% uncertainty in the 0-jet cross section.

Normalization uncertainties are assigned for each background process estimated with simulation to account for missing higher-order corrections and the PDF choice.
These are treated as uncorrelated between different processes, and are calculated following the same procedures as for the signal.
The missing higher-order correction uncertainties typically range from 2 to 3.5\%, and the PDF uncertainties are up to 4.8\%, depending on process.

\section{Results}\label{sec:results}
The results described in this section are obtained using statistical procedures developed by the ATLAS and CMS Collaborations, detailed in Refs.~\cite{LHC-HCG,ATLASCMSRun1} and implemented in the \textsc{RooFit}~\cite{Verkerke:2003ir} and \textsc{RooStats}~\cite{Moneta:2010pm} frameworks.

The parameters of interest \valpha for a particular result are estimated with their corresponding confidence intervals using a profile likelihood ratio test statistic \qStat~\cite{Cowan:2010st}, in which experimental and theoretical uncertainties are incorporated via nuisance parameters (NPs) \vtheta:
\begin{linenomath*}\begin{equation}\label{eq:LH}
  \qStat = -2\ln \left(\frac{\mathcal{L}\big(\valpha,\hvthetac\big)}
                                {\mathcal{L}(\hvalpha,\hvtheta)}\right).
\end{equation}\end{linenomath*}
The asymptotic approximation, whereby \qStat is assumed to follow a $\chi^{2}$ distribution, is used in the determination of the confidence intervals.
The likelihood functions in the numerator and denominator of Eq.~(\ref{eq:LH}) are constructed using products of binned signal and background probability density functions, as well as constraint terms for the NPs.
The quantities \hvalpha and \hvtheta denote the unconditional maximum likelihood estimates of the parameter values, while \hvthetac denotes the conditional maximum likelihood estimate for fixed values of the parameters of interest \valpha.
The binned likelihood function is expressed as
\begin{linenomath*}
\ifthenelse{\boolean{cms@external}}
{
\begin{multline}
\mathcal{L}(\text{data}\mid\valpha,\vtheta) \\ = \prod_{i}\text{Poisson}({n}_{i}\mid s_{i}(\valpha, \vtheta) + b_{i}(\vtheta)) p(\tvtheta\mid\vtheta),
\end{multline}
}
{
\begin{equation}
\mathcal{L}(\text{data}\mid\valpha,\vtheta) = \prod_{i}\text{Poisson}({n}_{i}\mid s_{i}(\valpha, \vtheta) + b_{i}(\vtheta)) p(\tvtheta\mid\vtheta),
\end{equation}
}
\end{linenomath*}
where $n_{i}$ is the number of observed events in each bin, $s$ and $b$ are the expected numbers of signal and background events, and \tvtheta represents the nominal values of the NPs, determined by external measurements.

For the measurement of a set of fiducial cross sections, $\valpha \equiv \sigj$, the signal expectation $s$ is constructed as
\begin{linenomath*}\begin{equation}\label{eqn:nsig_generic}
s_{i} = \sum_{j} \sigj \Rij(\vtheta) L + \sum_{k}\sigma_{k}^{\text{OOA}}  R_{ik}(\vtheta)L,
\end{equation}\end{linenomath*}
where $j$ runs over the set of \WGamma fiducial cross sections being measured; $k$ runs over a set of unmeasured fiducial cross section bins, referred to as out-of-acceptance (OOA) and discussed further below; $L$ is the integrated luminosity of a given data set; and \Rij is the response matrix, parameterized by \vtheta, giving the probability for an event in fiducial bin $j$ to be reconstructed in analysis bin $i$.

In this measurement a significant OOA contribution can come from events that are within the fiducial acceptance with the sole exception of the \ptmiss requirement, because \ptmiss is measured with poor resolution compared with the transverse momentum of the leptons and photons.
This component is therefore split into the same set of fiducial bins $j$ but with a modified \ptmiss selection.
Since it is not possible to measure these OOA cross sections \sigjmiss simultaneously, they are related to the \sigj using the predicted ratios from the \WGamma simulation: $\sigjmiss = (\sigjmissmc/\sigjmc) \sigj$.
The remaining OOA contribution \sigooa is small and is fixed to the value predicted by the simulation.
This includes events in which the electron or muon originates from a tau lepton decay.
Equation~(\ref{eqn:nsig_generic}) above can then be expressed as
\begin{linenomath*}
\ifthenelse{\boolean{cms@external}}
{
\begin{multline}\label{eqn:nsig_full}
s_{i} = \sum_{j} \left[ \sigj \Rij(\vtheta) L + \frac{\sigjmissmc}{\sigjmc} \sigj \Rij^{\text{miss}}(\vtheta) L \right] \\
 + \sigooa R^{\text{OOA}}_{i}(\vtheta)L.
\end{multline}
}
{
\begin{equation}\label{eqn:nsig_full}
s_{i} = \sum_{j} \left[ \sigj \Rij(\vtheta) L + \frac{\sigjmissmc}{\sigjmc} \sigj \Rij^{\text{miss}}(\vtheta) L \right] + \sigooa R^{\text{OOA}}_{i}(\vtheta)L.
\end{equation}
}
\end{linenomath*}

In addition to measuring the absolute \sigj, it is useful to present the results as fractional cross sections, $\fracj = \sigj/\sigtot$, where \sigtot is treated as a free parameter in the likelihood fit and $\sum{\fracj}$ is constrained to unity.
This facilitates comparisons between different predictions purely on the basis of observable shape, and reduces the impact of systematic uncertainties that are uniform across the distribution, for example from the integrated luminosity.

Tabulated versions of all the results in this section are provided in the HEPData record for this analysis~\cite{hepdata}.

\subsection{Differential cross sections}\label{sec:results-diff}
The differential cross sections $\sigj(\pp\to\PWpm\PGg\to\Pell^{\pm}\PGn\PGg)$, where \Pell denotes all three lepton flavors, are measured in the following fiducial region:
\begin{itemize}
\item $\ptL > 30\GeV$, $\abs{\etaL} < 2.5$,
\item $\ptG > 30\GeV$, $\abs{\etaG} < 2.5$,
\item $\ptmiss > 40\GeV$,
\item $\DRLG > 0.7$.
\end{itemize}

These selections are chosen to match as closely as possible with the reconstruction-level baseline described in Section~\ref{sec:event_sel}, with two exceptions: in the electron channel $\ptE > 35\GeV$ is required because of the higher trigger thresholds, and in the muon channel $\abs{\etaM} < 2.4$ is applied to reflect the acceptance of the muon system.

The lepton momenta are completed by adding the four-momenta of any photons with $\DRLG < 0.1$ to the four-momentum of the lepton.
A smooth-cone photon isolation selection is also applied.
This is necessary to ensure an infrared-safe cross section calculation in perturbative QCD.\@
It requires that for every cone in $\eta$--$\phi$ around the photon direction with radius $\delta < \delta_{0}$ the total scalar transverse energy sum \et must be smaller than $\et^{\text{max}}(\delta)$~\cite{Frixione:1998jh} where
\begin{linenomath*}\begin{equation}
\et^{\text{max}}(\delta) = \epsilon\ptG{\left( \frac{1-\cos{\delta}}{1-\cos{\delta_{0}}}\right)}^{n},
\end{equation}\end{linenomath*}
and the values of the tunable parameters are set as $\delta_{0}=0.4$, $\epsilon=1.0$, and $n=1$.

Differential cross sections are measured for \ptG, \etaG, \DRLG, \DetaLG, and \mTcluster under the baseline selection.
The \aphif observable is measured only under the higher-\pt EFT selection, given in Section~\ref{sec:results-eft}, where the resolution is sufficiently improved compared to the baseline selection.

The signal model is constructed as in Eq.~(\ref{eqn:nsig_full}), where the OOA contributions \sigjmiss are defined for events with $0 < \ptmiss < 40\GeV$, but otherwise within the fiducial region for $j$.
A simultaneous fit is performed to the reconstructed distributions of the observables in each year and for each of the electron and muon channels, where lepton flavor universality in the \PW boson decay is assumed.
These distributions use the same bin boundaries as the fiducial-level cross section bins.
The measurements are compared with the predictions from the NLO \MGvATNLO{}+\PYTHIA{} simulation (\MGPYSHORT) and NNLO predictions from the \MATRIX{}~\cite{Grazzini:2017mhc,Grazzini:2015nwa}, \MCFM{}~\cite{Campbell:2021mlr,Campbell:2011bn}, and \GENEVA{}~\cite{Cridge:2021hfr,Alioli:2012fc,Alioli:2015toa,Alioli:2021qbf} frameworks.

The \MATRIX{} calculation implements a dedicated \qT-subtraction formalism~\cite{Catani:2012qa,Catani:2007vq}, where \qT is the total transverse momentum of the colorless final state particles.
The framework utilizes amplitudes up to one-loop level from \textsc{OpenLoops}~\cite{Cascioli:2011va,Denner:2016kdg}, as well as dedicated two-loop calculations~\cite{Gehrmann:2011ab}.
The \MCFM{} prediction uses the $N$-jettiness slicing method, with all one- and two-loop amplitudes~\cite{Gehrmann:2011ab} calculated using analytic formulae.
It is given with and without the inclusion of NLO electroweak corrections, which have been shown~\cite{Denner:2014bna,Campbell:2021mlr} to modify \WGamma differential distributions by $\mathcal{O}(10\%)$.
The NLO electroweak correction to the $\Pq\Paq$-initiated process is applied multiplicatively to the NNLO QCD cross section, which is then summed with the contribution from the distinct photon-induced $\Pq\PGg$ channel.
The \GENEVA{} framework combines the fixed-order NNLO result with a next-to-next-to-leading logarithmic resummed calculation, in this case using soft-collinear effective theory to resum the 0-jettiness variable.
The output events are interfaced with \PYTHIA{}8~\cite{Sjostrand:2014zea} for parton showering, hadronization and underlying event simulation.

\begin{figure*}[hbtp]
\centering
\includegraphics[width=\cmsFigWidthU]{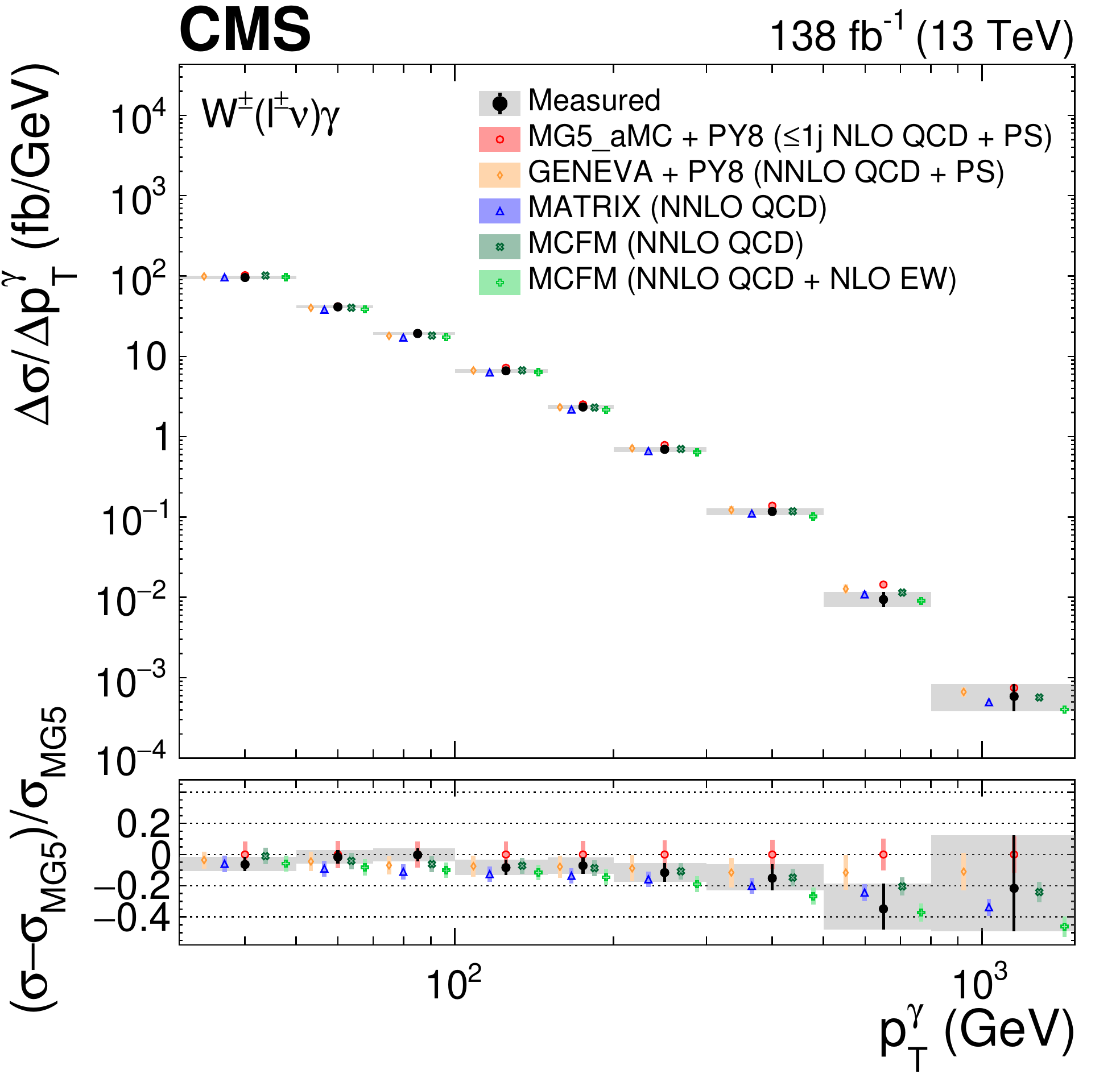}
\includegraphics[width=\cmsFigWidthU]{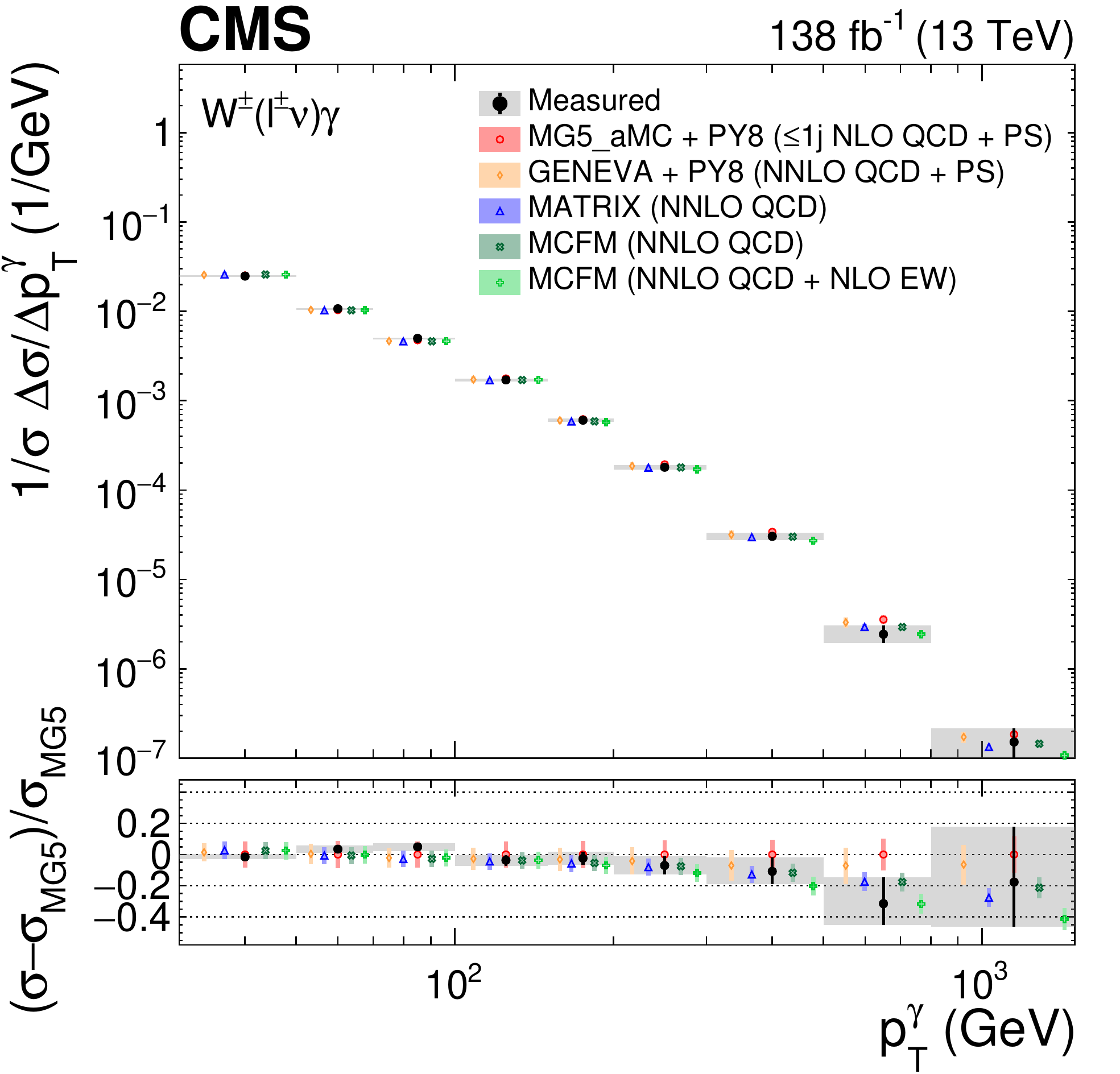} \\
\includegraphics[width=\cmsFigWidthU]{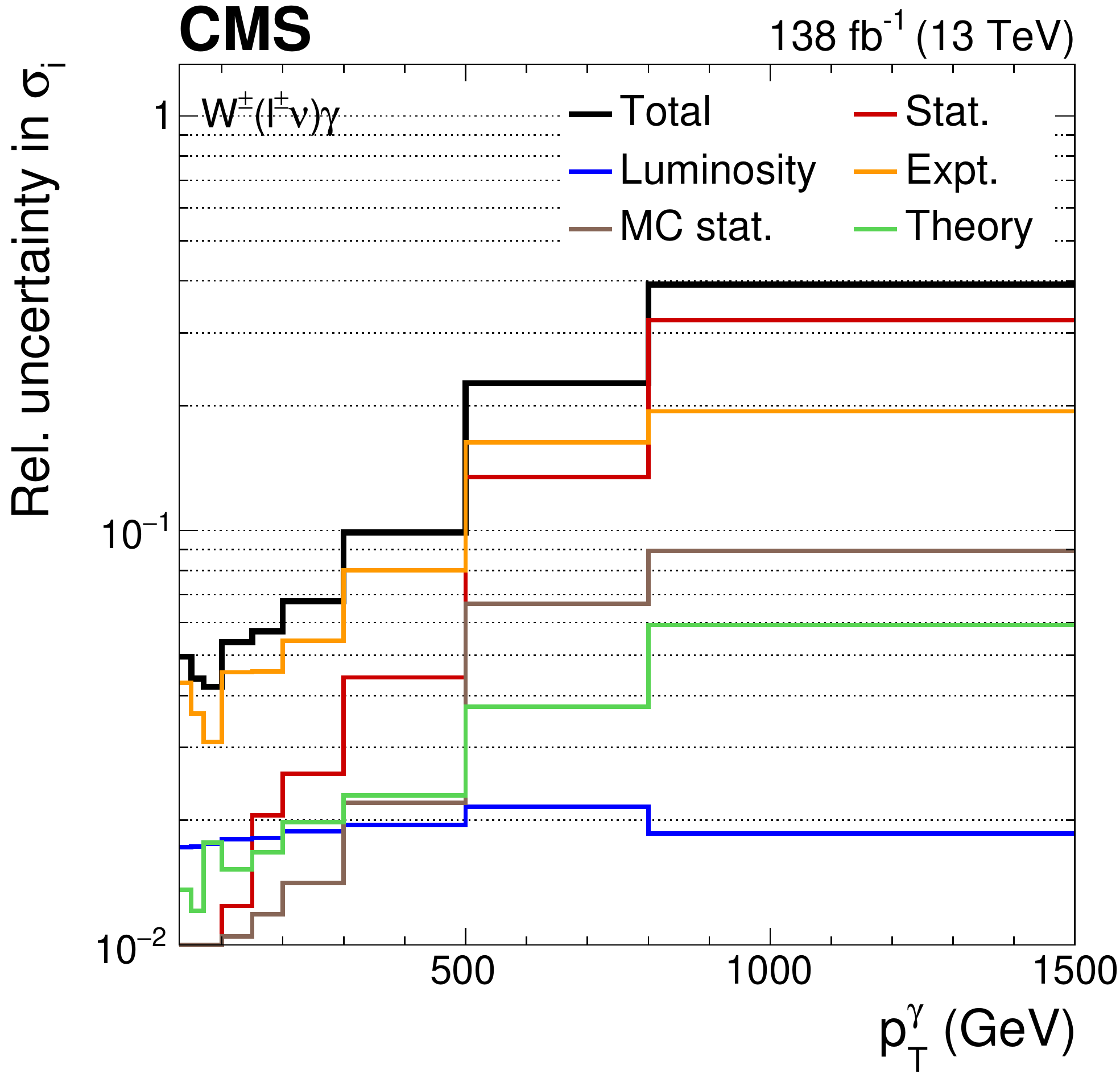}
\includegraphics[width=\cmsFigWidthU]{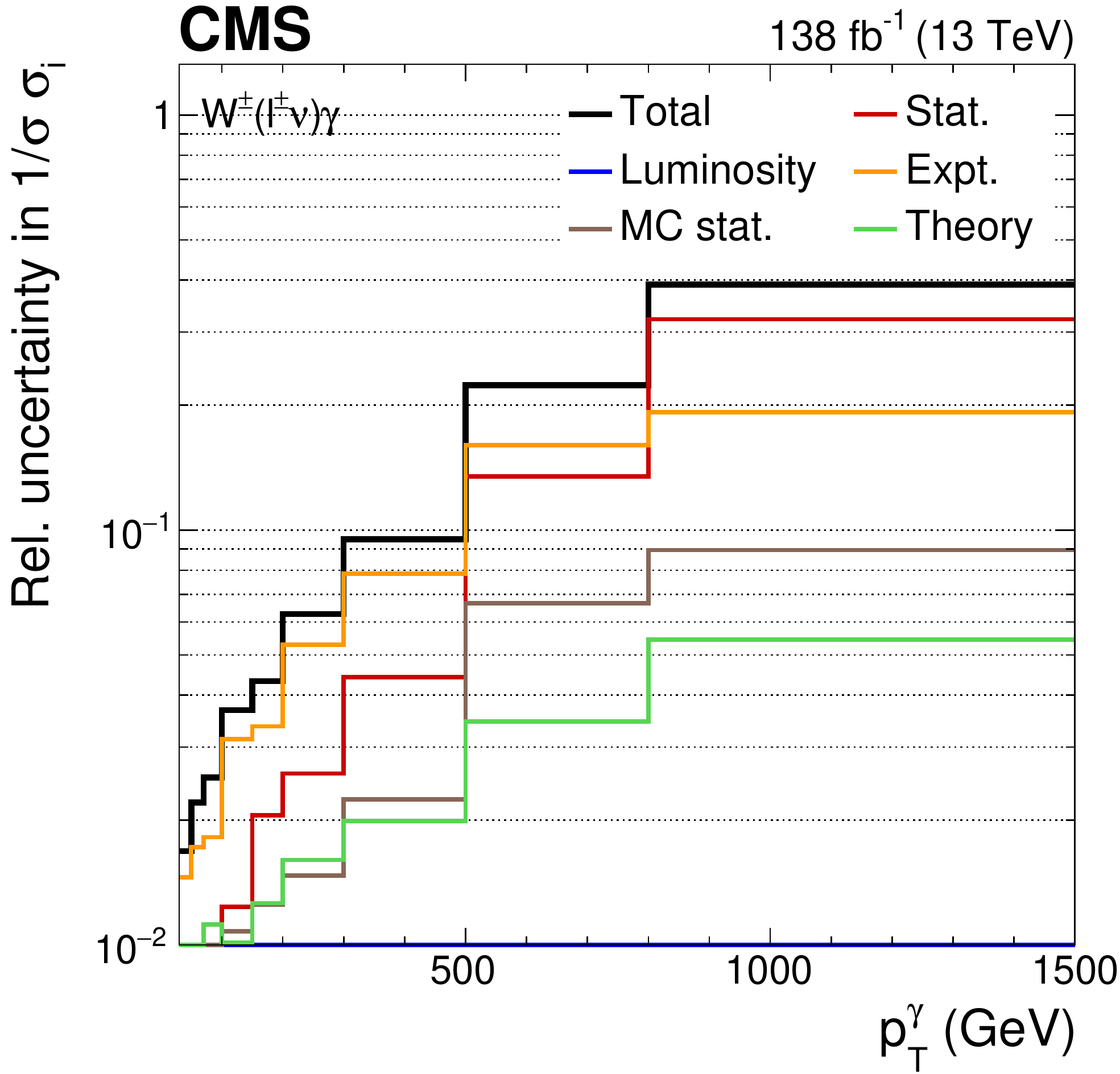} \\
\includegraphics[width=\cmsFigWidthU]{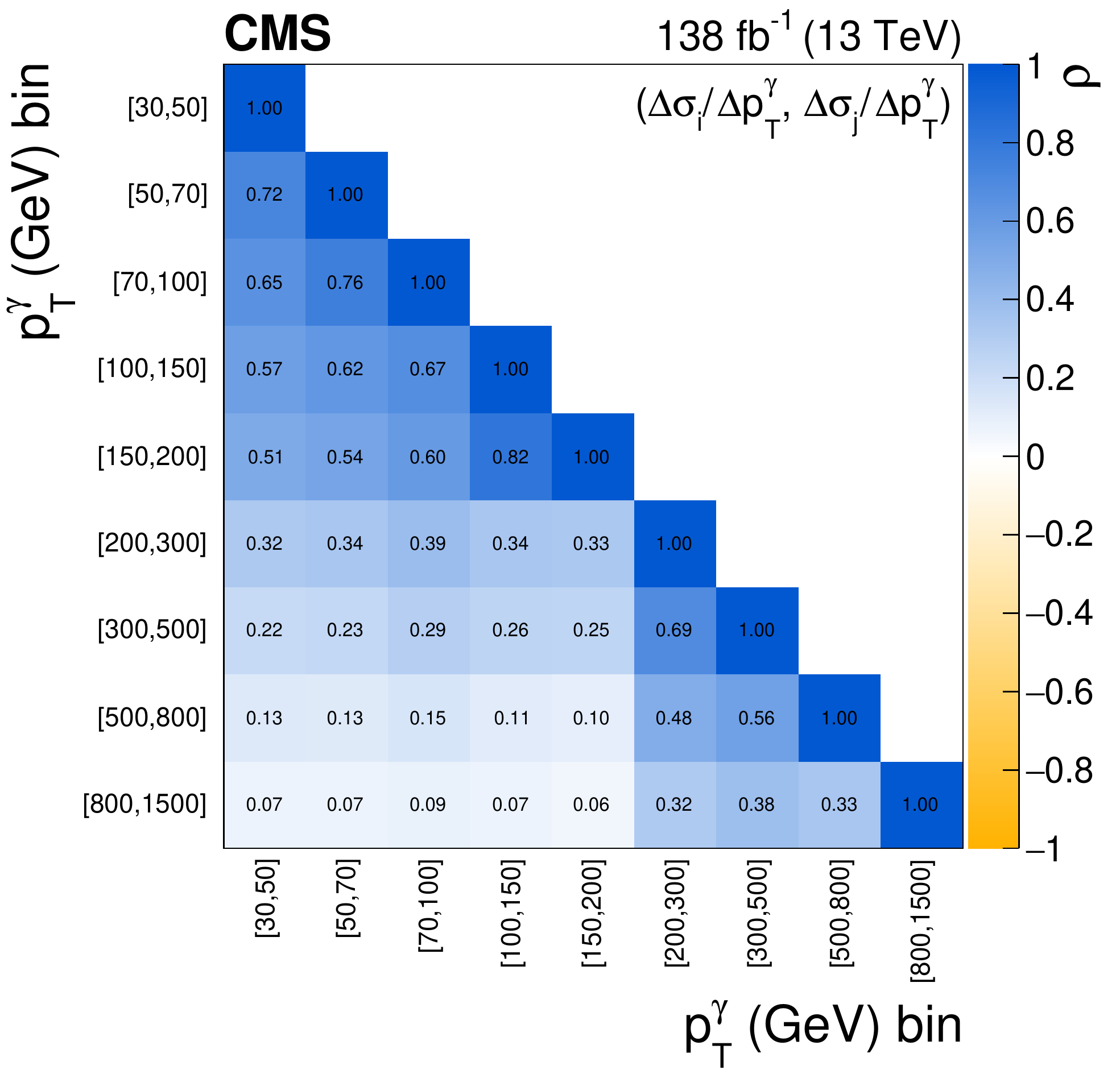}
\includegraphics[width=\cmsFigWidthU]{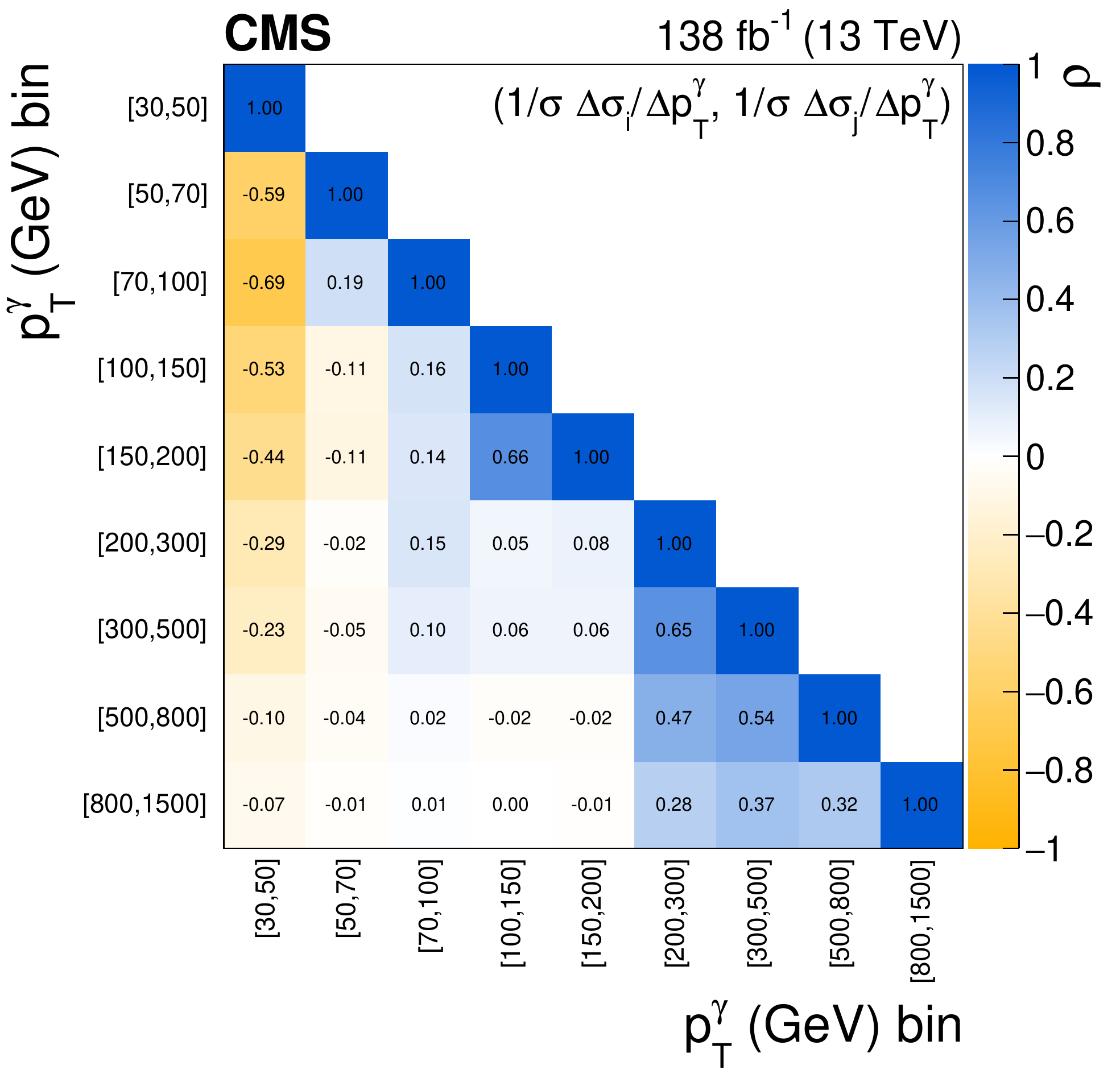} \\
\caption{The measured \ptG absolute (left) and fractional (right) differential cross sections (upper), compared with the \MGPYSHORT{}, \GENEVA{}, \MATRIX{}, and \MCFM{} predictions, and corresponding uncertainty decomposition (center) and correlation matrices (lower). In the upper figures, the black vertical bars give the total uncertainty on each measurement. The predictions are offset horizontally in each bin to improve visibility, and the corresponding vertical bars show the missing higher-order correction uncertainties.}\label{fig:res_1d_fid_pt_binned_2}
\end{figure*}

\begin{figure*}[hptb]
\centering
\includegraphics[width=\cmsFigWidthW]{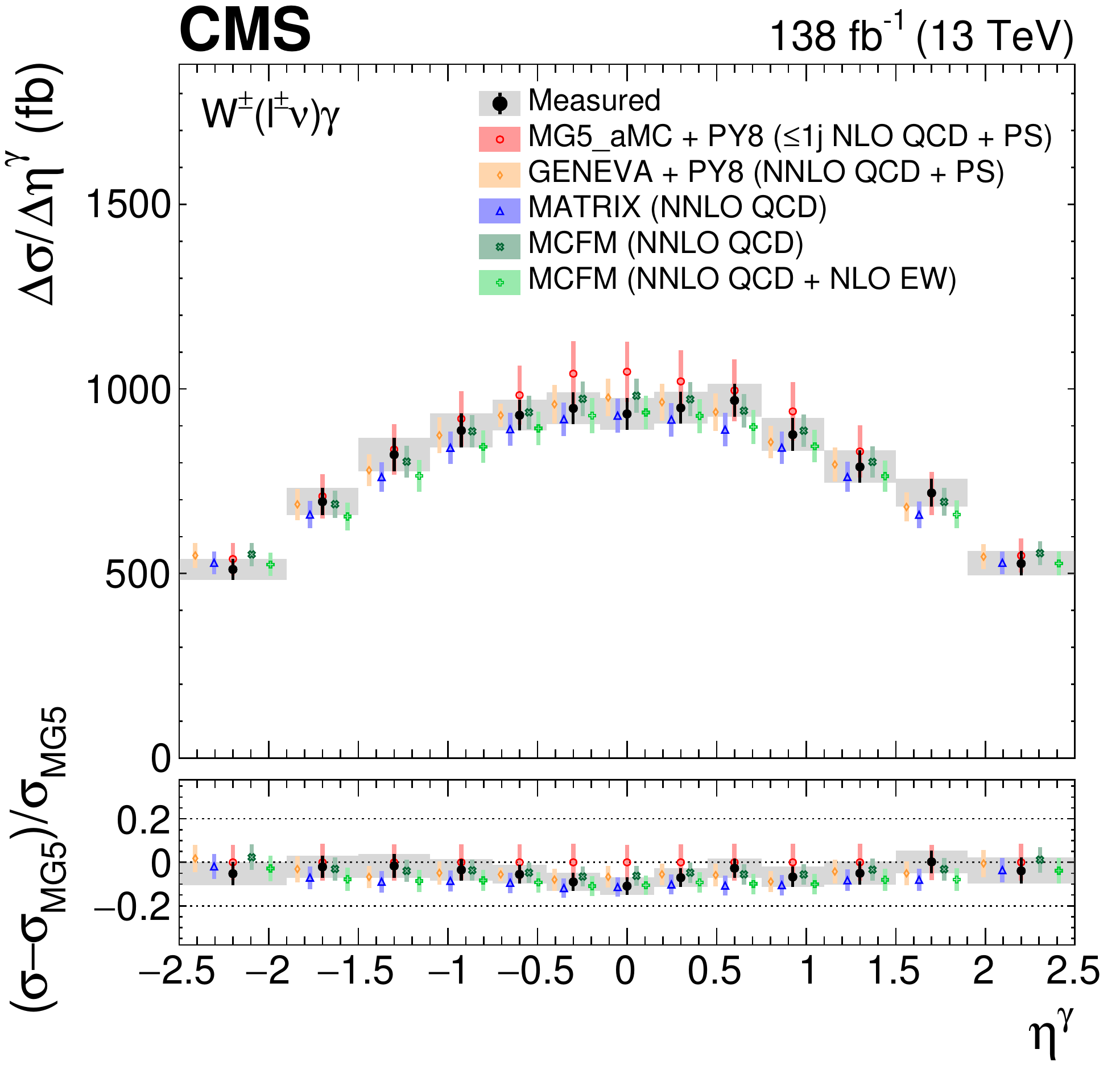}
\includegraphics[width=\cmsFigWidthW]{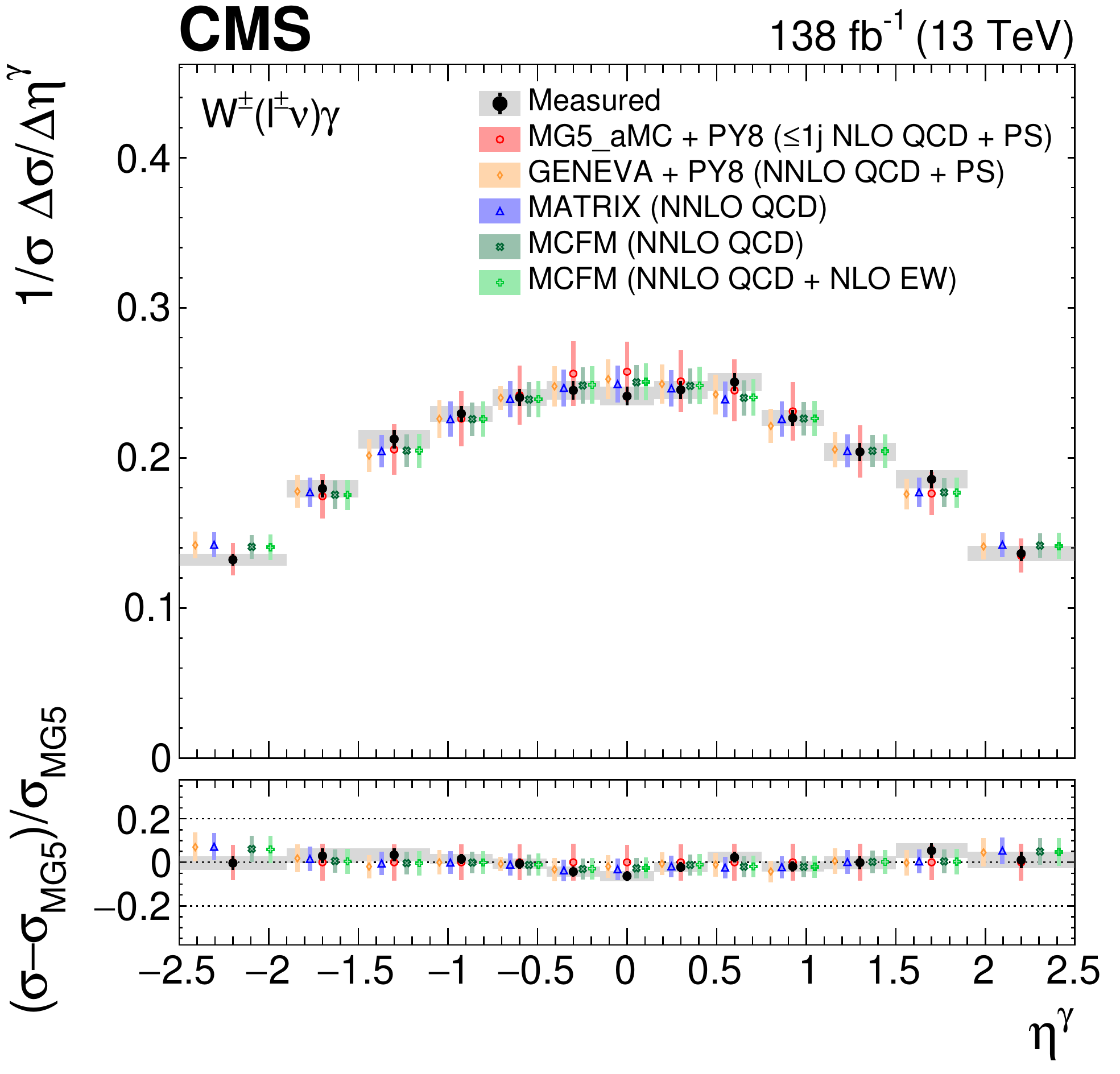} \\
\includegraphics[width=\cmsFigWidthW]{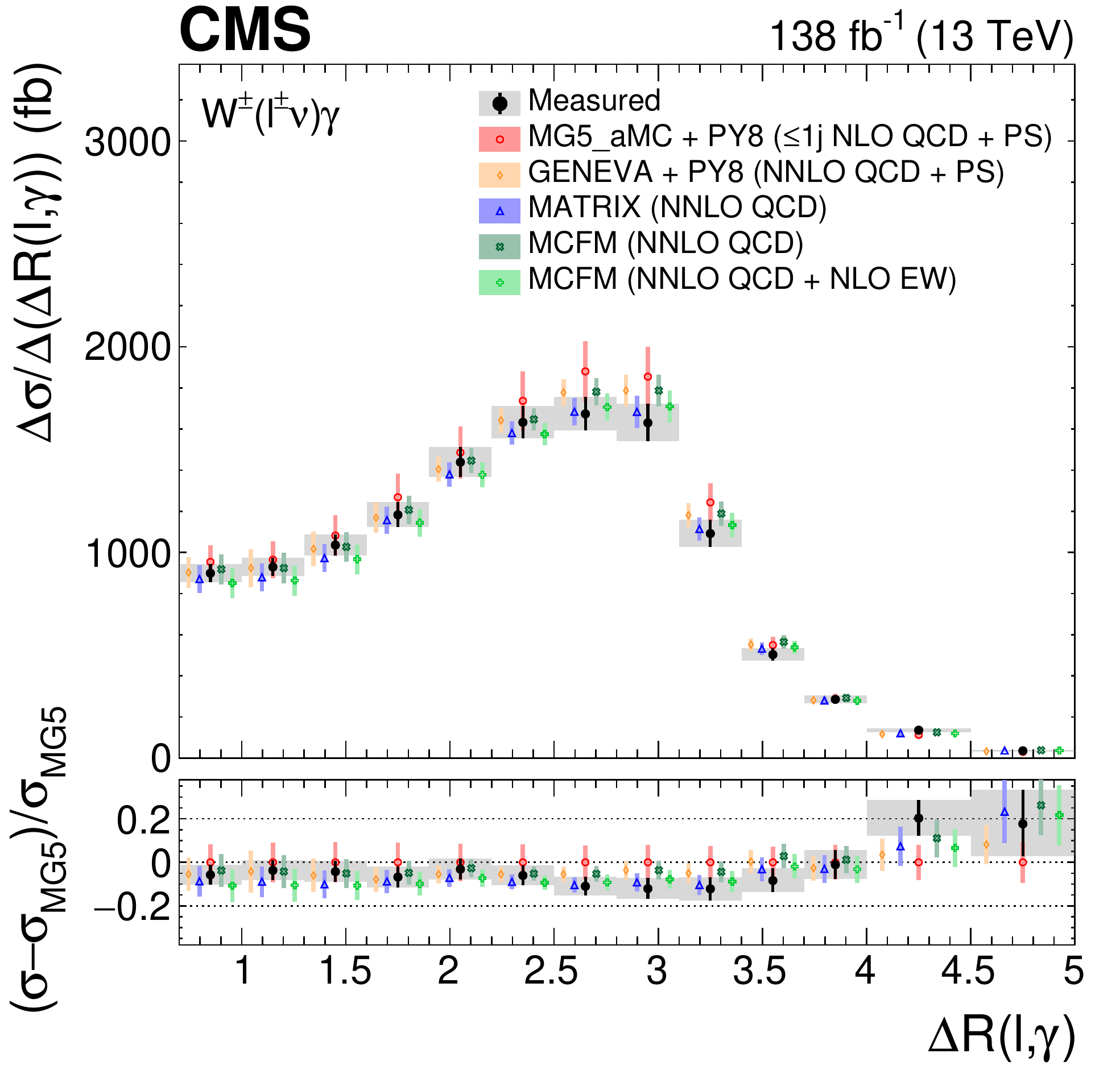}
\includegraphics[width=\cmsFigWidthW]{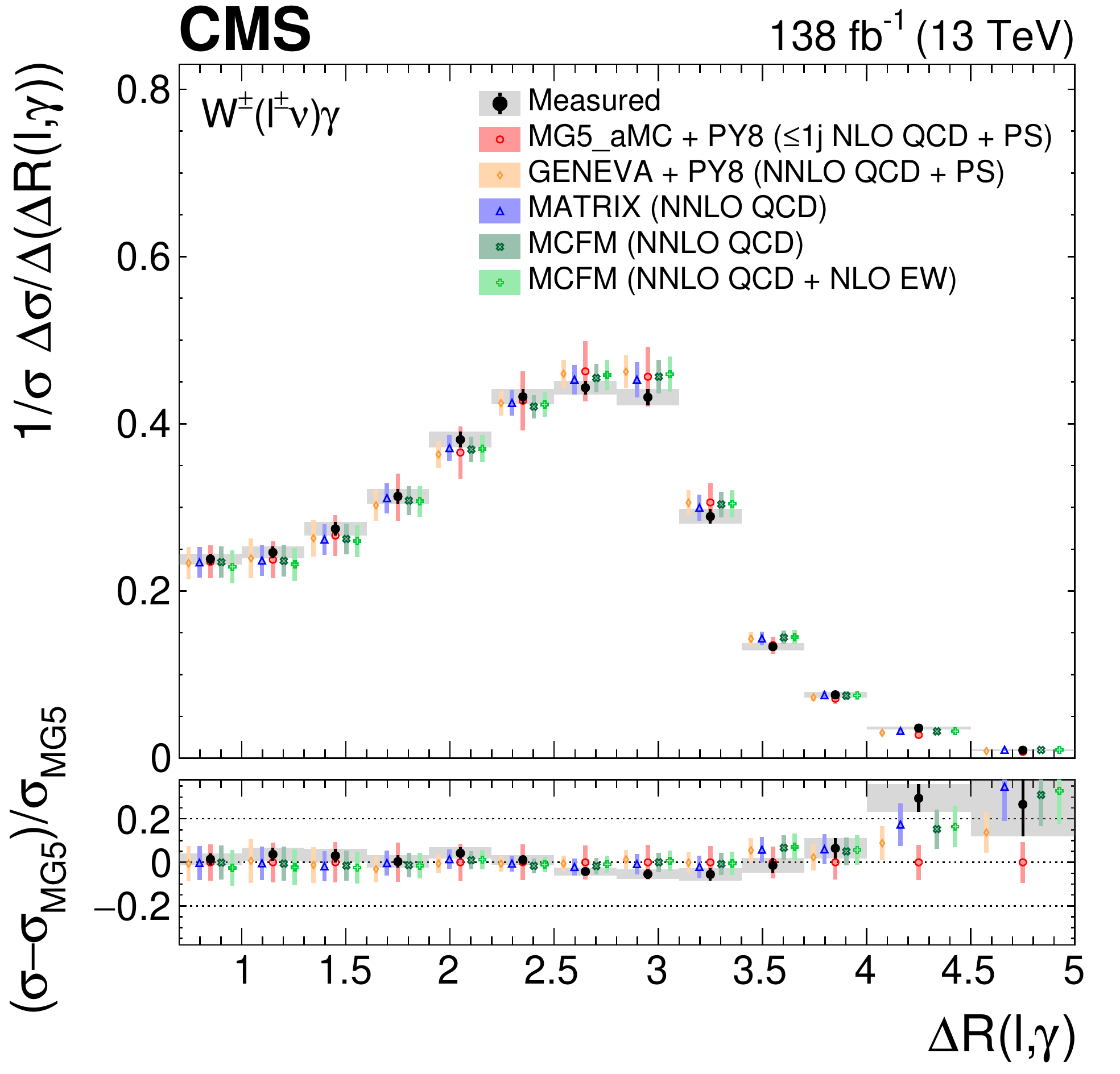} \\
\caption{The measured absolute (left) and fractional (right) differential cross sections for \etaG (upper) and \DRLG (lower), compared with the \MGPYSHORT{}, \GENEVA{}, \MATRIX{}, and \MCFM{} predictions. The black vertical bars give the total uncertainty on each measurement. The predictions are offset horizontally in each bin to improve visibility, and the corresponding vertical bars show the missing higher-order correction uncertainties.}\label{fig:res_1d_fid_l0p0_dr_binned_2}
\end{figure*}

\begin{figure*}[hptb]
\centering
\includegraphics[width=\cmsFigWidthW]{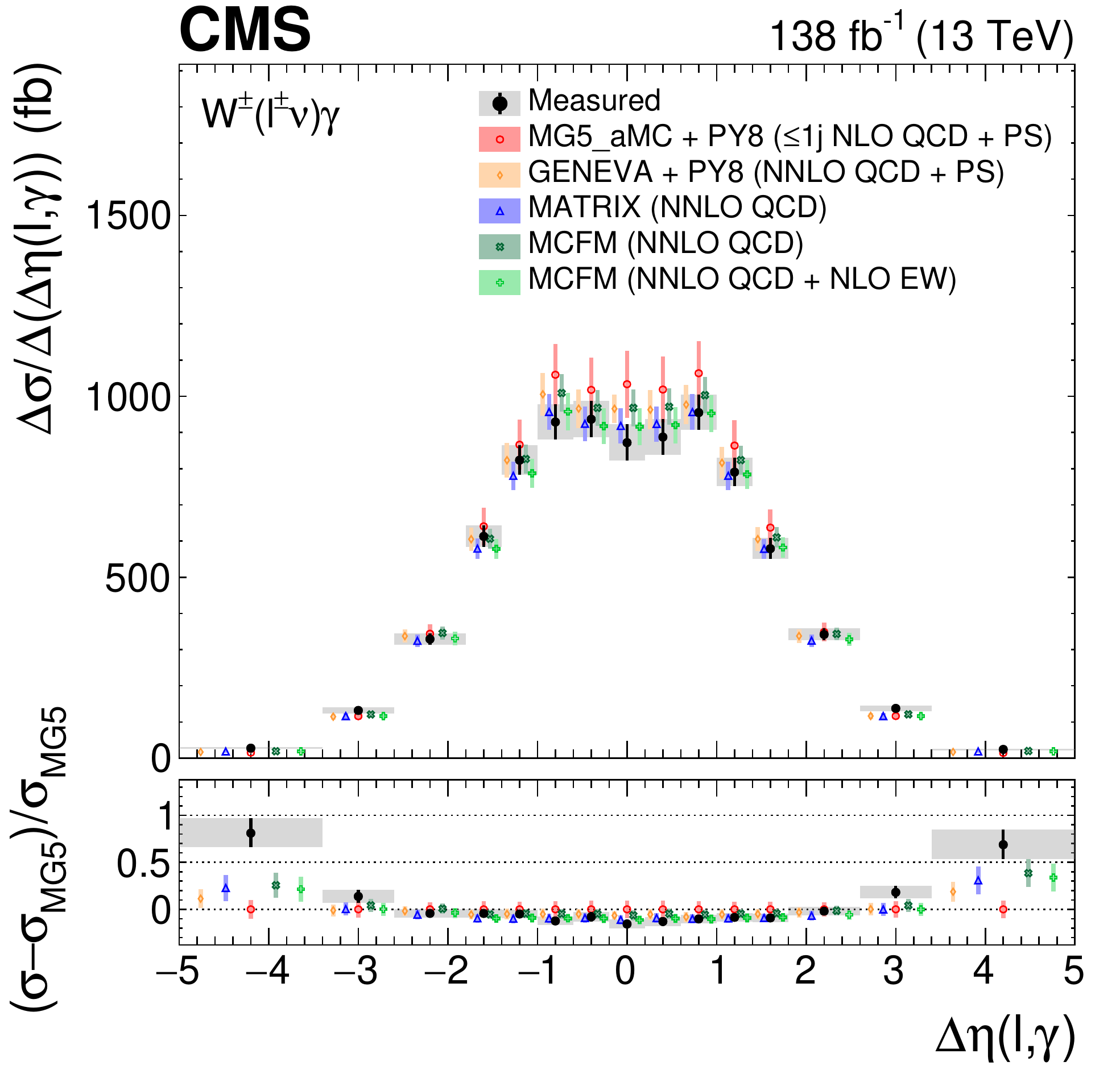}
\includegraphics[width=\cmsFigWidthW]{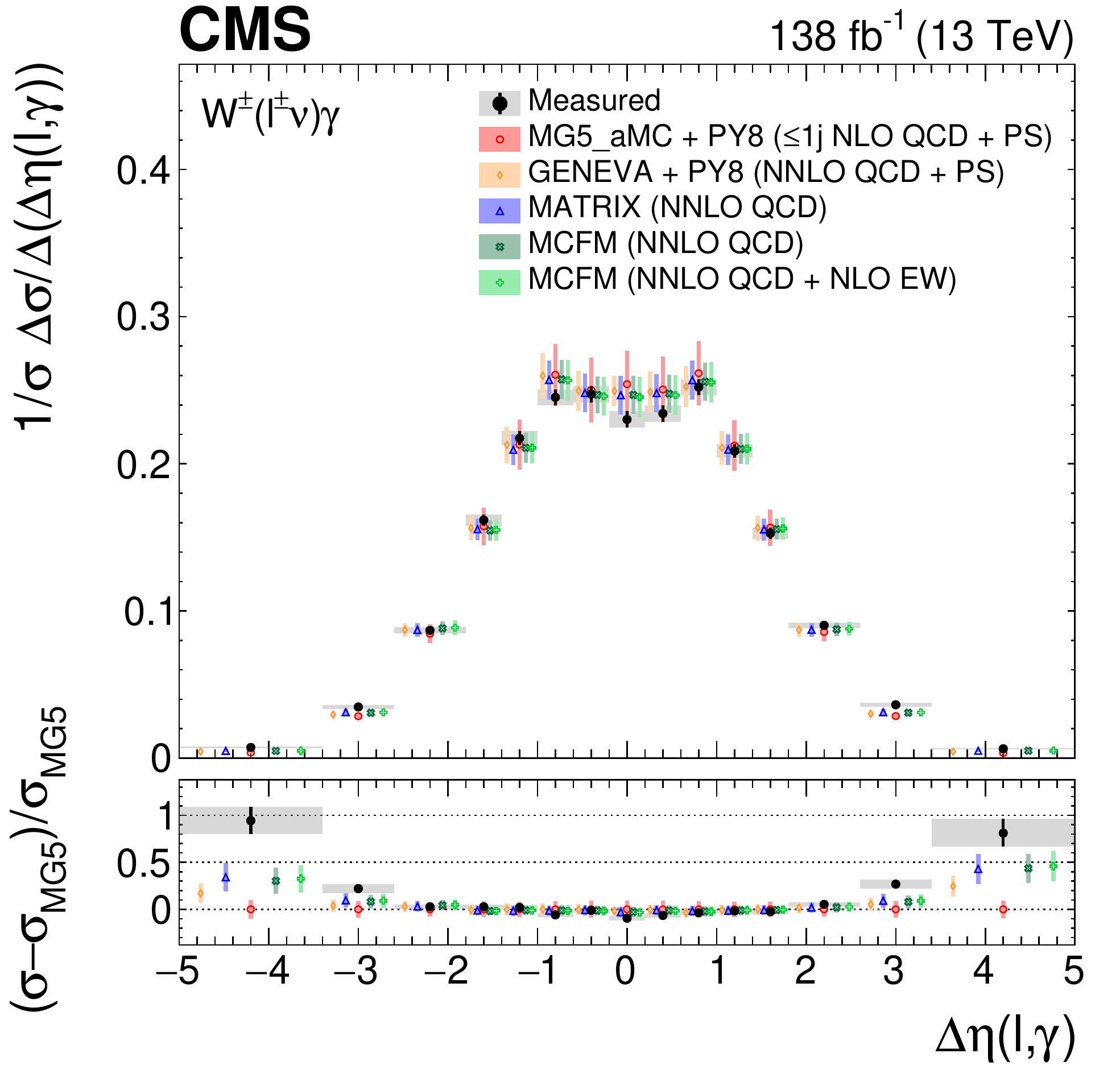} \\
\includegraphics[width=\cmsFigWidthW]{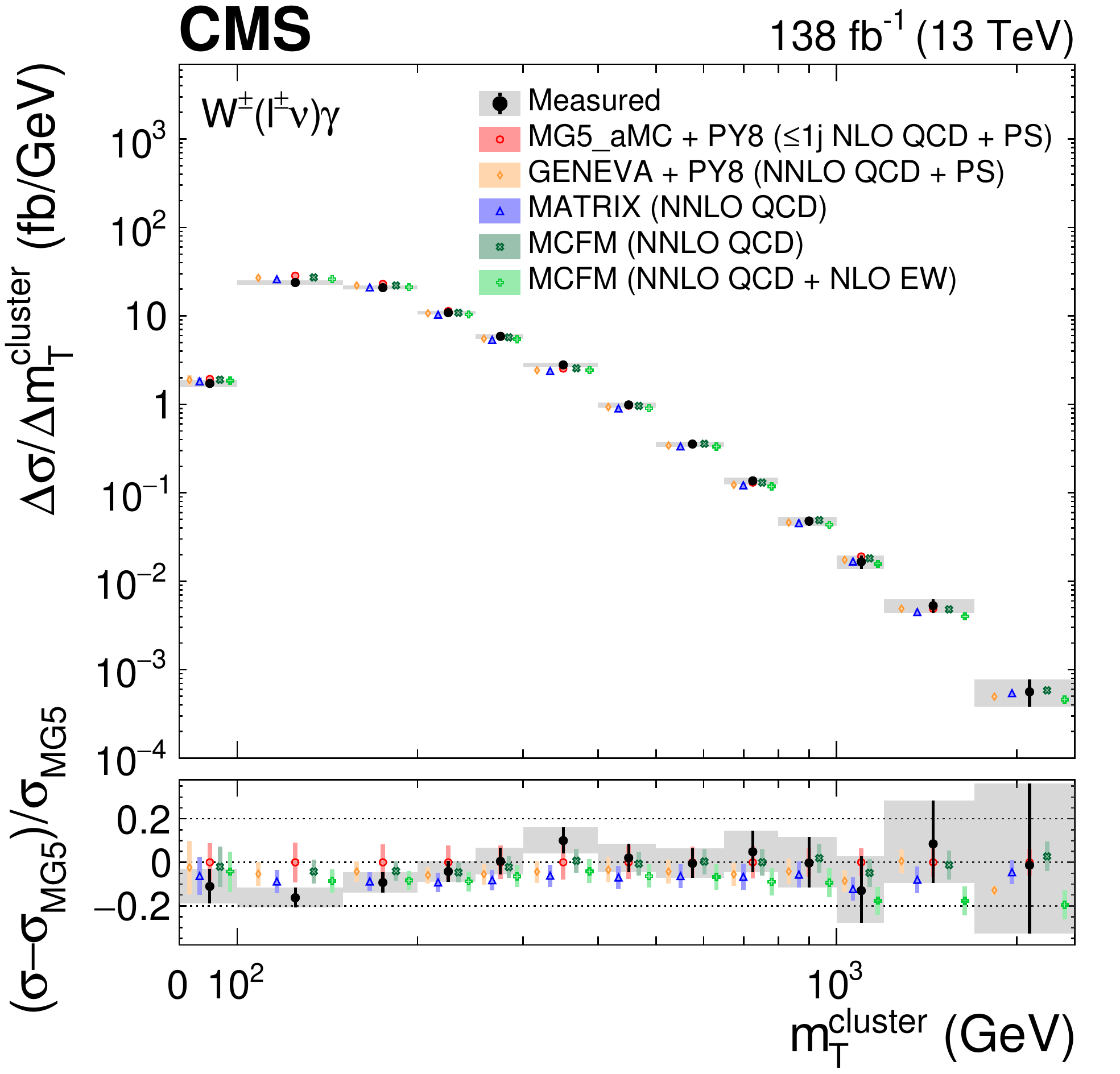}
\includegraphics[width=\cmsFigWidthW]{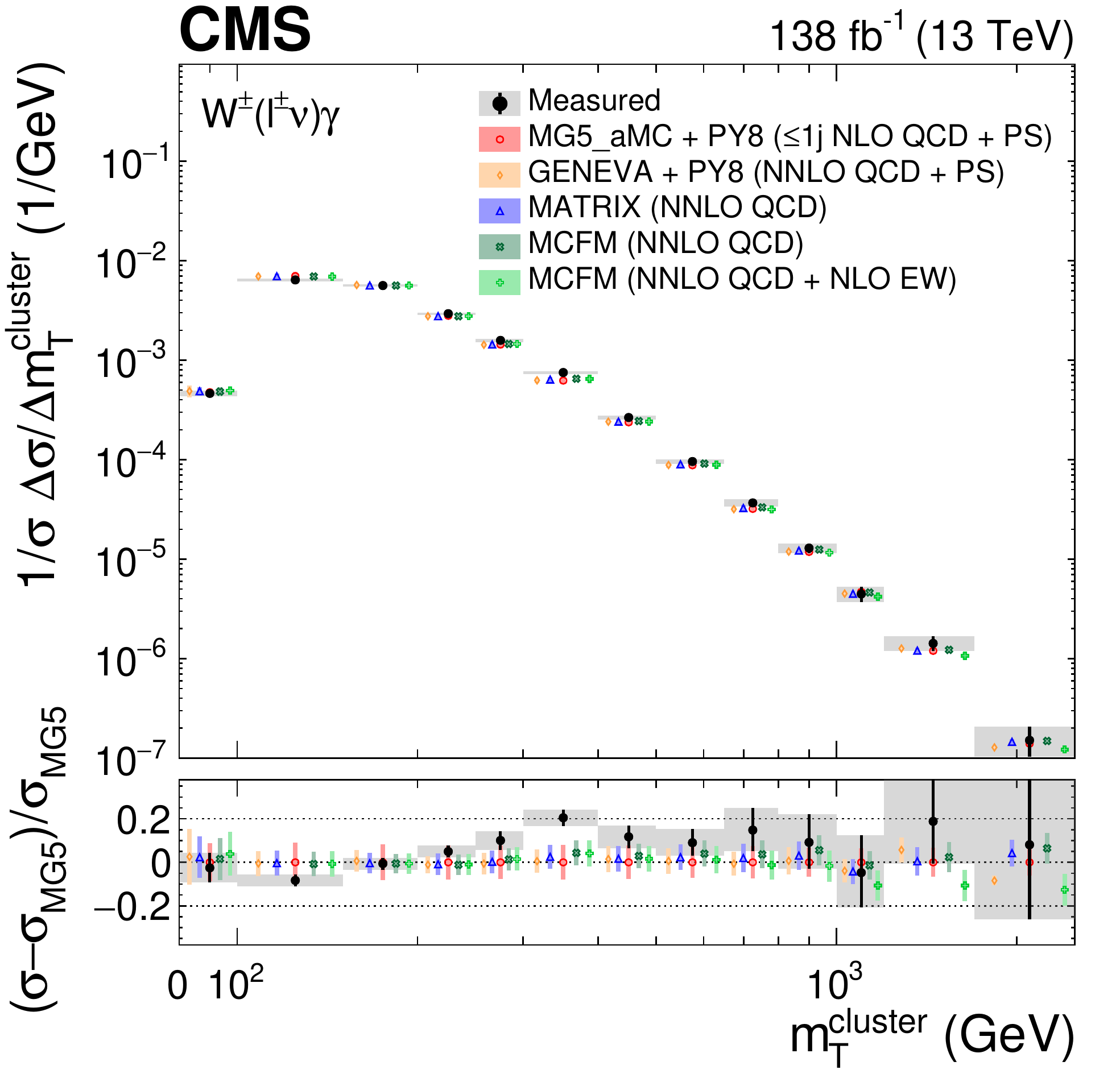} \\
\caption{The measured absolute (left) and fractional (right) differential cross sections for \DetaLG (upper) and \mTcluster (lower), compared with the \MGPYSHORT{}, \GENEVA{}, \MATRIX{}, and \MCFM{} predictions. The black vertical bars give the total uncertainty on each measurement. The predictions are offset horizontally in each bin to improve visibility, and the corresponding vertical bars show the missing higher-order correction uncertainties.}\label{fig:res_1d_fid_l0p0_deta_binned_2}
\end{figure*}

Figures~\ref{fig:res_1d_fid_pt_binned_2}--\ref{fig:res_1d_fid_l0p0_deta_binned_2} show the resulting measurements for each observable.
For \ptG in Fig.~\ref{fig:res_1d_fid_pt_binned_2}, the uncertainty composition and correlation matrices are also given.

The absolute cross section, as a function of \ptG, is measured with a precision of 4--5\% at low \pt, dominated by experimental systematic uncertainties, to 40\% in the highest bin, dominated by the statistical uncertainty.
The fractional cross section uncertainty falls to 2--3\% at low \pt, owing to a cancellation of systematic uncertainties.
Both measurements show a tendency towards the lower values of the NNLO predictions at high \pt.
Although the NLO electroweak correction is predicted to reduce the cross section in this region by up to 30\%, the current precision is not sufficient to distinguish this effect.
The correlation matrices illustrate how large positive correlations between bins in the absolute measurement are reduced significantly when using the fractional parameterization.

The \DRLG measurement is in agreement with all the predictions in the bulk of the distribution, but favors the higher NNLO predictions in the $\DR > 4.0$ region.
The \DetaLG measurement shows significant enhancement in the highest $\abs{\Deta}$ bins compared with the \MGPYSHORT{} prediction.
The NNLO predictions feature a similar, though less pronounced effect.
The \mTcluster distribution is broadly in agreement with the predictions, however with a moderate disagreement in the shape of the peak in the 100--300\GeV region.

\subsection{Cross section as a function of the jet multiplicity}\label{sec:results-jets}
\begin{figure*}[hbtp]
\centering
\includegraphics[width=\cmsFigWidthW]{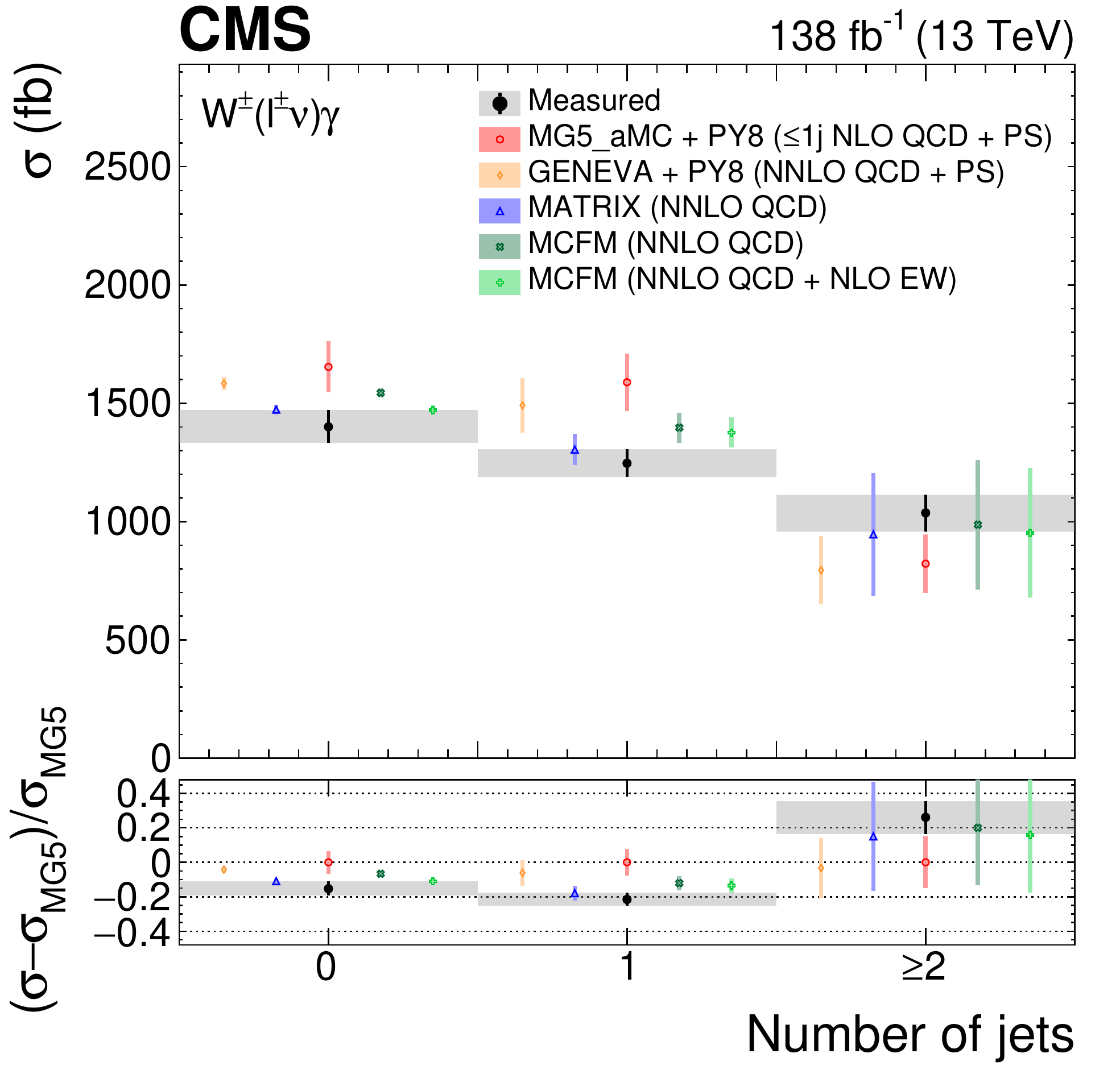}
\includegraphics[width=\cmsFigWidthW]{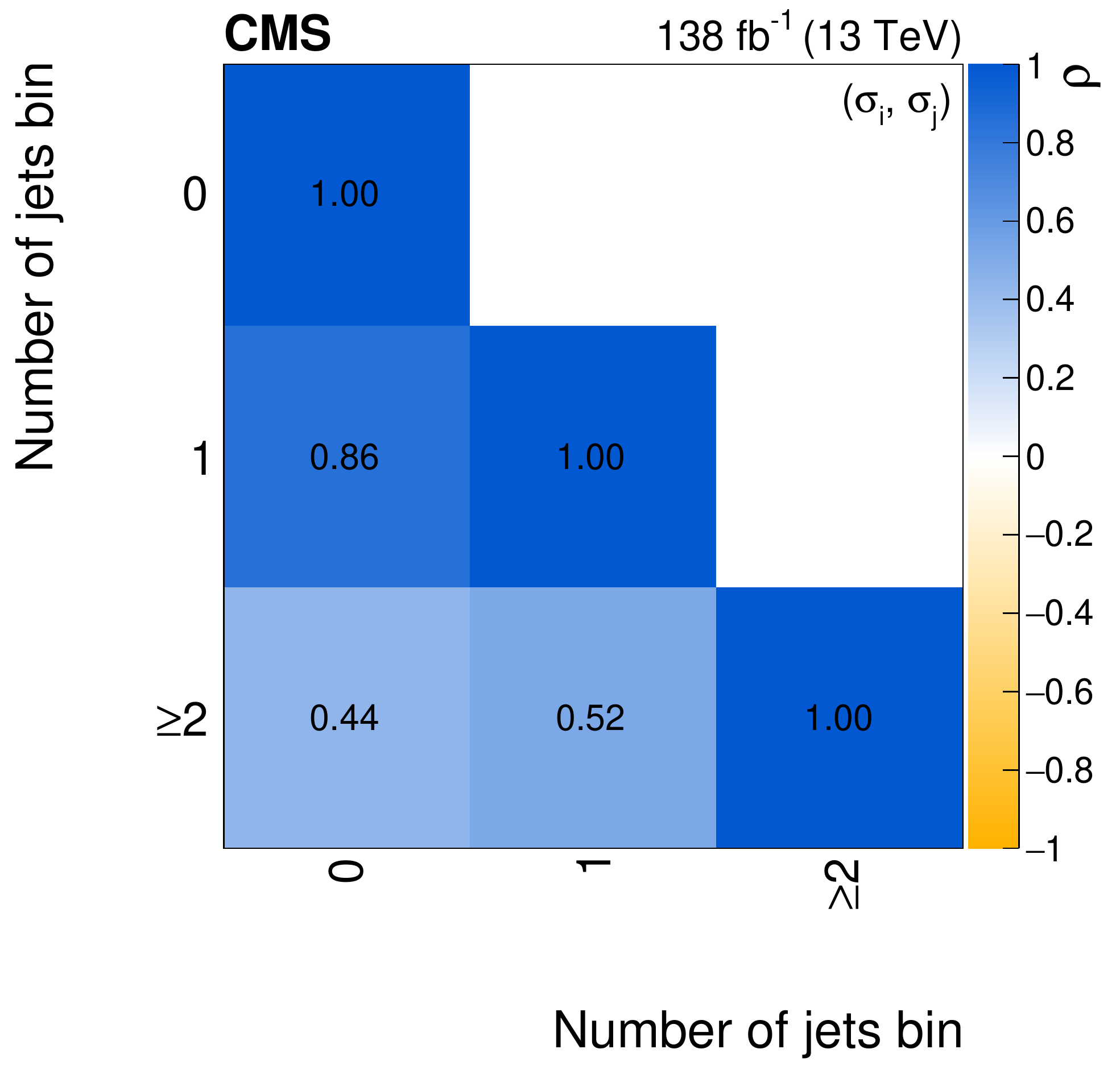}
\caption{The measured jet multiplicity cross sections (left) and corresponding correlation matrix (right). In the left figure, the black vertical bars give the total uncertainty on each measurement. The predictions are offset horizontally in each bin to improve visibility, and the corresponding vertical bars show the missing higher-order correction uncertainties.}\label{fig:res_1d_fid_njet_binned}
\end{figure*}

The cross section is measured as a function of the jet multiplicity using the same fiducial region as the previous results.
The fiducial selection for the jets is chosen to match the experimental selection.
Jets are clustered from final state particles, excluding neutrinos, with the anti-\kt algorithm and a distance parameter of 0.4.
They are required to have $\pt > 30\GeV$, $\aeta < 2.5$, and be separated from both the lepton and photon by $\DR > 0.4$.
The measured cross sections and their correlation matrix are shown in Fig.~\ref{fig:res_1d_fid_njet_binned}, and the cross section values are listed in Table~\ref{tab:res_xsec_1d_fit_njet_binned}.

\begin{table*}[hbt]
\centering
\topcaption{Cross sections measured in bins of jet multiplicity and comparison with the \MGPYSHORT{}, \GENEVA{}, \MATRIX{}, and \MCFM{} predictions. The \MCFM{} column marked (EW) is the NNLO QCD prediction combined with NLO electroweak corrections, as described in the text.}
\cmsTable{
\renewcommand{\arraystretch}{1.3}
\begin{scotch}{l c c c c c c c c}
 & \multicolumn{8}{c}{$\sigma (\fbns)$}  \\
Number of jets  &   Best fit &   Stat &  Syst &   \MGPYSHORT{} & \MATRIX{} & \MCFM{} & \MCFM{} (EW) & \GENEVA{} \\
\hline
$= 0$     & $1400_{-67}^{+71}$ & $ _{-11}^{+11}$ & $ _{-67}^{+70}$ & $ 1650 \pm 110 $ & $ 1473 \pm 19 $ & $ 1544 \pm 18 $ & $ 1471 \pm 18 $ & $ 1584 \pm 26 $ \\
$= 1$     & $1246_{-58}^{+61}$ & $ _{-11}^{+11}$ & $ _{-57}^{+60}$ & $ 1590 \pm 120 $ & $ 1304 \pm 64 $ & $ 1397 \pm 62 $ & $ 1376 \pm 62 $ & $ 1490 \pm 110 $ \\
$\geq2$   & $1037_{-79}^{+78}$ & $ _{-10}^{+10}$ & $ _{-78}^{+77}$ & $ 820 \pm 120 $  & $ 950 \pm 260 $ & $ 990 \pm 270 $ & $ 950 \pm 270 $ & $ 790 \pm 140 $ \\
\end{scotch}}\label{tab:res_xsec_1d_fit_njet_binned}
\end{table*}

The measured 0-jet cross section is 5--15\% smaller than the various predictions.
A similar trend is observed in the 1-jet measurement, whereas the $\geq$2-jet measurement is higher than the predictions.
Each measurement is positively correlated with the other two, with the strongest correlation of $+0.86$ between the 0-jet and 1-jet bins.

\subsection{The radiation amplitude zero effect}\label{sec:results-raz}
In \pp collisions the RAZ effect is observed~\cite{Chatrchyan:2013fya} in the rapidity difference between the charged lepton and the photon; it manifests itself as a suppression of the cross section near $\DetaLG = 0$.
This region is of interest because new physics effects could make the minimum in this region less pronounced.
Several models with resonances that would exhibit this feature are discussed in Ref.~\cite{Capdevilla:2019zbx}.
In the EFT framework this enhancement is generated by \CWWW, however, the \DetaLG observable is significantly less sensitive than the approach pursued in this paper, as described in Section~\ref{sec:intres}.

\begin{figure*}[hbtp]
\centering
\includegraphics[width=\cmsFigWidthW]{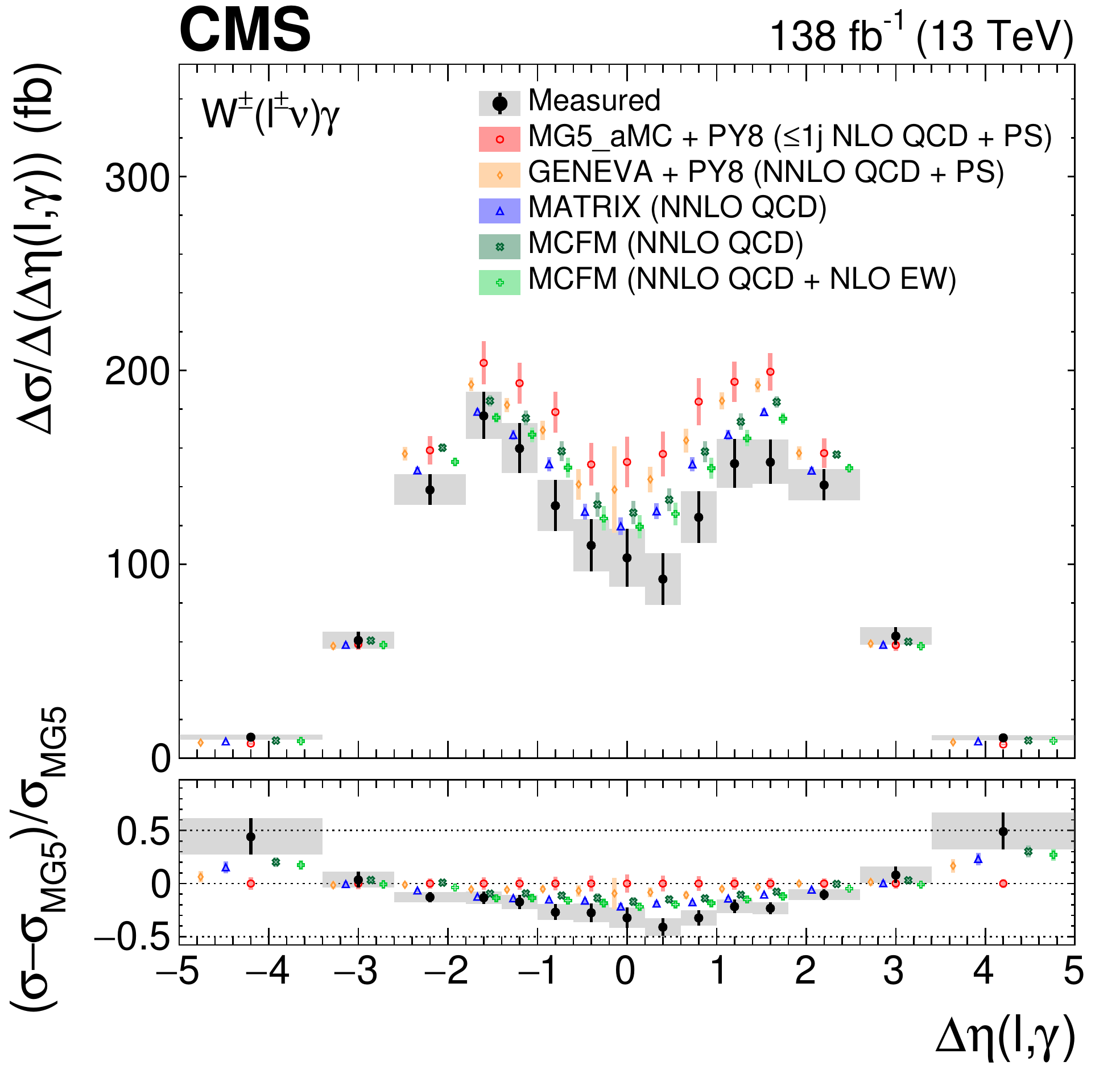}
\includegraphics[width=\cmsFigWidthW]{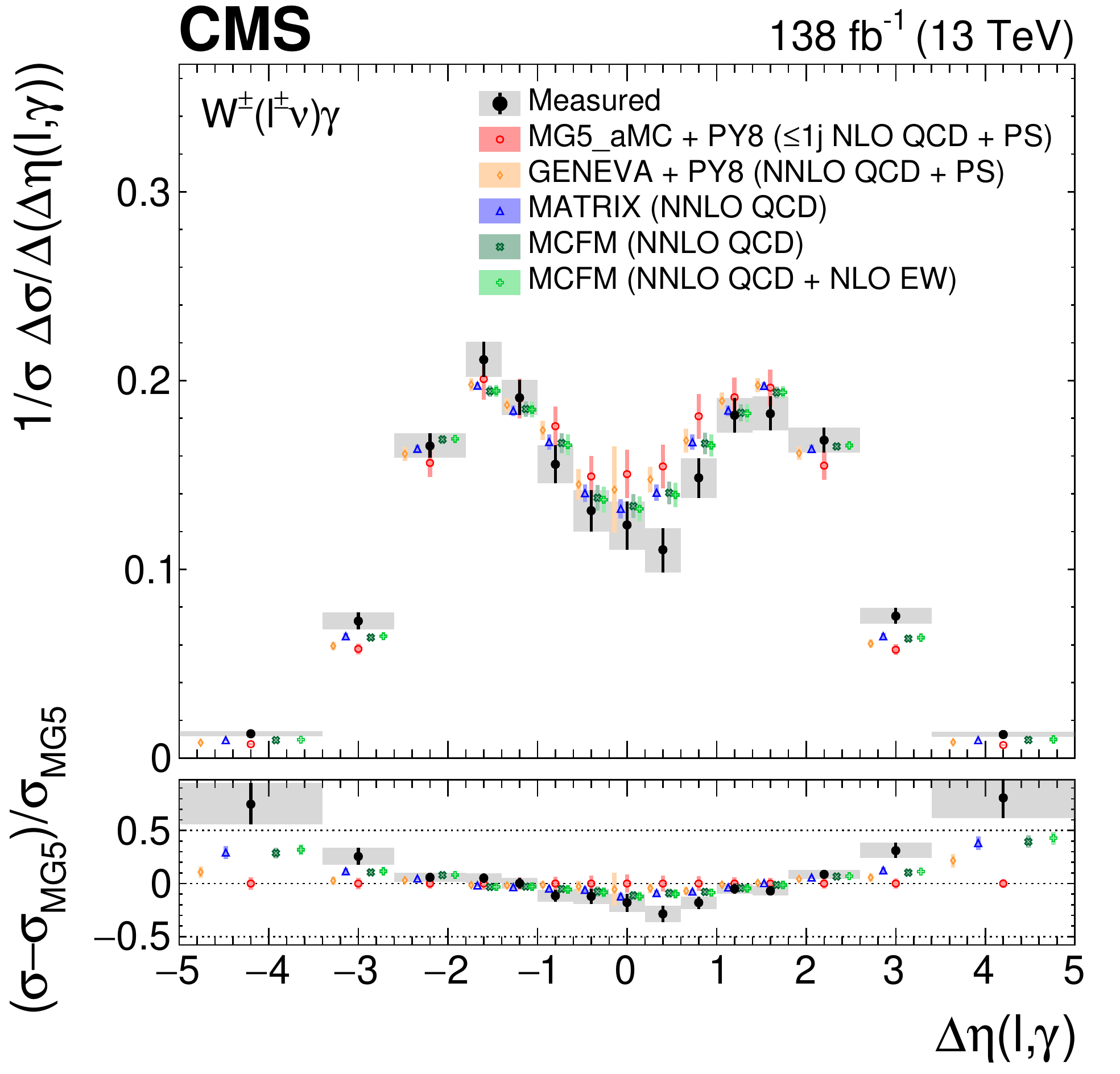}
\caption{The measured \DetaLG absolute (left) and fractional (right) differential cross sections, compared with the \MGPYSHORT{}, \GENEVA{}, \MATRIX{}, and \MCFM{} predictions. The black vertical bars give the total uncertainty on each measurement. The predictions are offset horizontally in each bin to improve visibility, and the corresponding vertical bars show the missing higher-order correction uncertainties.}\label{fig:res_1d_fid_l0p0_deta_binned_jvetomt_2}
\end{figure*}

To observe the RAZ effect in data it is necessary to add further requirements to the baseline selection.
A veto on the presence of any jet with $\pt>30\GeV$ and $\aeta<2.5$ is applied to preferentially select events in the Born configuration, and a requirement that $\mTcluster > 150\GeV$.
Figure~\ref{fig:res_1d_fid_l0p0_deta_binned_jvetomt_2} shows the differential cross section measurements, where a pronounced dip at $\DetaLG = 0$ is observed.
The jet veto in this selection gives a similar difference in normalization between measurement and predictions as for the 0-jet cross section in the previous section.
In addition, the dip at $\DetaLG = 0$ is larger than the predictions, though closer to the NNLO calculations, and the enhancement in the high $\abs{\DetaLG}$ region is similar to that observed under the baseline selection in Fig.~\ref{fig:res_1d_fid_l0p0_deta_binned_2}.

\subsection{EFT constraints}\label{sec:results-eft}
In this section, constraints on the \CWWW coefficient are derived via a parameterization of the fiducial cross section in \ptG and \aphif, under the EFT selection described in Section~\ref{sec:event_sel}.
The bin boundaries are [150, 200, 300, 500, 800, 1500] $\GeV$ and [0, $\pi/6$, $\pi/3$, $\pi/2$], respectively.
For the corresponding 2D differential cross section measurement the two highest \pt bins are merged, owing to the small number of events expected in the highest bin.
As for the one-dimensional (1D) differential measurements the signal model is constructed following Eq.~(\ref{eqn:nsig_full}), where $j$ runs over all \ptG and \aphif bins.

Constraints on \CWWW are determined by parameterizing the fiducial cross section in each bin according to Eq.~(\ref{eqn:xsec}).
Table~\ref{tab:c3w_scaling_rel} shows the corresponding values of the \sigInt and \sigBSM terms, relative to \sigSM.

\begin{table*}[htbp]
\centering
\topcaption{Coefficients of the fiducial cross section scaling terms in all $\ptG{\times}\aphif$ bins. Values are given relative to the SM prediction: $\muInt = \sigInt / \sigSM$ and $\muBSM = \sigBSM / \sigSM$, with the total relative normalization defined as $1 + \CWWW\muInt + \CWWW^{2}\muBSM$.}
\begin{scotch}{l c c c c c c}
\multirow{2}{*}{$\ptG$ bin ($\GeVns$) } & \multicolumn{2}{c}{$0 \leq \aphif < \pi/6$} & \multicolumn{2}{c}{$\pi/6 \leq \aphif < \pi/3$} & \multicolumn{2}{c}{$\pi/3 \leq \aphif < \pi/2$} \\
 &  \muInt  & \muBSM & \muInt & \muBSM & \muInt & \muBSM \\
\hline
150--200  & $-0.19 $ & 0.52   & 0.03 & 0.50   & 0.23 & 0.44  \\
200--300  & $-0.38 $ & 2.5   & 0.02 & 2.1   & 0.43 & 1.9  \\
300--500  & $-0.95 $ & 10.7  & 0.06 & 10.3  & 1.0 & 11.0 \\
500--800  & $-2.2 $ & 83.0  & 0.07 & 82.5  & 2.4 & 81.6 \\
800--1500 & $-4.9 $ & 688.5 & 0.02 & 651.7 & 4.9 & 646.2\\
\end{scotch}\label{tab:c3w_scaling_rel}
\end{table*}

\begin{figure*}[htbp]
\centering
\includegraphics[width=\cmsFigWidthT]{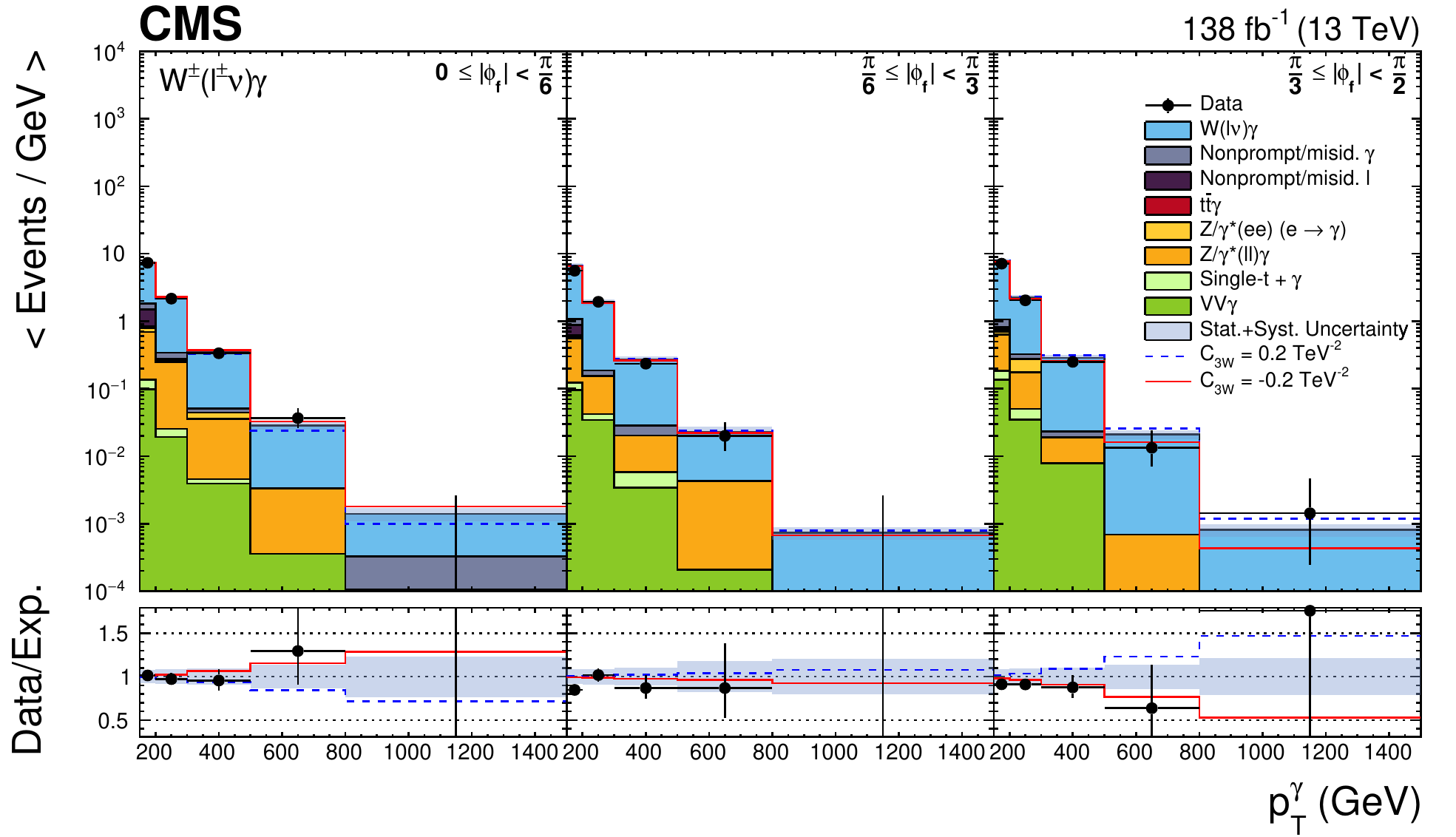}
\caption{The expected and observed \ptG distribution in each \aphif region before the maximum likelihood fit is performed, combining the electron and muon channels. The horizontal and vertical bars associated to the data points correspond to the bin widths and statistical uncertainties, respectively. The shaded uncertainty band incorporates all statistical and systematic uncertainties. The red and blue lines show how the total expectation changes when \CWWW is set to $-0.2$ and $0.2\TeV^{-2}$, respectively. Only the SM and interference terms are included in this example.}\label{fig:res_prefit_2d}
\end{figure*}

Figure~\ref{fig:res_prefit_2d} shows the expected and observed distributions, summed over the electron and muon channels, before the maximum likelihood fit is performed.
The change in the expected distribution when \CWWW is set to either $-0.2$ or $0.2\TeV^{-2}$ is also shown, considering only the SM and interference terms in Eq.~(\ref{eqn:xsec}).

\begin{figure}[htbp]
\centering
\includegraphics[width=\cmsFigWidthS]{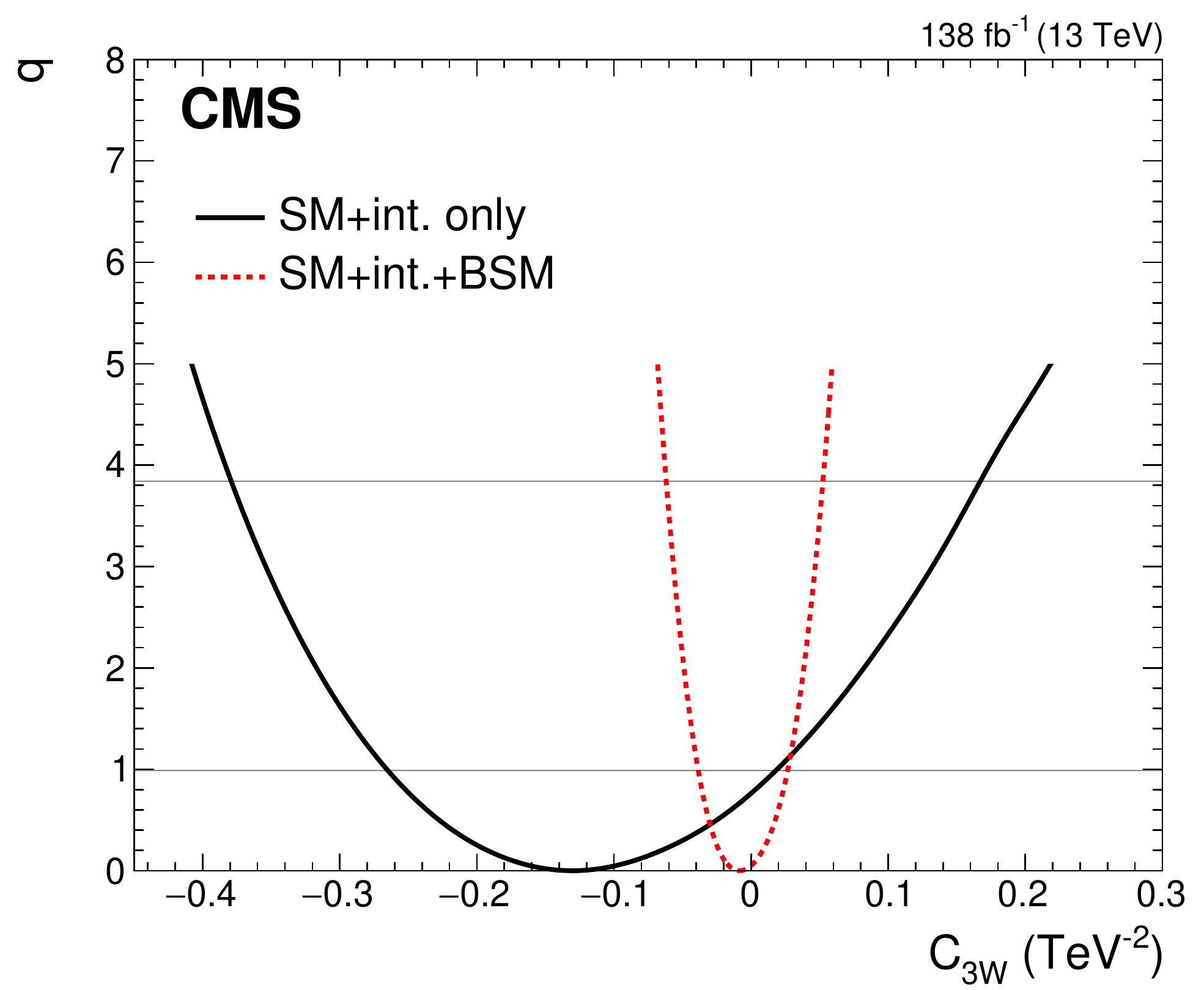}
\caption{Scans of the profile likelihood test statistic $q$ as a function of \CWWW, given with and without the pure BSM term by the dashed red and solid black lines, respectively. The full set of \ptG and \aphif bins, described in the text, are included for these scans.}\label{fig:res_nll_scan}
\end{figure}

\begin{figure}[htbp]
\centering
\includegraphics[width=0.48\textwidth]{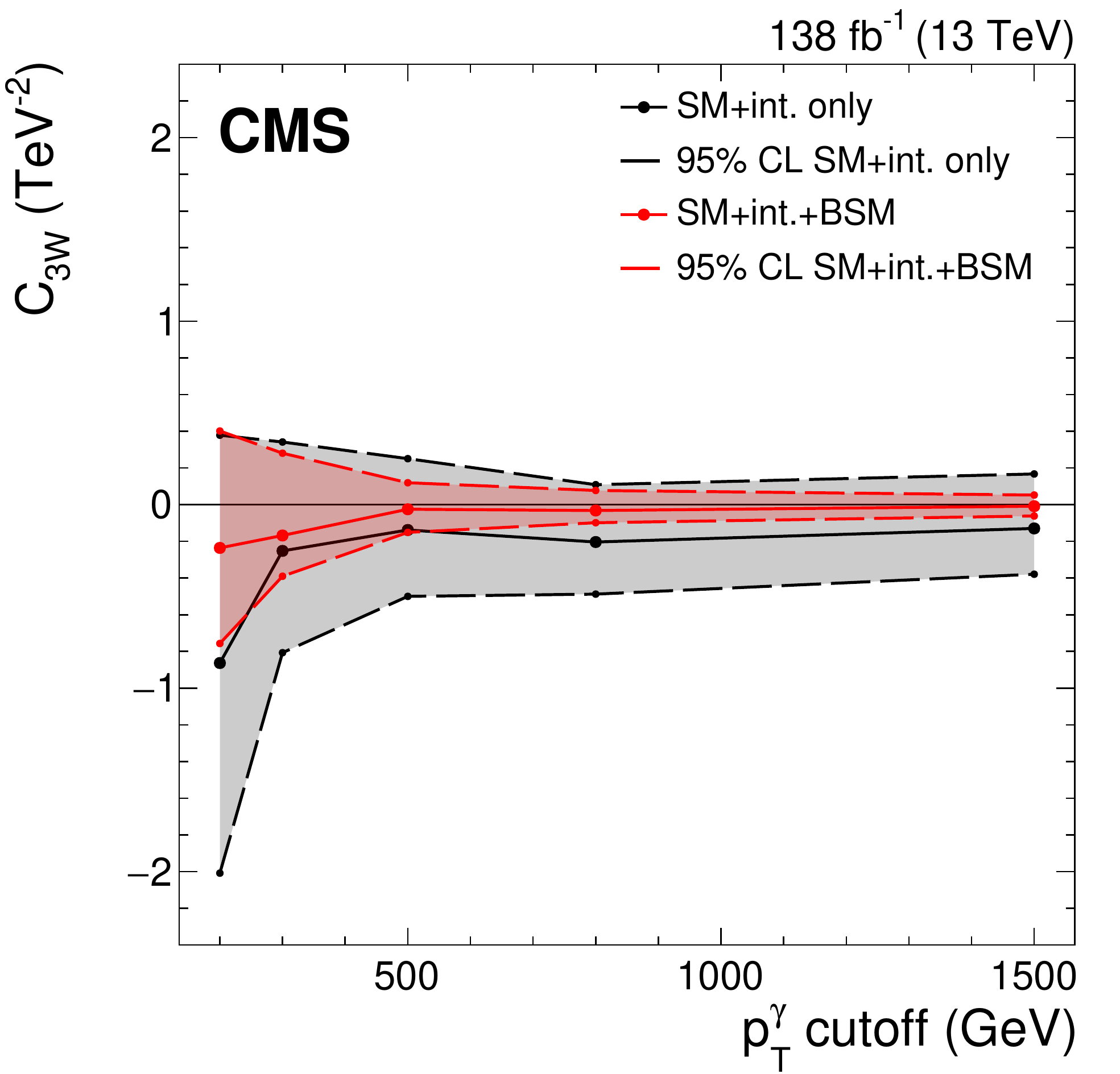}
\includegraphics[width=0.48\textwidth]{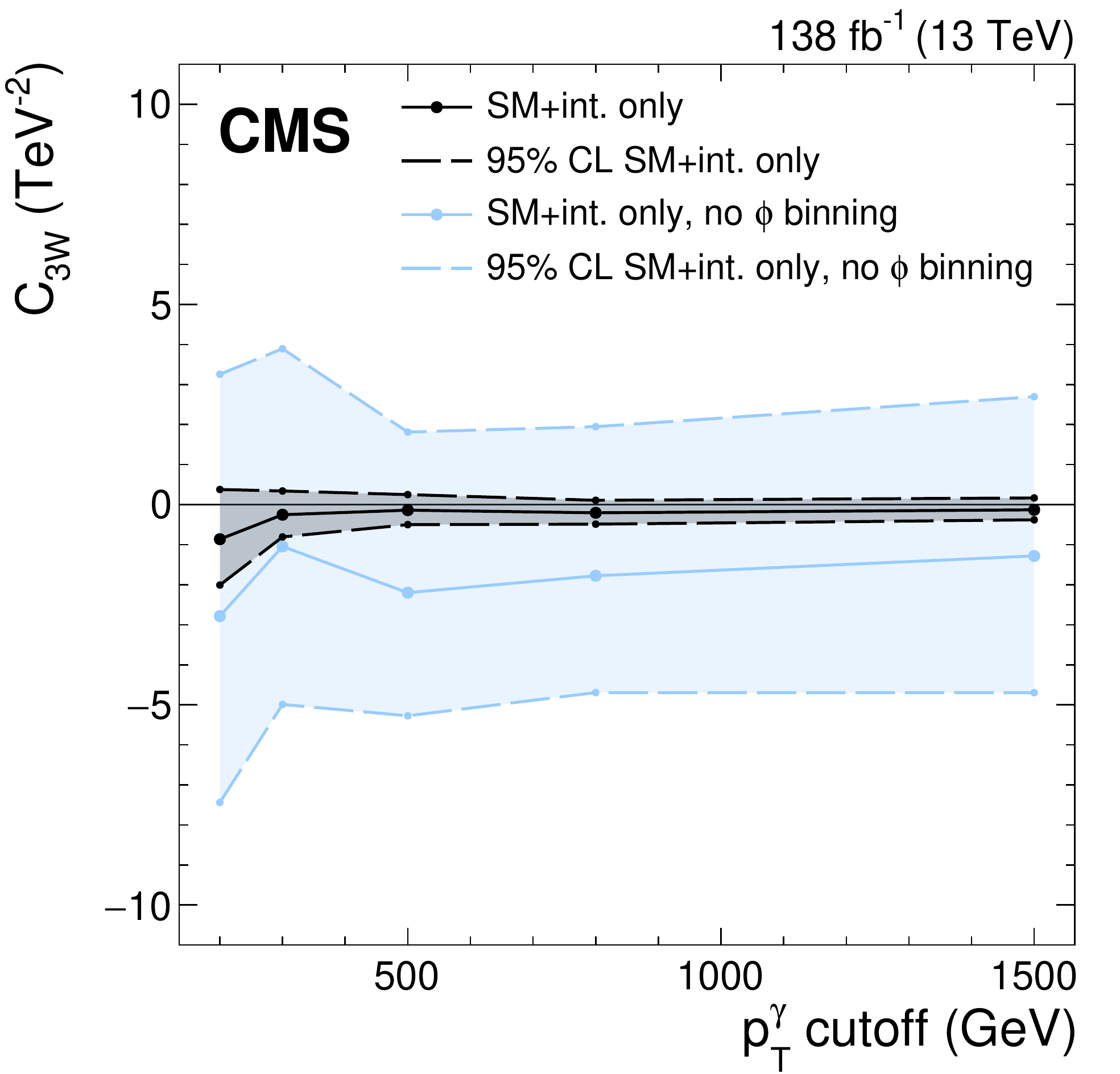}
\caption{Best-fit values of \CWWW and the corresponding 95\% \CL confidence intervals as a function of the maximum \ptG bin included in the fit (\cmsLeft). Measurements with and without the pure BSM term are given by the black and red lines, respectively. The limits without the pure BSM term given with and without the binning in \aphif are also shown (\cmsRight), with black and blue lines, respectively. Please note the different vertical scales; the black lines in both figures correspond to the same limits.}\label{fig:res_c3w_limits}
\end{figure}

The 95\% confidence level (\CL) intervals on \CWWW are calculated using the asymptotic properties of the profile likelihood ratio of Eq.~(\ref{eq:LH}).
Figure~\ref{fig:res_nll_scan} shows scans of this ratio as a function of \CWWW.
The intervals are $[-0.062, 0.052] \TeV^{-2}$ with the inclusion of the pure BSM term in Eq.~(\ref{eqn:xsec}) and $[-0.38,  0.17] \TeV^{-2}$ with only the SM and interference terms.
The observation is therefore compatible with the SM prediction of $\CWWW = 0$.
Constraints are also determined as a function of the maximum \ptG bin that is included in the fit, shown in Fig.~\ref{fig:res_c3w_limits} (\cmsLeft) and Table~\ref{tab:res_c3w_limits}.
Such a presentation is useful for interpretations where the highest bins may be beyond the validity of the EFT or specific BSM model being tested~\cite{Contino:2016jqw}.
These constraints are also determined with and without the inclusion of the pure BSM term.
This shows that with the current amount of data the pure BSM term dominates for all values of the cutoff.
However, the sensitivity to the interference term is comparable at lower values of \ptG.

\begin{table*}[htbp]
\centering
\topcaption{Best fit values of \CWWW and corresponding 95\% \CL confidence intervals as a function of the maximum \ptG bin included in the fit.}
\cmsTable{
\begin{scotch}{l c c c c c c}
\ptG cutoff ($\GeVns$)      &   \multicolumn{2}{c}{Best fit \CWWW (${\TeVns}^{-2}$)}  &  \multicolumn{2}{c}{Observed 95\% \CL (${\TeVns}^{-2}$)} & \multicolumn{2}{c}{Expected 95\% \CL (${\TeVns}^{-2}$)} \\
       & SM+int.\ only & SM+int.+BSM &  SM+int.\ only &  SM+int.+BSM &  SM+int.\ only &  SM+int.+BSM \\
\hline
 200     &  $-0.86$ &  $-0.24$   &  $[-2.01,  0.38]$ &  $[-0.76, 0.40]$    &  $[-1.16,  1.27]$ &  $[-0.81, 0.71]$  \\
 300     &  $-0.25$ &  $-0.17$   &  $[-0.81,  0.34]$ &  $[-0.39, 0.28]$    &  $[-0.56,  0.60]$ &  $[-0.33, 0.33]$  \\
 500     &  $-0.13$ &  $-0.025$  &  $[-0.50,  0.25]$ &  $[-0.15, 0.12]$    &  $[-0.35,  0.38]$ &  $[-0.17, 0.16]$  \\
 800     &  $-0.20$ &  $-0.033$  &  $[-0.49,  0.11]$ &  $[-0.10, 0.08]$    &  $[-0.29,  0.31]$ &  $[-0.097, 0.095]$  \\
1500     &  $-0.13$ &  $-0.009$  &  $[-0.38,  0.17]$ &  $[-0.062, 0.052]$  &  $[-0.27,  0.29]$ &  $[-0.066, 0.065]$  \\
\end{scotch}}\label{tab:res_c3w_limits}
\end{table*}

The constraint including the pure BSM term is similar in value to that of Ref.~\cite{CMS:2021foa}, after accounting for a difference in the normalization of \CWWW due to the use of an alternative EFT basis.
However, the main feature of this result is the ability to constrain \CWWW when only the interference term is considered.
To demonstrate this, the constraints are also determined without the binning in \aphif applied.
The same set of \ptG bins are used, but the three bins in \aphif are replaced by a single bin integrating over the full $\aphif$ range.
The result is shown in Fig.~\ref{fig:res_c3w_limits} (\cmsRight), without the inclusion of the pure BSM term.
This demonstrates the significant improvement, by up to a factor of ten, that the \aphif binning gives to this constraint.

The interference-only confidence intervals on $\CWWW$ can be compared with measurements of other processes that are sensitive to the \OWWW operator.
An ATLAS measurement of $\PW\PW$ production~\cite{ATLAS:2021jgw}, in which the coefficient is denoted $c_{W}$, uses an alternative technique to improve sensitivity to the interference.
The presence of an additional high-\pt jet is required, which partially mitigates the helicity suppression effect~\cite{Azatov:2017kzw}, and gives an interference-only sensitivity that is around a factor of eight lower.
The ATLAS measurement of electroweak \PZ boson production in association with two jets~\cite{ATLAS:2020nzk} gives comparable sensitivity to our result.
It exploits the distribution of the azimuthal angle between the jets, which is sensitive to the interference contribution.

\begin{figure}[htbp]
\centering
\includegraphics[width=\cmsFigWidthR]{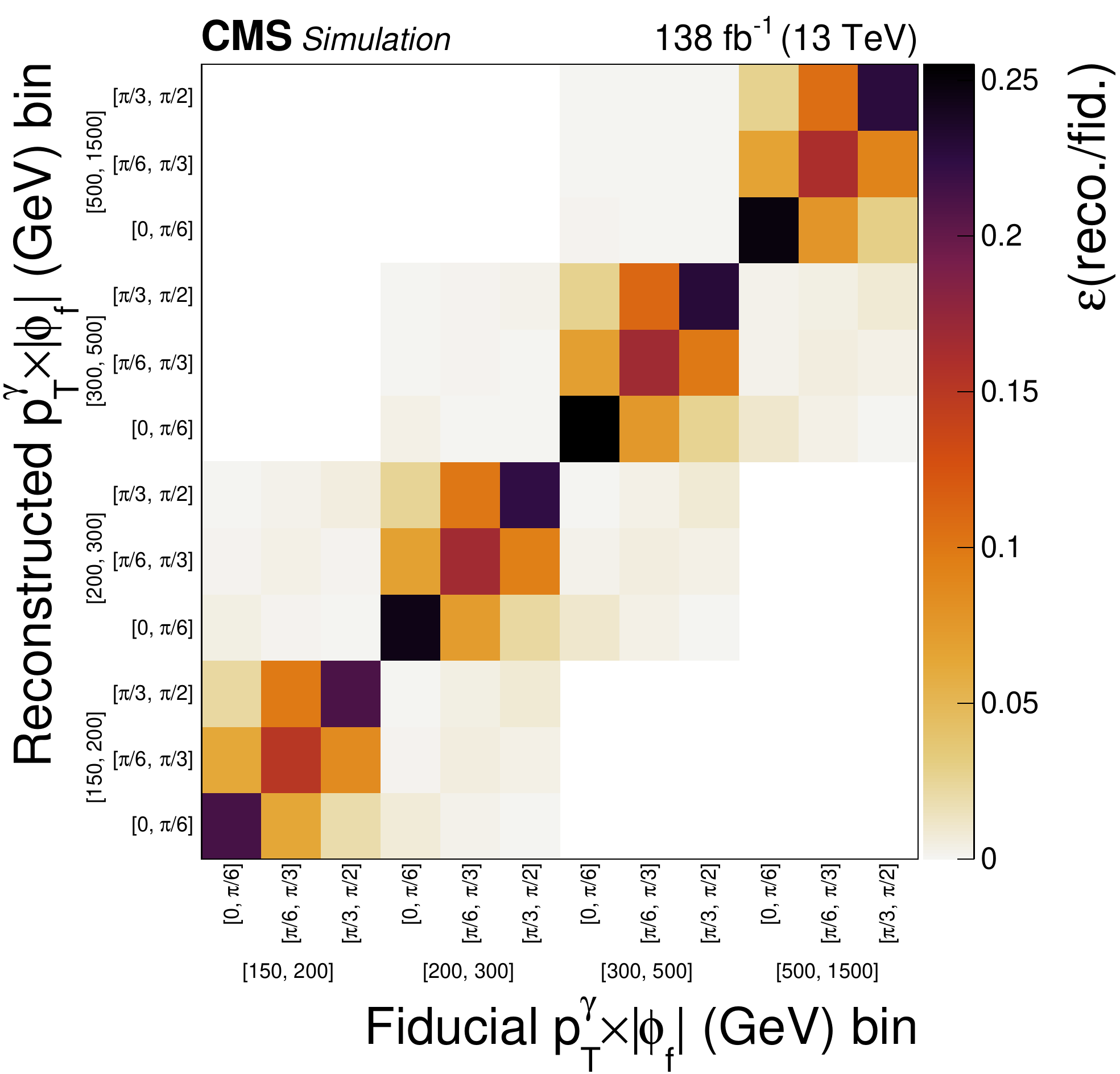}
\caption{Response matrix for the differential $\ptG{\times}\aphif$ cross section measurement. The entry in each bin gives the probability for an event of a given particle-level fiducial bin to be reconstructed in one of the corresponding reconstruction-level bins. The inner labels give the \aphif bin and the outer labels indicate the \ptG bin.}\label{fig:res_response_2d}
\end{figure}

The response matrix \Rij for the 2D differential cross section measurement of \ptG and \aphif is shown in Fig.~\ref{fig:res_response_2d}.
Although the migration between \ptG bins is small, the migration between \aphif bins is larger, owing both to the limited \ptmiss resolution and the fundamental limitations of the method used to reconstruct the neutrino four-momentum via the \mW pole mass constraint.

\begin{figure*}[!htbp]
\centering
\includegraphics[width=\cmsFigWidthQ]{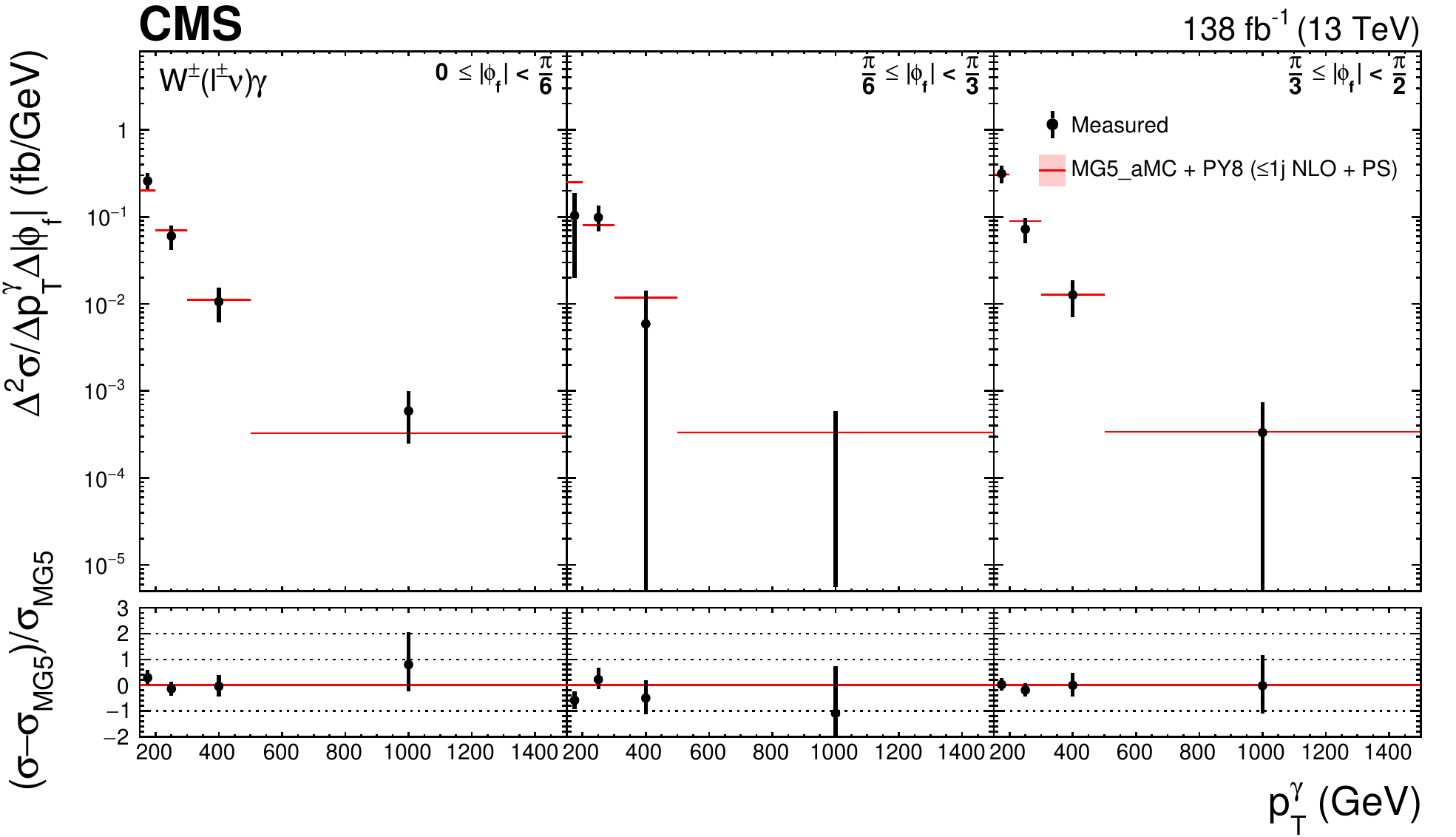}
\caption{Measured double-differential $\ptG{\times}\aphif$ cross section and comparison to the \MGPYSHORT{} NLO prediction. The black vertical bars give the total uncertainty on each measurement. The red shaded bands give the missing higher-order correction uncertainties in the prediction.}\label{fig:res_xsec_2d}
\end{figure*}

\begin{figure*}[!htbp]
\centering
\includegraphics[width=\cmsFigWidthP]{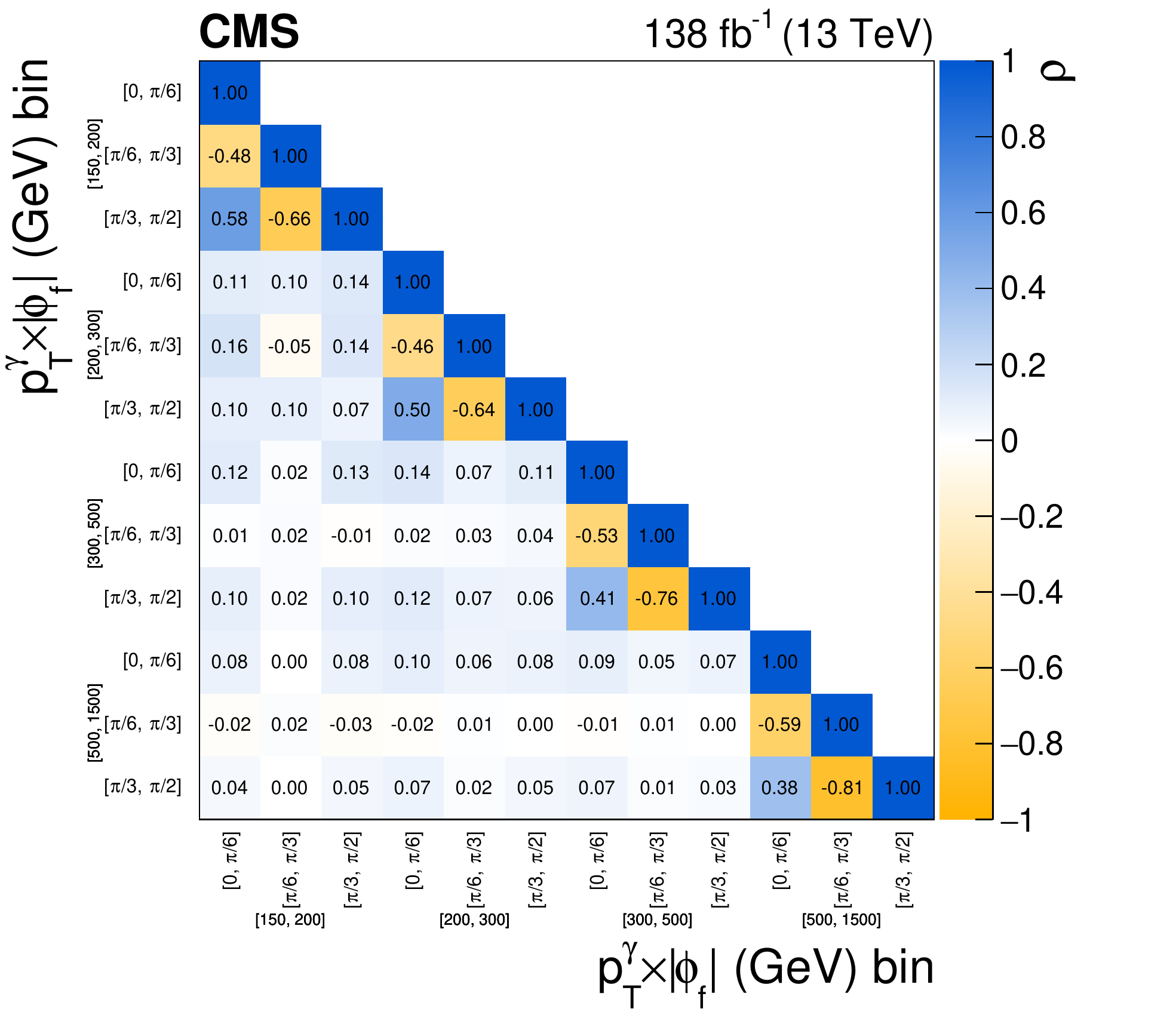}
\caption{Correlation matrix for the measured $\ptG{\times}\aphif$ cross sections.}\label{fig:res_corr_2d}
\end{figure*}

The resulting cross section measurements are shown in Fig.~\ref{fig:res_xsec_2d}.
The measured values are compared with the prediction from the NLO \MGPYSHORT{} simulation.
The correlation matrix is presented in Fig.~\ref{fig:res_corr_2d}.
Unlike the 1D \ptG cross section, the correlations between different \ptG bins are relatively small, since these measurements at high \ptG are much more dominated by the statistical uncertainties.
For a given \ptG bin the (anti-)correlation between \aphif bins is larger, owing to the migration in the response matrix discussed previously.

\section{Summary}\label{sec:summary}
This paper has presented an analysis of \WGamma production in $\sqrt{s} = 13\TeV$ proton-proton collisions using data recorded with the CMS detector at the LHC, corresponding to an integrated luminosity of 138\fbinv.
Differential cross sections have been measured for several observables and compared with standard model (SM) predictions computed at next-to-leading and next-to-next-to-leading orders in perturbative quantum chromodynamics.
The radiation amplitude zero effect, caused by interference between \WGamma production diagrams, has been studied via a measurement of the pseudorapidity difference between the lepton and the photon.

Constraints on the presence of \TeVns-scale new physics affecting the \WWGamma vertex have been determined using an effective field theory framework.
Confidence intervals on the Wilson coefficient \CWWW, determined at the 95\% confidence level, are $[-0.062, 0.052] \TeV^{-2}$ with the inclusion of the interference and pure beyond the SM contributions, and $[-0.38,  0.17] \TeV^{-2}$ when only the interference is considered.
A novel two-dimensional approach is used with the simultaneous measurement of the photon transverse momentum and the azimuthal angle of the charged lepton in a special reference frame.
With this method, the sensitivity to the interference between the SM and the \OWWW operator is enhanced by up to a factor of ten compared to a measurement using the transverse momentum alone.
This improves the validity of the result, as the dependence on the missing higher-order contributions in the EFT expansion is reduced.
The technique will also be valuable in the future when sufficiently small values of \CWWW are probed such that the interference contribution will be dominant.

\begin{acknowledgments}
    We congratulate our colleagues in the CERN accelerator departments for the excellent performance of the LHC and thank the technical and administrative staffs at CERN and at other CMS institutes for their contributions to the success of the CMS effort. In addition, we gratefully acknowledge the computing   centers and personnel of the Worldwide LHC Computing Grid and other   centers for delivering so effectively the computing infrastructure essential to our analyses. Finally, we acknowledge the enduring support for the construction and operation of the LHC, the CMS detector, and the supporting computing infrastructure provided by the following funding agencies: BMBWF and FWF (Austria); FNRS and FWO (Belgium); CNPq, CAPES, FAPERJ, FAPERGS, and FAPESP (Brazil); MES and BNSF (Bulgaria); CERN; CAS, MoST, and NSFC (China); MINCIENCIAS (Colombia); MSES and CSF (Croatia); RIF (Cyprus); SENESCYT (Ecuador); MoER, ERC PUT and ERDF (Estonia); Academy of Finland, MEC, and HIP (Finland); CEA and CNRS/IN2P3 (France); BMBF, DFG, and HGF (Germany); GSRI (Greece); NKFIA (Hungary); DAE and DST (India); IPM (Iran); SFI (Ireland); INFN (Italy); MSIP and NRF (Republic of Korea); MES (Latvia); LAS (Lithuania); MOE and UM (Malaysia); BUAP, CINVESTAV, CONACYT, LNS, SEP, and UASLP-FAI (Mexico); MOS (Montenegro); MBIE (New Zealand); PAEC (Pakistan); MSHE and NSC (Poland); FCT (Portugal); JINR (Dubna); MON, RosAtom, RAS, RFBR, and NRC KI (Russia); MESTD (Serbia); SEIDI, CPAN, PCTI, and FEDER (Spain); MOSTR (Sri Lanka); Swiss Funding Agencies (Switzerland); MST (Taipei); ThEPCenter, IPST, STAR, and NSTDA (Thailand); TUBITAK and TAEK (Turkey); NASU (Ukraine); STFC (United Kingdom); DOE and NSF (USA).
    
    \hyphenation{Rachada-pisek} Individuals have received support from the Marie-Curie program and the European Research Council and Horizon 2020 Grant, contract Nos.\ 675440, 724704, 752730, 758316, 765710, 824093, 884104, and COST Action CA16108 (European Union); the Leventis Foundation; the Alfred P.\ Sloan Foundation; the Alexander von Humboldt Foundation; the Belgian Federal Science Policy Office; the Fonds pour la Formation \`a la Recherche dans l'Industrie et dans l'Agriculture (FRIA-Belgium); the Agentschap voor Innovatie door Wetenschap en Technologie (IWT-Belgium); the F.R.S.-FNRS and FWO (Belgium) under the ``Excellence of Science -- EOS" -- be.h project n.\ 30820817; the Beijing Municipal Science \& Technology Commission, No. Z191100007219010; the Ministry of Education, Youth and Sports (MEYS) of the Czech Republic; the Deutsche Forschungsgemeinschaft (DFG), under Germany's Excellence Strategy -- EXC 2121 ``Quantum Universe" -- 390833306, and under project number 400140256 - GRK2497; the Lend\"ulet (``Momentum") Program and the J\'anos Bolyai Research Scholarship of the Hungarian Academy of Sciences, the New National Excellence Program \'UNKP, the NKFIA research grants 123842, 123959, 124845, 124850, 125105, 128713, 128786, and 129058 (Hungary); the Council of Science and Industrial Research, India; the Latvian Council of Science; the Ministry of Science and Higher Education and the National Science Center, contracts Opus 2014/15/B/ST2/03998 and 2015/19/B/ST2/02861 (Poland); the Funda\c{c}\~ao para a Ci\^encia e a Tecnologia, grant CEECIND/01334/2018 (Portugal); the National Priorities Research Program by Qatar National Research Fund; the Ministry of Science and Higher Education, projects no. 14.W03.31.0026 and no. FSWW-2020-0008, and the Russian Foundation for Basic Research, project No.19-42-703014 (Russia); the Programa Estatal de Fomento de la Investigaci{\'o}n Cient{\'i}fica y T{\'e}cnica de Excelencia Mar\'{\i}a de Maeztu, grant MDM-2015-0509 and the Programa Severo Ochoa del Principado de Asturias; the Stavros Niarchos Foundation (Greece); the Rachadapisek Sompot Fund for Postdoctoral Fellowship, Chulalongkorn University and the Chulalongkorn Academic into Its 2nd Century Project Advancement Project (Thailand); the Kavli Foundation; the Nvidia Corporation; the SuperMicro Corporation; the Welch Foundation, contract C-1845; and the Weston Havens Foundation (USA).
\end{acknowledgments}

\bibliography{auto_generated}
\cleardoublepage \appendix\section{The CMS Collaboration \label{app:collab}}\begin{sloppypar}\hyphenpenalty=5000\widowpenalty=500\clubpenalty=5000\cmsinstitute{Yerevan~Physics~Institute, Yerevan, Armenia}
A.~Tumasyan
\cmsinstitute{Institut~f\"{u}r~Hochenergiephysik, Vienna, Austria}
W.~Adam\cmsorcid{0000-0001-9099-4341}, J.W.~Andrejkovic, T.~Bergauer\cmsorcid{0000-0002-5786-0293}, S.~Chatterjee\cmsorcid{0000-0003-2660-0349}, M.~Dragicevic\cmsorcid{0000-0003-1967-6783}, A.~Escalante~Del~Valle\cmsorcid{0000-0002-9702-6359}, R.~Fr\"{u}hwirth\cmsAuthorMark{1}, M.~Jeitler\cmsAuthorMark{1}\cmsorcid{0000-0002-5141-9560}, N.~Krammer, L.~Lechner\cmsorcid{0000-0002-3065-1141}, D.~Liko, I.~Mikulec, P.~Paulitsch, F.M.~Pitters, J.~Schieck\cmsAuthorMark{1}\cmsorcid{0000-0002-1058-8093}, R.~Sch\"{o}fbeck\cmsorcid{0000-0002-2332-8784}, M.~Spanring\cmsorcid{0000-0001-6328-7887}, S.~Templ\cmsorcid{0000-0003-3137-5692}, W.~Waltenberger\cmsorcid{0000-0002-6215-7228}, C.-E.~Wulz\cmsAuthorMark{1}\cmsorcid{0000-0001-9226-5812}
% ...
\vskip\cmsinstskip
% ...
1:~Also at TU~Wien, Wien, Austria\\
% ...
\end{sloppypar}
\end{document}